\documentclass[preprint2]{aastex62}
\usepackage{amsmath}
\usepackage{color}
\begin{document}

\title{Utilizing Small Telescopes Operated by Citizen Scientists for Transiting Exoplanet Follow-up}

\email{rzellem@jpl.nasa.gov}
\author[0000-0001-7547-0398]{Robert T. Zellem}
\affil{Jet Propulsion Laboratory, California Institute of Technology, 4800 Oak Grove Drive, Pasadena, California, 91109, USA}

\author[0000-0002-5785-9073]{Kyle A. Pearson}
\affil{Lunar and Planetary Laboratory, University of Arizona, 1629 E University Blvd, Tucson, Arizona 85721, USA}
\affil{Jet Propulsion Laboratory, California Institute of Technology, 4800 Oak Grove Drive, Pasadena, California, 91109, USA}

\author{Ethan Blaser}
\affil{University of Virginia, Charlottesville, Virginia 22904, USA}
\affil{Jet Propulsion Laboratory, California Institute of Technology, 4800 Oak Grove Drive, Pasadena, California, 91109, USA}

\author{Martin Fowler}
\affil{Citizen Scientist, Les Rocquettes, Orchard Road, South Wonston, Winchester SO21 3EX, UK}

\author[0000-0002-5741-3047]{David R. Ciardi}
\affil{NASA Exoplanet Science Institute/California Institute of Technology, MC 314-6, 1200 E California Blvd, Pasadena, California 91125, USA}

\author{Anya Biferno}
\affil{Jet Propulsion Laboratory, California Institute of Technology, 4800 Oak Grove Drive, Pasadena, California, 91109, USA}


\author{Bob Massey}
\affil{American Association of Variable Star Observers, 49 Bay State Rd, Cambridge, Massachusetts 02138, USA}

\author[0000-0001-7016-7277]{Franck Marchis}
\affil{SETI Institute, 189 Bernardo Ave, Suite 200, Mountain View, California, 94043, USA}
\affil{Unistellar, 19 rue Vacon, 13001 Marseille, France}

\author{Robert Baer}
\affil{Southern Illinois University Carbondale, MC 4401, 1245 Lincoln Dr,  Carbondale, Illinois 62901, USA}
\affil{Citizen CATE Experiment}

\author{Conley Ball}
\affil{Laguna Blanca School, 4125 Paloma Dr, Santa Barbara, CA 93110, USA}

\author{Mike Chasin}
\affil{Boyce Research Initiatives and Education Foundation}
\affil{San Diego Astronomy Association}

\author{Mike Conley}
\affil{Citizen CATE Experiment}

\author{Scott Dixon}
\affil{Boyce Research Initiatives and Education Foundation}
\affil{San Diego Astronomy Association}

\author{Elizabeth Fletcher}
\affil{Towson University, 8000 York Rd, Towson, Maryland, 21252, USA}

\author{Saneyda Hernandez}
\affil{Towson University, 8000 York Rd, Towson, Maryland, 21252, USA}

\author{Sujay Nair}
\affil{Stanford Online High School, 415 Broadway Academy Hall, Floor 2, 8853, Redwood City, CA 94063}

\author{Quinn Perian}
\affil{Stanford Online High School, 415 Broadway Academy Hall, Floor 2, 8853, Redwood City, CA 94063}

\author{Frank Sienkiewicz}
\affil{The Center for Astrophysics $\mid$ Harvard \& Smithsonian, 60 Garden Street, Cambridge, Massachusetts 02138, USA}

\author{Kal\'{e}e Tock}
\affil{Stanford Online High School, 415 Broadway Academy Hall, Floor 2, 8853, Redwood City, CA 94063}

\author{Vivek Vijayakumar }
\affil{Boyce Research Initiatives and Education Foundation}

\author[0000-0002-0919-4468]{Mark R. Swain}
\affil{Jet Propulsion Laboratory, California Institute of Technology, 4800 Oak Grove Drive, Pasadena, California, 91109, USA}

\author[0000-0002-7402-7797]{Gael M. Roudier}
\affil{Jet Propulsion Laboratory, California Institute of Technology, 4800 Oak Grove Drive, Pasadena, California, 91109, USA}

\author{Geoffrey Bryden}
\affil{Jet Propulsion Laboratory, California Institute of Technology, 4800 Oak Grove Drive, Pasadena, California, 91109, USA}

\author{Dennis M. Conti}
\affil{American Association of Variable Star Observers, 49 Bay State Rd, Cambridge, Massachusetts 02138, USA}

\author{Dolores H. Hill}
\affil{Lunar and Planetary Laboratory, University of Arizona, 1629 E University Blvd, Tucson, Arizona 85721, USA}

\author{Carl W. Hergenrother}
\affil{Lunar and Planetary Laboratory, University of Arizona, 1629 E University Blvd, Tucson, Arizona 85721, USA}

\author[0000-0002-8625-6474]{Mary Dussault}
\affil{The Center for Astrophysics $\mid$ Harvard \& Smithsonian, 60 Garden Street, Cambridge, Massachusetts 02138, USA}

\author[0000-0002-7084-0529]{Stephen~R.~Kane}
\affil{Department of Earth and Planetary Sciences, University of California, Riverside, CA 92521, USA}

\author{Michael Fitzgerald}
\affil{Edith Cowan University, 270 Joondalup Drive, Joondalup, WA 6027, Australia}

\author{Pat Boyce}
\affil{Boyce Research Initiatives and Education Foundation}

\author{Laura Peticolas}
\affil{Sonoma State University, 1801 East Cotati Ave, Rohnert Park, California 94928, USA}

\author[0000-0002-2931-7605]{Wilfred Gee}
\affil{Macquarie University, Macquarie University, Sydney, New South Wales 2109, AUS}

\author[0000-0003-2073-1065]{Lynn Cominsky}
\affil{Sonoma State University, 1801 East Cotati Ave, Rohnert Park, California 94928, USA}

\author{Rachel Zimmerman-Brachman}
\affil{Jet Propulsion Laboratory, California Institute of Technology, 4800 Oak Grove Drive, Pasadena, California, 91109, USA}

\author[0000-0001-6807-5015]{Denise Smith}
\affil{Space Telescope Science Institute, 3700 San Martin Drive, Baltimore, Maryland 21218, USA}

\author{Michelle J. Creech-Eakman}
\affil{Department of Physics, New Mexico Institute of Mining and Technology, 801 Leroy Place, Socorro, NM 87801, USA}

\author{John Engelke}
\affil{Raytheon Intelligence, Information, and Services, 300 N Lake Ave, Suite 1120, Pasadena, CA 91101}
\affil{Jet Propulsion Laboratory, California Institute of Technology, 4800 Oak Grove Drive, Pasadena, California, 91109, USA}

\author{Alexandra Iturralde}
\affil{The University of New Mexico, Albuquerque, New Mexico, 87131, USA}
\affil{Jet Propulsion Laboratory, California Institute of Technology, 4800 Oak Grove Drive, Pasadena, California, 91109, USA}

\author[0000-0003-2313-467X]{Diana Dragomir}
\affil{Massachusetts Institute of Technology, 77 Massachusetts Ave, Cambridge, Massachusetts 02139, USA}
\affil{NASA Hubble Fellow}
\affil{Department of Physics and Astronomy, University of New Mexico, Albuquerque, NM, USA}

\author{Nemanja Jovanovic}
\affil{California Institute of Technology, 1200 East California Boulevard, Pasadena, California 91125, USA}

\author[0000-0002-5972-9555]{Brandon Lawton}
\affil{Space Telescope Science Institute, 3700 San Martin Drive, Baltimore, Maryland 21218, USA}

\author{Emmanuel Arbouch}
\affil{Unistellar, 19 rue Vacon, 13001 Marseille, France}

\author[0000-0002-2387-5489]{Marc Kuchner}
\affil{NASA Goddard Space Flight Center, 8800 Greenbelt Rd, Greenbelt, Maryland 20771, USA}

\author{Arnaud Malvache}
\affil{Unistellar, 19 rue Vacon, 13001 Marseille, France}

\vspace{12pt}
\begin{abstract}
Due to the efforts by numerous ground-based surveys and NASA's Kepler and TESS, there will be hundreds, if not thousands, of transiting exoplanets ideal for atmospheric characterization via {{spectroscopy with large platforms such as JWST and ARIEL}}. However their {{next predicted mid-transit time}} could become so increasingly uncertain over time that significant overhead would be required to ensure the detection of the entire transit. As a result, follow-up observations to characterize these exoplanetary atmospheres would require less-efficient use of an observatory's time---which is an issue for large platforms where minimizing observing overheads is a necessity. Here we demonstrate the power of citizen scientists operating smaller observatories ($\le$1-m) to keep ephemerides ``fresh'', defined here as when the {{1$\sigma$}} uncertainty in the mid-transit time is less than half the transit duration. We advocate for the creation of a community-wide effort to perform ephemeris maintenance on transiting exoplanets by citizen scientists. Such observations can be conducted with even a 6-inch telescope, which has the potential to save up to $\sim$10,000~days for a 1000-planet survey. {{Based on a preliminary analysis of 14 transits from a single 6-inch MicroObservatory telescope, we empirically estimate the ability of small telescopes to benefit the community. Observations with a small-telescope network operated by citizen scientists are capable of resolving stellar blends to within 5''/pixel, can follow-up long period transits in short-baseline TESS fields, monitor epoch-to-epoch stellar variability at a precision 0.67\%$\pm$0.12\%  for a 11.3 V-mag star, and search for new planets or constrain the masses of known planets with transit timing variations greater than two minutes.}}

\end{abstract}

\vspace{12pt}
\section{Introduction}
NASA's Transiting Exoplanet Survey Satellite \citep[TESS;][]{ricker14} is predicted to discover 10,000+ transiting exoplanets in its full-frame images \citep{barclay18}, crucially providing hundreds of bright targets with large scale heights that are ideal for detailed follow-up spectroscopic characterization by Hubble, JWST, ARIEL, and other next-generation platforms \citep{zellem17, zellem19b, kempton18}. However, TESS will sample some of its targets for only 27 days and, even in the continuous viewing zone, TESS’s nominal mission will end before the launch of JWST, ARIEL, and an Astro2020 Decadal Mission (2021, 2028, and $\sim$2030, respectively). As a result, the ephemerides of many of these targets and all previously-discovered transiting exoplanets could become ``stale'', i.e., where the {{1$\sigma$}} uncertainty in the mid-transit time $\Delta T_{mid}$ exceeds half the transit duration $t_{dur}$ (assuming an observing strategy where the planet is observed the same amount of time in-transit as out-of-transit{{; Fig.~\ref{fig:lcsim}}}). Since this uncertainty increases with time, a significant amount of observing time must be invested to recover the complete transit. For example, a planet with an uncertainty of just 1~minute in both its orbital period and mid-transit time will have an uncertainty of $\sim$15~hours in its mid-transit time in 10~years {{(see Eqn.~\ref{eqn:dt_next};}} see also \citet{dragomir19}).

In some cases, the transit duration can be relatively long and observed from the ground provided that the uncertainty in the mid-transit time is much less than the transit duration. {{Even}} observing just the ingress or egress is still extremely valuable for bounding the solution space. {{Therefore even partial observations of}} long transit durations can still {{serve to help refine a transit's ephemeris}}, as has been done for the hot Jupiter HD~80606b \citep{fossey09, garciamelendo09}.

Given that JWST will study tens to 100--200 transiting exoplanets in detail \citep{cowan15, kempton18} and that ARIEL will survey $\sim$1000 transiting exoplanets \citep{tinetti16, tinetti18, edwards19, zellem19b}, ephemerides maintenance is required to keep predicted transit times ``fresh'' ($\Delta T_{mid} \leq 0.5t_{dur}$). Combating this problem requires follow-up observations from either ground- \citep[e.g.,][]{hellier19, mallonn19} or space-based platforms \citep[e.g., with Spitzer;][]{benneke17} to reduce the uncertainties on a planet’s orbital period and mid-transit time. Here we examine how the transit uncertainties increase as a function of time for the current TESS Objects of Interest (TOIs), confirmed TESS targets, and currently-known transiting exoplanets and illustrate the power of a network of small telescopes ($\leq1$~m) operated by citizen scientists to keep ephemerides fresh, provide comparatively-higher spatial resolution, help confirm long-period planets, monitor epoch-to-epoch stellar variability, and leverage transit timing variations to measure planetary masses or discover new planets.

\section{Transit Maintenance Is Critical for Atmospheric Characterization}
\subsection{Analytic Derivation of Mid-transit Uncertainty}
The time of an exoplanet's next transit or eclipse can be calculated via:
\begin{align}
T_{mid} = n_{orbit} \cdot P+T_{0}
\label{eqn:t_next}
\end{align}
where $T_{mid}$ is the time of an upcoming transit $n$ exoplanetary orbits $n_{orbit}$ in the future, $P$ is the orbital period of the planet, and $T_{0}$ is the planet’s mid-transit or mid-eclipse time. Performing error propagation on this equation, one finds:
\begin{align}
\label{eqn:dt_next_FFC}
\mathrm{var}(T_{mid}) =& n_{orbit}^{2} \cdot \mathrm{var}(P)\\ \nonumber 
&+ 2n_{orbit} \cdot \mathrm{cov}(P,T_{0})\\ \nonumber
&+\mathrm{var}(T_{0}) \\ \nonumber
\end{align}
and, after linearization:
\begin{align}
\label{eqn:dt_next}
\Delta T_{mid} =& \big(n_{orbit}^{2} \cdot \Delta P^{2}\\ \nonumber 
&+ 2n_{orbit} \cdot \Delta P \Delta T_{0}\\ \nonumber
&+\Delta T_{0}^{2}\big)^{1/2}. \\ \nonumber
\end{align}
Due to the $n_{orbit}$ term, the uncertainty in measuring the next transit $\Delta T_{mid}$ is mostly dependent upon the uncertainty in the orbital period $\Delta P$ compared to the uncertainty in the transit ephemeris $\Delta T_{0}$. In addition, short-period planets run the risk of becoming stale faster than long-period planets {{with similar uncertainties in their orbital period $\Delta P$ and mid-transit time $\Delta T_{0}$,}} as the number of elapsed orbits $n_{orbit}$ in a given amount of time is inversely proportional to the orbital period. Therefore, if an exoplanet has a sufficiently-large uncertainty in its mid-transit time or orbital period, then it will have a large uncertainty in its next expected mid-transit or mid-eclipse time. A similar expression has been independently derived by \citet{kane09}, \citet{dragomir19}, and \citet{mallonn19}. 

\begin{figure}[htbp]
		\includegraphics[width=1\columnwidth]{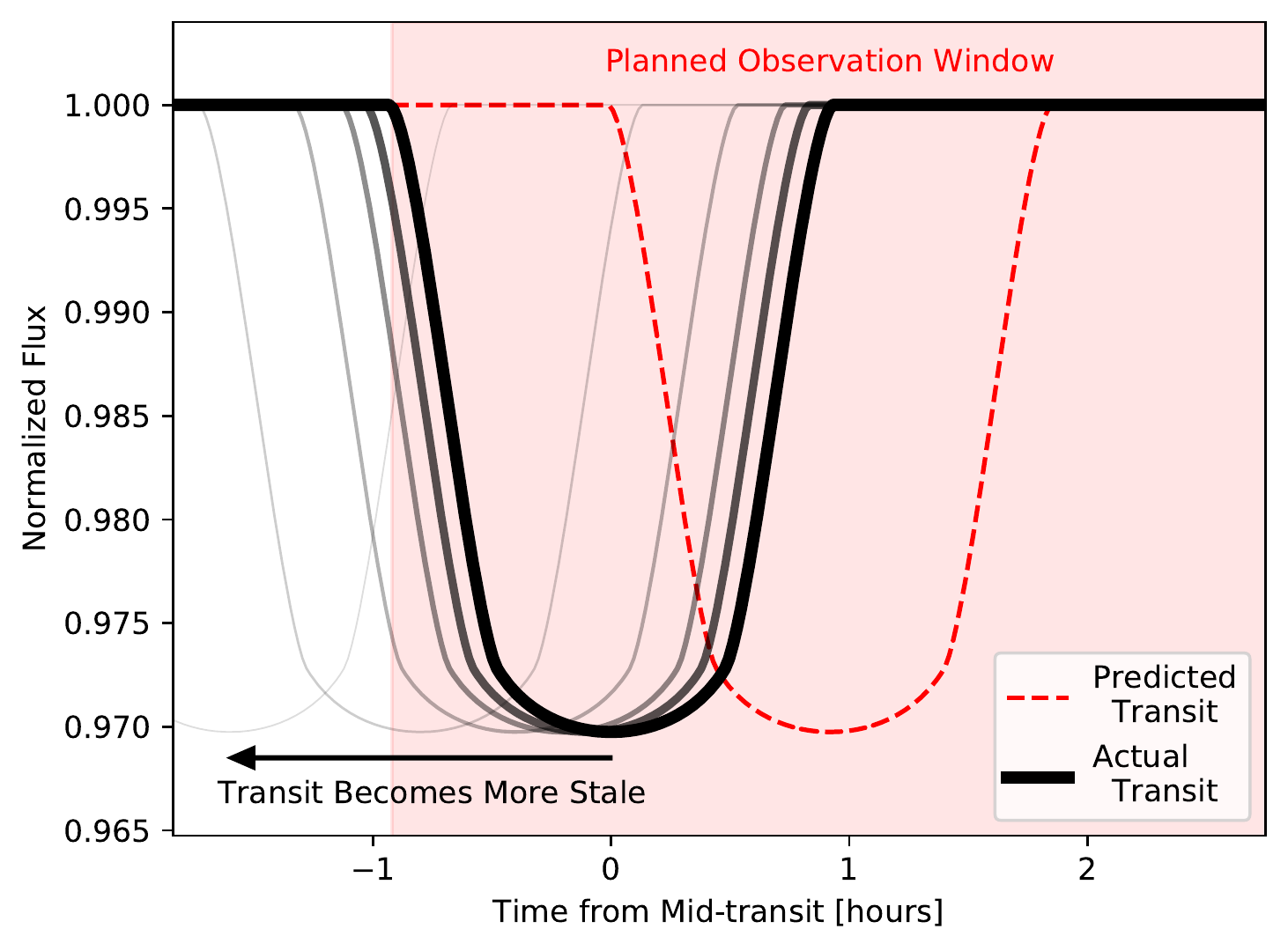}
		\caption{{{If the uncertainty in the mid-transit time is greater than or equal to half of the transit duration, then one runs the risk of not observing any pre-transit baseline (thereby making the identification and removal of systematic errors more difficult) or missing part or all of the transit. In the figure above, an astronomer has planned an observing run (red shaded region) based on a predicted transit time (dashed red line). However, due to uncertainties in the planet's period and mid-transit time, the midpoint of the upcoming transit actually occurs half a transit duration earlier than anticipated (solid black line), resulting in the lack of pre-transit baseline.}}}
	\label{fig:lcsim}
\end{figure}

Assuming an observing strategy where an equal amount of time is spent integrating both in- and out-of-transit, then we define here that a transiting exoplanet's ephemeris will become ``stale'' when the {{1$\sigma$}} mid-transit uncertainty $\Delta T_{mid}$ exceeds half the transit duration $t_{dur}$\footnote{Note that we base our definition on the 1$\sigma$ mid-transit uncertainty. However, if a publication has underestimated their 1$\sigma$ uncertainties, then this definition will falsely define a transiting exoplanet as fresh rather than stale; thus one could more conservatively adopt the 3$\sigma$ mid-transit uncertainties to identify stale targets instead. We proceed with the 1$\sigma$ mid-transit uncertainty definition for the rest of this study.}. Such a scenario would run the risk of either partially or completely missing the transit {{(Fig.~\ref{fig:lcsim})}}. We can further expand Equation~\ref{eqn:dt_next}:
\begin{align}
\Delta T_{mid} =& \big(n_{orbit}^{2} \cdot \Delta P^{2}\\ \nonumber 
&+ 2n_{orbit} \cdot \Delta P \Delta T_{0}\\ \nonumber
&+\Delta T_{0}^{2}\big)^{1/2} \\ \nonumber
&\geq 0.5t_{dur}
\end{align}
and thus derive a figure of merit (FOM) that allows one to quickly determine when a particular transit runs the risk of becoming stale, requiring follow-up observations to reduce the uncertainties in the orbital period $\Delta P$ and mid-transit time $\Delta T_{0}$.

One can also implement the figure of merit that ranks targets ideal for detailed spectral characterization, as independently derived by \citet{cowan15}, \citet{zellem17}, \citet{goyal18}, \citet{kempton18}, and \citet{morgan18}:
\begin{align}
\mathrm{FOM_{transit}} &= \frac{2H_{s}R_{p}R_{s}^{-2}}{10^{0.2 H\text{-}mag}}.
\label{eqn:transit_FOM}
\end{align}
where $H_{s}$ is the planet's scale height, $R_{p}$ and $R_{s}$ are the planet and host star's radii, respectively, and $H\text{-}mag$ is the host star's H-band magnitude. $\mathrm{FOM_{transit}}$ prioritizes planets with large atmospheric spectral modulation and bright host stars and are most likely to be followed-up for detailed atmospheric spectral characterization with future missions; since these targets typically have larger radii $R_{p}$ and brighter host stars, they are also ideal targets for smaller telescopes. By adding in a term that also prioritizes targets that are stale, where $\Delta T_{mid} \geq 0.5t_{dur}$, we define:
\begin{align}
\mathrm{FOM_{maint}} &= \frac{2H_{s}R_{p}R_{s}^{-2} \cdot \Delta T_{mid} (0.5t_{dur})^{-1}}{10^{0.2 H\text{-}mag}}.
\label{eqn:maint_FOM}
\end{align}
$\mathrm{FOM_{maint}}$ prioritizes not only larger planets with large scale heights around bright host stars (and are therefore conducive to atmospheric characterization) but also those that have the greatest risk of becoming stale, thus requiring ephemeris maintenance. By using Equation~\ref{eqn:dt_next} to derive $\Delta T_{mid}$, then this figure of merit also takes into account a planet's uncertainty in its ephemeris $\Delta T_{0}$ and orbital period $\Delta P$.

\subsection{Current Ephemerides Uncertainties}
From the $\sim$4000 confirmed transiting exoplanets listed on the NASA Exoplanet Archive \citep{akeson13}, we rank the top 20 planets requiring ephemeris maintenance at three different epochs (JWST's launch, ARIEL's launch, and an Astro2020 Decadal mission's launch) with $\mathrm{FOM_{maint}}$ (Eqn.~\ref{eqn:maint_FOM}) in Tables~\ref{tab:deltaTestimates_JWST}--\ref{tab:deltaTestimates_Astro2020}. Also included are each planet's ``observational likelihood'' as calculated by $\mathrm{FOM_{transit}}$ (Eqn.~\ref{eqn:transit_FOM}), which estimates how likely a particular planet is to be observed for future atmospheric characterization. {{To provide a probability for a target to be spectroscopically characterized in the future, we rank all of the currently-known transiting exoplanets via $\mathrm{FOM_{transit}}$ and then calculate the percentile rank of each planet so that a planet with an observational likelihood of 100\% is the top-ranked planet (and most likely to be observed in the future) while one with a likelihood of 0\% is the ``worst-ranked'' planet.}}

{{As indiciated in Tables~\ref{tab:deltaTestimates_JWST}--\ref{tab:deltaTestimates_Astro2020}, there are planets that have high probability to be observed with future platforms (as indicated by their high observational likelihood), yet}} they are at high risk  of going stale or will have substantial uncertainties in their mid-transit times. Therefore, ephemeris maintenance of these high-priority targets is necessary to ensure the efficient use of large observatory time. {{Based on the planet-star area ratio and V-magnitude of the host star, a transit detection significance is estimated for a single 6~inch telescope and presented in these tables. However some targets are rather small or the star is dim, requiring multiple telescopes to statistically resolve the signal to $\ge3\sigma$. The number of 6-inch telescopes required to make a 3$\sigma$ detection is also shown in these tables. For a detailed description on how the values for these two columns were calculated, please see Section~\ref{sec:network} and Figure~\ref{fig:MO_survey}.}}

\begin{table*}
  \caption{Top 20 Planets With Large Spectral Modulation Also Requiring Ephemeris Maintenance for JWST Launch (30 March 2021)}
  \label{tab:deltaTestimates_JWST}
  \begin{center}
    \leavevmode
    \resizebox{\textwidth}{!}{%
\begin{tabular}{l|c|c|c|c|c|c|c|c|c|c}
\hline
\hline
 & & & & & & & & & \textbf{Estimated V-band Detection} & \textbf{Number of 6-inch Telescopes}\\
 & & & \textbf{Transit} & \textbf{$\Delta T_{mid}$} & \textbf{Transit} & \textbf{Orbital} & \textbf{Scale Height} & \textbf{Observational} & \textbf{Significance with a}& \textbf{Needed for a}\\
\textbf{Planet Name} & \textbf{V-mag} & \textbf{H-mag} & \textbf{Depth [\%]} & \textbf{[min]} & \textbf{Duration [min]} & \textbf{Period [days]} & \textbf{[km]} & \textbf{Likelihood [\%]} & \textbf{Single 6-inch Telescope [$\sigma$]}& \textbf{3$\sigma$ Detection}\\
\hline
KIC 5951458 b&N/A&11.382&0.1558&71434382.4&2261.69&1320.1&111.28&61.27&N/A&N/A\\

HIP 41378 e&8.93&7.786&0.1392&1493281.44&220.09&131.0&87.73&84.25&0.87&12\\

KIC 3558849 b&N/A&12.819&0.3891&2224569.6&1116.53&1322.3&81.44&58.22&N/A&N/A\\

KIC 8540376 c&N/A&13.014&0.0305&2240668.8&337.55&75.2&93.05&4.6&N/A&N/A\\

K2-13 b&12.901&11.26&0.0475&29470.48&234.6&39.91488&108.84&63.79&0.11&739\\

TRAPPIST-1 h&18.8&10.718&0.3403&507.69&83.52&18.767&78.21&93.39&0.18&275\\

HD 97658 b&7.71&5.821&0.0884&678.24&183.57&9.4909&119.27&94.57&0.75&16\\

Qatar-8 b&11.526&10.005&1.0153&426.02&242.33&3.71495&888.49&98.34&3.33&1\\

K2-141 c&11.389&8.524&0.8858&65.12&180.06&7.7485&1022.57&99.9&3.01&1\\

TOI 216.02&12.324&10.332&0.6944&922.46&255.26&17.089&275.39&92.45&1.86&3\\

KELT-8 b&10.833&9.269&1.2927&173.12&282.38&3.24406&911.62&99.07&5.04&1\\

GJ 1132 b&13.49&8.666&0.2744&51.08&50.06&1.628931&111.96&97.4&0.55&30\\

WASP-117 b&10.15&8.86&0.7982&220.72&287.28&10.02165&529.51&98.13&3.69&1\\

K2-155 c&12.806&9.686&0.1712&936.72&244.91&13.85&140.16&88.2&0.41&55\\

K2-155 b&12.806&9.686&0.0761&679.82&161.55&6.342&144.86&85.43&0.18&275\\

CoRoT-24 c&N/A&13.046&0.2738&3521.81&315.29&11.759&169.78&72.55&N/A&N/A\\

HD 219134 c&5.57&3.469&0.0322&130.05&174.95&6.76458&78.9&95.19&0.47&42\\

Kepler-32 f&16.452&12.901&0.0228&589.59&65.77&0.74296&395.35&75.29&0.02&18915\\

HAT-P-3 b&11.577&9.542&1.2431&107.74&147.3&2.8997&217.3&94.88&4.02&1\\

Kepler-42 d&16.124&11.685&0.0865&47.61&45.08&1.865169&293.8&92.42&0.09&1116\\
 
\end{tabular}}
\end{center}
\end{table*}

\begin{table*}
  \caption{Top 20 Planets With Large Spectral Modulation Also Requiring Ephemeris Maintenance for Estimated ARIEL Launch (1 January 2028)}
  \label{tab:deltaTestimates_ARIEL}
  \begin{center}
    \leavevmode
    \resizebox{\textwidth}{!}{%
\begin{tabular}{l|c|c|c|c|c|c|c|c|c|c}
\hline
\hline
 & & & & & & & & & \textbf{Estimated V-band Detection} & \textbf{Number of 6-inch Telescopes}\\
 & & & \textbf{Transit} & \textbf{$\Delta T_{mid}$} & \textbf{Transit} & \textbf{Orbital} & \textbf{Scale Height} & \textbf{Observational} & \textbf{Significance with a}& \textbf{Needed for a}\\
\textbf{Planet Name} & \textbf{V-mag} & \textbf{H-mag} & \textbf{Depth [\%]} & \textbf{[min]} & \textbf{Duration [min]} & \textbf{Period [days]} & \textbf{[km]} & \textbf{Likelihood [\%]} & \textbf{Single 6-inch Telescope [$\sigma$]}& \textbf{3$\sigma$ Detection}\\
\hline
HIP 41378 e&8.93&7.786&0.1392&3162241.44&220.09&131.0&83.48&84.34&0.87&12\\

KIC 5951458 b&N/A&11.382&0.1558&89292974.4&2261.69&1320.1&108.9&62.29&N/A&N/A\\

KIC 3558849 b&N/A&12.819&0.3891&3336537.6&1116.53&1322.3&75.32&58.02&N/A&N/A\\

KIC 8540376 c&N/A&13.014&0.0305&4089196.8&337.55&75.2&85.5&4.24&N/A&N/A\\

K2-13 b&12.901&11.26&0.0475&58465.94&234.6&39.91488&103.68&64.34&0.11&739\\

Qatar-8 b&11.526&10.005&1.0153&1382.18&242.33&3.71495&874.28&98.37&3.33&1\\

TOI 216.02&12.324&10.332&0.6944&3203.42&255.26&17.089&255.48&91.98&1.86&3\\

TRAPPIST-1 h&18.8&10.718&0.3403&1262.25&83.52&18.767&47.37&90.59&0.18&275\\

K2-141 c&11.389&8.524&0.8858&165.86&180.06&7.7485&940.06&99.86&3.01&1\\

HD 97658 b&7.71&5.821&0.0884&1277.28&183.57&9.4909&112.41&94.34&0.75&16\\

KELT-8 b&10.833&9.269&1.2927&348.22&282.38&3.24406&896.07&99.06&5.04&1\\

K2-155 c&12.806&9.686&0.1712&2474.64&244.91&13.85&130.73&87.95&0.41&55\\

GJ 1132 b&13.49&8.666&0.2744&109.99&50.06&1.628931&102.45&97.01&0.55&30\\

WASP-117 b&10.15&8.86&0.7982&415.56&287.28&10.02165&521.62&98.16&3.69&1\\

K2-155 b&12.806&9.686&0.0761&1800.14&161.55&6.342&136.42&85.38&0.18&275\\

TOI 216.01&12.324&10.332&1.5035&2016.98&334.39&34.556&102.58&88.99&4.04&1\\

HD 219666 b&N/A&8.254&0.1743&527.96&217.42&6.03607&325.51&94.1&N/A&N/A\\

pi Men c&5.67&4.424&0.0298&366.05&218.64&6.2679&184.11&95.49&0.42&51\\

HD 219134 c&5.57&3.469&0.0322&303.49&174.95&6.76458&71.57&94.86&0.47&42\\

K2-266 b&11.808&9.041&0.1837&146.71&79.62&0.658524&326.95&94.38&0.56&29\\
\end{tabular}}
\end{center}
\end{table*}

\begin{table*}
  \caption{Top 20 Planets With Large Spectral Modulation Also Requiring Ephemeris Maintenance for Estimated Astro2020 Launch (1 January 2030)}
  \label{tab:deltaTestimates_Astro2020}
  \begin{center}
    \leavevmode
    \resizebox{\textwidth}{!}{%
\begin{tabular}{l|c|c|c|c|c|c|c|c|c|c}
\hline
\hline
 & & & & & & & & & \textbf{Estimated V-band Detection} & \textbf{Number of 6-inch Telescopes}\\
 & & & \textbf{Transit} & \textbf{$\Delta T_{mid}$} & \textbf{Transit} & \textbf{Orbital} & \textbf{Scale Height} & \textbf{Observational} & \textbf{Significance with a}& \textbf{Needed for a}\\
\textbf{Planet Name} & \textbf{V-mag} & \textbf{H-mag} & \textbf{Depth [\%]} & \textbf{[min]} & \textbf{Duration [min]} & \textbf{Period [days]} & \textbf{[km]} & \textbf{Likelihood [\%]} & \textbf{Single 6-inch Telescope [$\sigma$]}& \textbf{3$\sigma$ Detection}\\
\hline
HIP 41378 e&8.93&7.786&0.1392&3601441.44&220.09&131.0&83.48&84.34&0.87&12\\

KIC 5951458 b&N/A&11.382&0.1558&107151566.4&2261.69&1320.1&108.9&62.29&N/A&N/A\\

KIC 8540376 c&N/A&13.014&0.0305&4593340.8&337.55&75.2&85.5&4.24&N/A&N/A\\

KIC 3558849 b&N/A&12.819&0.3891&3336537.6&1116.53&1322.3&75.32&58.02&N/A&N/A\\

K2-13 b&12.901&11.26&0.0475&66883.98&234.6&39.91488&103.68&64.34&0.11&739\\

Qatar-8 b&11.526&10.005&1.0153&1665.86&242.33&3.71495&874.28&98.37&3.33&1\\

TOI 216.02&12.324&10.332&0.6944&3884.54&255.26&17.089&255.48&91.98&1.86&3\\

TRAPPIST-1 h&18.8&10.718&0.3403&1486.89&83.52&18.767&47.37&90.59&0.18&275\\

K2-141 c&11.389&8.524&0.8858&195.96&180.06&7.7485&940.06&99.86&3.01&1\\

HD 97658 b&7.71&5.821&0.0884&1454.69&183.57&9.4909&112.41&94.34&0.75&16\\

KELT-8 b&10.833&9.269&1.2927&400.29&282.38&3.24406&896.07&99.06&5.04&1\\

K2-155 c&12.806&9.686&0.1712&2932.56&244.91&13.85&130.73&87.95&0.41&55\\

GJ 1132 b&13.49&8.666&0.2744&127.4&50.06&1.628931&102.45&97.01&0.55&30\\

WASP-117 b&10.15&8.86&0.7982&473.37&287.28&10.02165&521.62&98.16&3.69&1\\

K2-155 b&12.806&9.686&0.0761&2131.34&161.55&6.342&136.42&85.38&0.18&275\\

TOI 216.01&12.324&10.332&1.5035&2440.34&334.39&34.556&102.58&88.99&4.04&1\\

HD 219666 b&N/A&8.254&0.1743&639.48&217.42&6.03607&325.51&94.1&N/A&N/A\\

pi Men c&5.67&4.424&0.0298&443.55&218.64&6.2679&184.11&95.49&0.42&51\\

K2-266 b&11.808&9.041&0.1837&173.89&79.62&0.658524&326.95&94.38&0.56&29\\

HD 219134 c&5.57&3.469&0.0322&354.82&174.95&6.76458&71.57&94.86&0.47&42\\
\end{tabular}}
\end{center}
\end{table*}

{{To more easily place the planets discovered by TESS into context with the $\sim$4000 currently-known exoplanets, we construct ``representative planets''---planets that adopt median values of the properties of their larger sample---of the currently-known exoplanets listed on the NASA Exoplanet Archive, the TESS-discovered exoplanets, and the TOIs. The Representative Known Planet adopts}} {{a median orbital period $P = 12.06$~days {{and median uncertainty}} $\Delta P = 8.12$~sec, a median mid-transit ephemeris $T_{0} = 2454985.88$~JD {{and median uncertainty}} $\Delta T_{0} = 220.28$~sec, and a median V-mag = 13.716 (Table~\ref{tab:rep_planets})}}. {{At the time of this study, thirty-seven}} TESS-discovered exoplanets have been confirmed \citep{bakos18, huang18, gandolfi18, gandolfi19, canas19, dawson19, dragomir19b, dumusque19, esposito19, gunther19, huber19, jones19, kipping19, kostov19, luque19, newton19, nielsen19, quinn19, rodriguez19,trifonov19, vanderburg19, vanderspek19, wang19, winters19}; {{the Representative TESS Planet adopts their}} {{median orbital period $P = 5.97$~days {{and median uncertainty}} $\Delta P = 26.35$~sec, a median mid-transit ephemeris $T_{0} = 2458366.17$~JD {{and median uncertainty}} $\Delta T_{0} = 64.66$~sec, and a median V-mag = 9.724}}. Of the current {{1604 TESS Objects of Interest\footnote{https://tess.mit.edu/publications/} (TOIs), 1555 have reported mid-transit and orbital period uncertainties; these TOIs}} have mid-transit and orbital period uncertainties that range from $\sim$0.5~seconds to tens of minutes. {{{{The Representative TOI Planet adopts their}} median orbital period $P = 3.85$~days {{and median uncertainty}} $\Delta P = 34.56$~sec, median mid-transit ephemeris $T_{0} = 2458517.96$~JD {{and median uncertainty}} $\Delta T_{0} = 181.87$~sec, and TESS magnitude of 10.261.}}

\begin{table*}
  \caption{{{The Representative Planets Adopted for this Study}}}
  \label{tab:rep_planets}
  \begin{center}
    \resizebox{1\textwidth}{!}{%
\begin{tabular}{l|c|c|c|c|c}
\hline
\hline
\textbf{Representative} & \textbf{Host Star} & \textbf{Orbital Period} & \textbf{Period Uncertainty]} & \textbf{Mid-transit Time} & \textbf{Mid-transit Uncertainty} \\
\textbf{Planet} & \textbf{Magnitude} & \textbf{P [days]} & \textbf{$\Delta$P [sec]} & \textbf{T$_{mid}$ [JD]} & \textbf{$\Delta$T$_{mid}$ [sec]} \\
\hline
TOI & 10.261 TESS-mag & 3.85 & 34.56 & 2458517.96 & 181.87 \\
TESS & 9.724 V-mag & 5.97 & 26.35 & 2458366.17 & 64.66 \\
Known & 13.716 V-mag & 12.06 & 8.12 & 2454985.88 & 220.28 \\
\end{tabular}}
\end{center}
\end{table*}

\begin{figure}
\centering
\includegraphics[width=1\columnwidth]{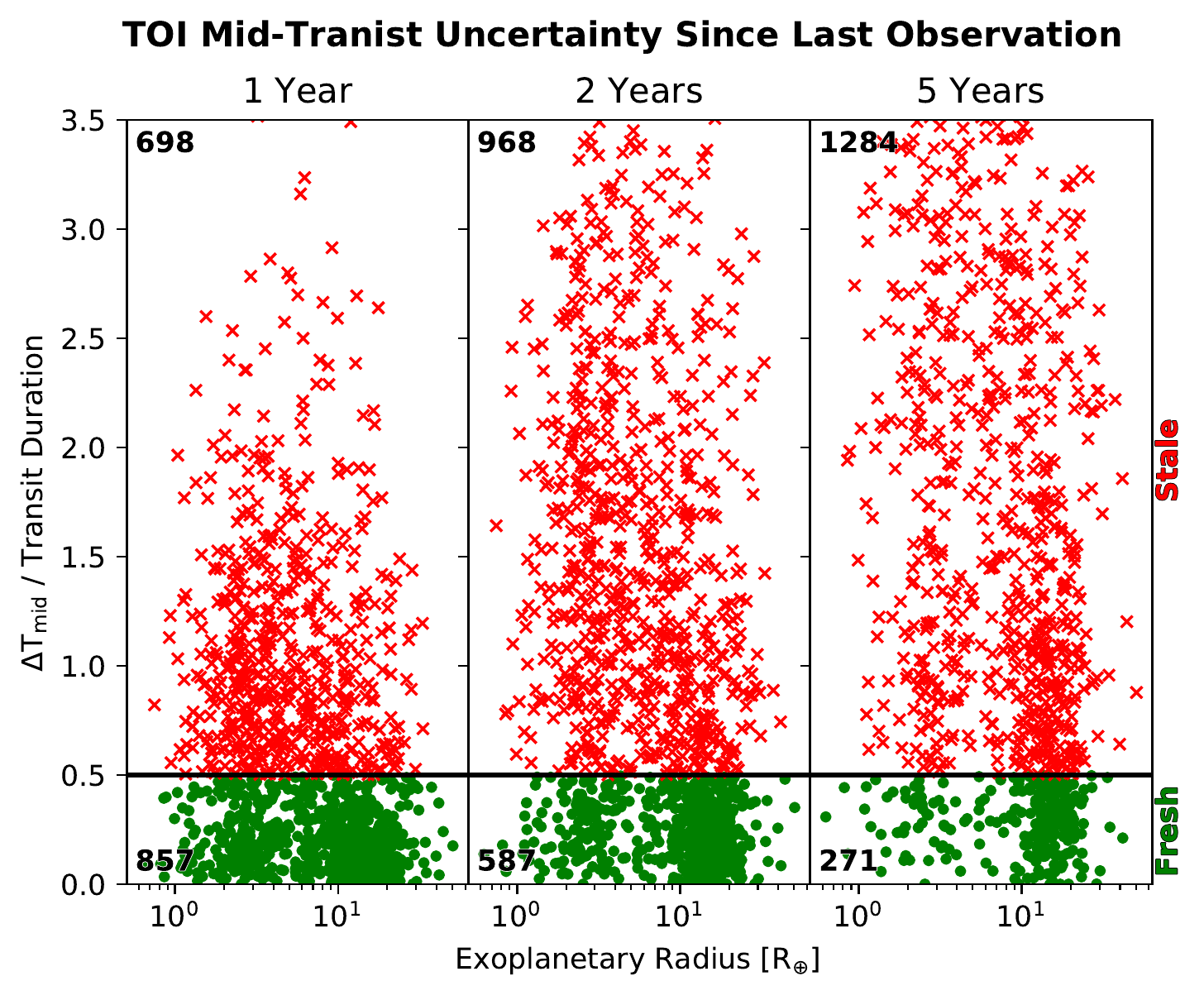}
\caption{The number of TOIs that transition from fresh (green dots) to stale (red Xs; where the uncertainty in the mid-transit time $\Delta T_{mid}$ exceeds half the transit duration) as a function of time since their last transit observation. We find that after one year, two years, and five years 45\%, 62\%, and 83\% of the TOI population, respectively, will become stale, requiring ephemerides maintenance.}
\label{fig:stale_vs_fresh}
\end{figure}

The TESS Follow-up Observation Program\footnote{https://heasarc.gsfc.nasa.gov/docs/tess/followup.html} (TFOP) has done an excellent job to refine the mid-transit and orbital period uncertainties of TOIs in the process of confirming these targets, as evidenced by currently-reported high precisions of the confirmed TESS-discovered planets. We conservatively proceed with the TOIs for the rest of the analysis presented here, not only to de-bias our analysis from potential small-number statistics {{(versus the confirmed TESS-discovered planets)}}, but also in case future confirmed TESS-discovered exoplanets do not achieve similarly-high timing precisions. Using the currently-reported TOI mid-transit and period uncertainties and Equation~\ref{eqn:dt_next}, we estimate how quickly the TOIs would become stale, assuming no additional observations are conducted. Because the covariance term $\mathrm{cov}(P,T_{0})$ is not currently supplied for each TOI, we estimate a global covariance term via:
\begin{align}
\mathrm{cov}(P,T_{0}) = \frac{\Sigma^{n}_{i=1}(\Delta T_{mid,i}-\overline{\Delta T_{mid}})(\Delta P_{i}-\overline{\Delta P})}{n-1}
\label{eqn:cov}
\end{align}
where $n$ is the number of TOIs, $\Delta T_{mid,i}$ is the mid-transit uncertainty of TOI $i$, $\overline{\Delta T_{mid}}$ is the mean mid-transit uncertainty of all $n$ TOIs, $\Delta P_{i}$ is the uncertainty of the orbital period of TOI $i$, and $\overline{\Delta P}$ is the mean orbital period uncertainty of all $n$ TOIs. Using this equation, we find $\mathrm{cov}(P,T_{0}) = 0.18$~sec. We assume that this covariance term is global in that it applies to each individual currently-known TOI object. In this manner, we estimate that 698 TOIs (45\% total) will become at risk one year after their last observation, 968 TOIs (62\% total) will become stale in two years, and 1284 TOIs (83\% total) will become stale in five years (Fig.~\ref{fig:stale_vs_fresh}). 

\begin{figure}[htbp]
		\includegraphics[width=1\columnwidth]{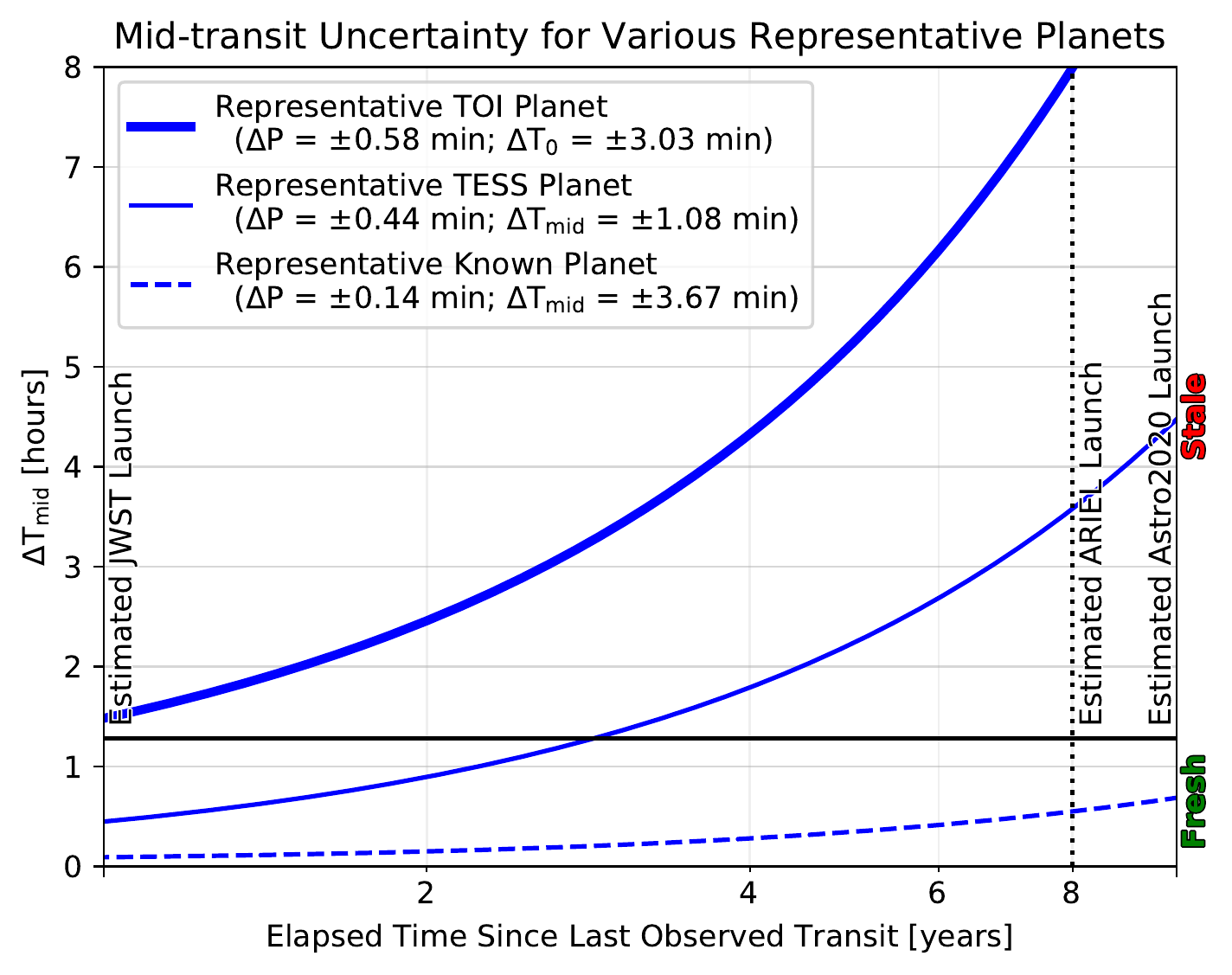}
		\caption{The mid-transit uncertainty as a function of the elapsed time since the last observation of the Representative TESS Planet (blue line) and Known Planet (dashed blue line). The Representative TOI Planet becomes stale on the order of $\sim$3~years, requiring additional observations to keep them fresh for follow-up observations with large telescopes.}
	\label{fig:multirepplanets}
\end{figure}

{{We can similarly calculate how long the timing of the three Representative Planets (Table~\ref{tab:rep_planets}) will remain fresh (Fig.~\ref{fig:multirepplanets}). We find that the}} Representative TOI Planet would become stale after $\sim$0.8~years and have a mid-{{transit}} uncertainty of 54.69~minutes after 1~year and 4.55~hours after 5~years, while the Representative Confirmed TESS Planet would become stale after $\sim$3~years, and the Representative Known Planet could remain fresh for over 15~years.

Thus, while a planet may now have a precise estimated upcoming transit time, it can still accrue large uncertainties, especially after many orbits (Eqn.~\ref{eqn:dt_next}). Therefore, orbital period and mid-transit maintenance (``ephemerides maintenance'') through regular monitoring is necessary to keep transit time uncertainties small so transiting exoplanets can be efficiently followed-up {{for atmospheric characterization.}}

A community-wide effort to monitor transiting exoplanets with ground-based telescopes, particularly smaller telescopes ($\le$1-m) used by citizen scientists, could support future missions by alleviating observing overhead. A study by \citet{mallonn19}, for example, highlights the capability of 0.3-m through 2.2-m telescopes to greatly reduce ephemeris uncertainties for {{21}} hot Jupiters. {{Similarly, the 200-mm (7.87-in) WASP-South survey demonstrated the capability of smaller telescopes by recovering the transit of the hot-Neptune HD~219666b ($V-mag=9.9$; transit depth {{(R$_{\mathrm{p}}$/R$_{\mathrm{s}}$)$^{2}$}}$=0.2$\%); when combined with TESS observations, WASP-South was able to revise the uncertainty in calculating future transit events to 1 minute per year \citep{hellier19}.}}

{{Here we explore the potential of observations of transiting exoplanets with small telescopes ($\le$1-m) operated by citizen scientists. We ultimately find that, despite their relatively small aperture sizes, these platforms can produce high quality transit observations (Figs.~\ref{fig:MOscopes1}--\ref{fig:smallscopes}), successfully keep transit times fresh, and enable additional science explored in the following sections.}}

\section{EXOTIC: The EXOplanet Transit Interpretation Code}\label{sec:EXOTIC}
\begin{figure*}[htbp] 
  \begin{minipage}[b]{0.5\linewidth}
    \centering
    \includegraphics[width=1\columnwidth]{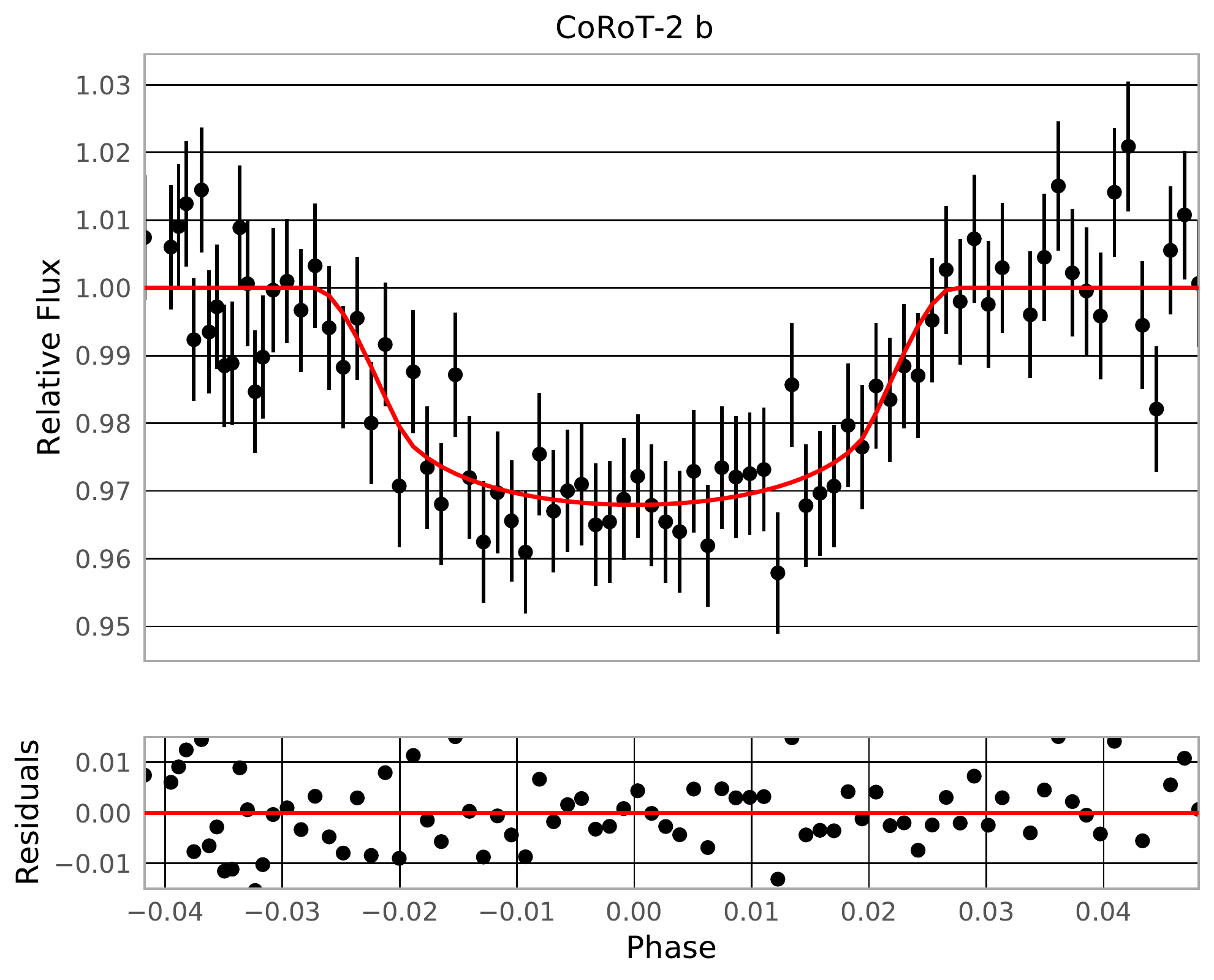}
  \end{minipage}
  \begin{minipage}[b]{0.5\linewidth}
    \centering
    \includegraphics[width=1\columnwidth]{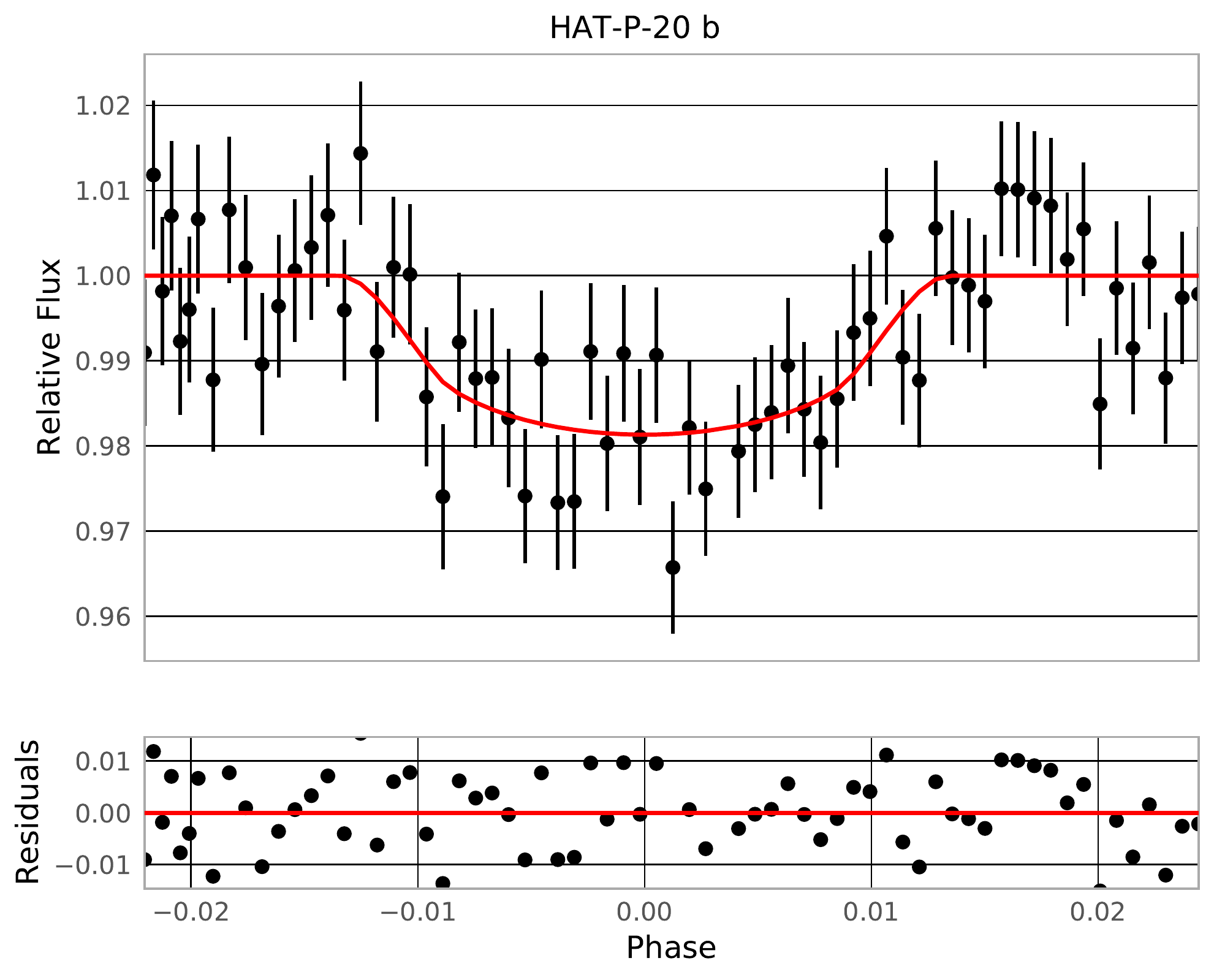}
  \end{minipage} 
  \begin{minipage}[b]{0.5\linewidth}
    \centering
    \includegraphics[width=1\columnwidth]{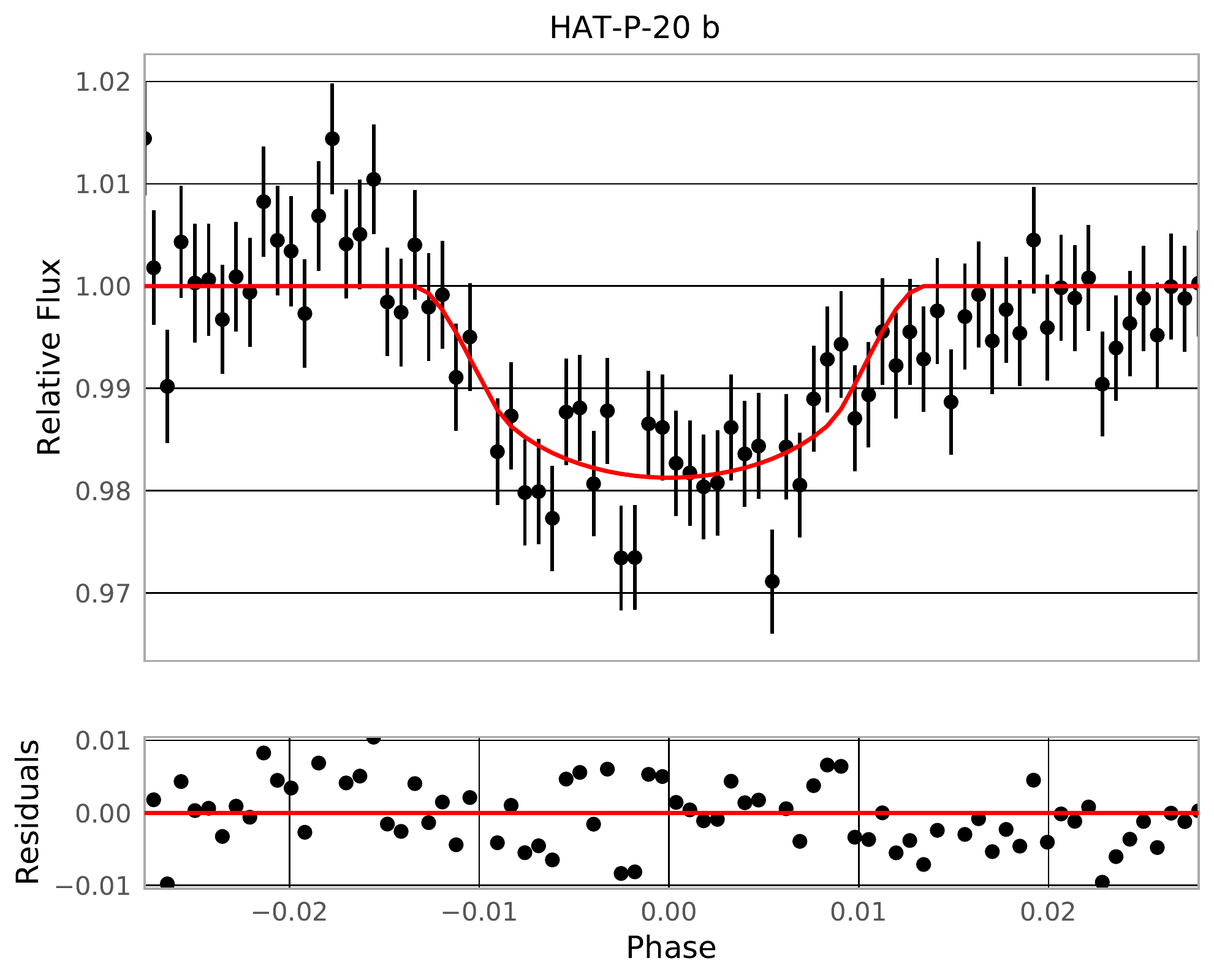}
  \end{minipage}
  \begin{minipage}[b]{0.5\linewidth}
    \centering
    \includegraphics[width=1\columnwidth]{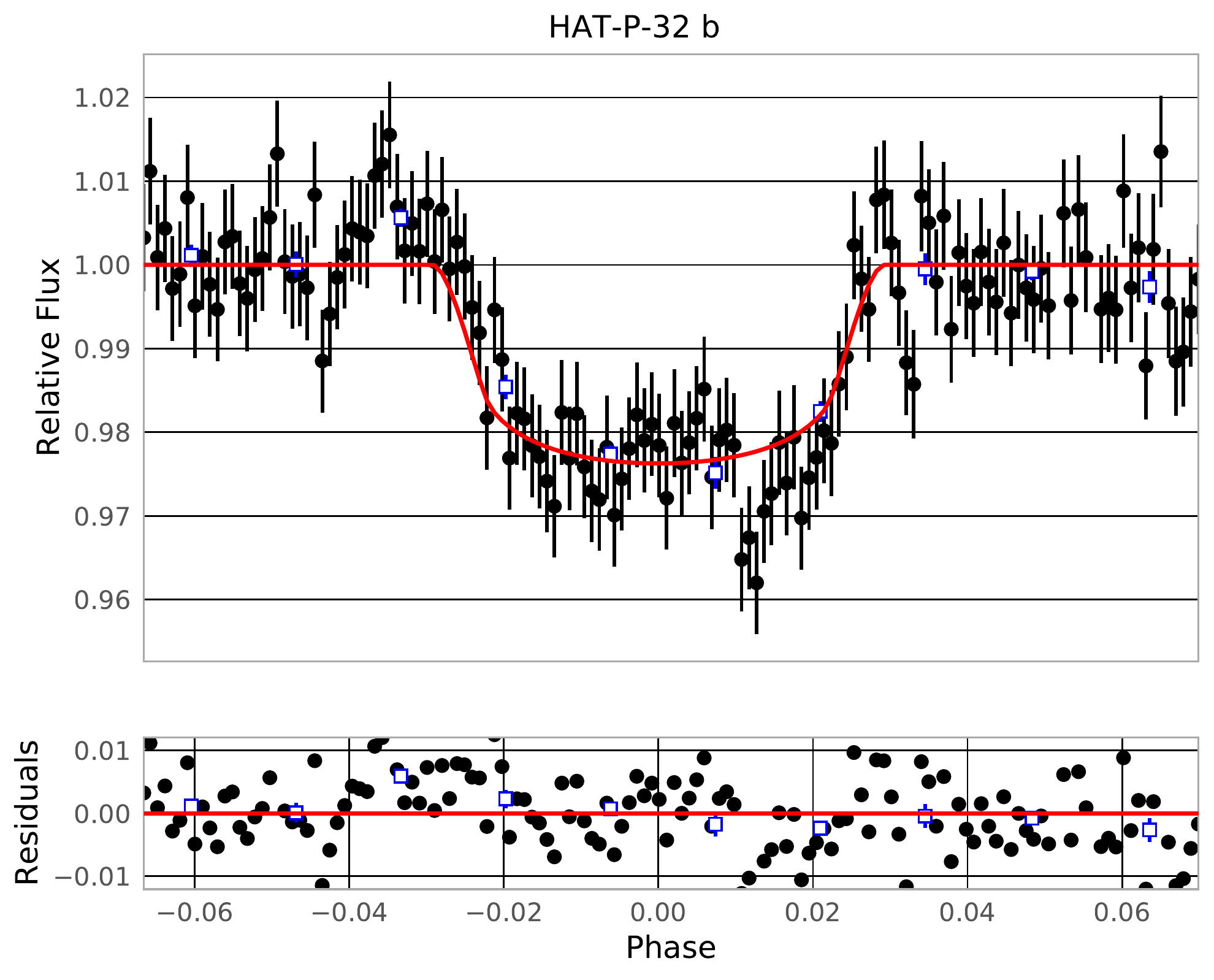}
  \end{minipage} 
    \begin{minipage}[b]{0.5\linewidth}
    \centering
    \includegraphics[width=1\columnwidth]{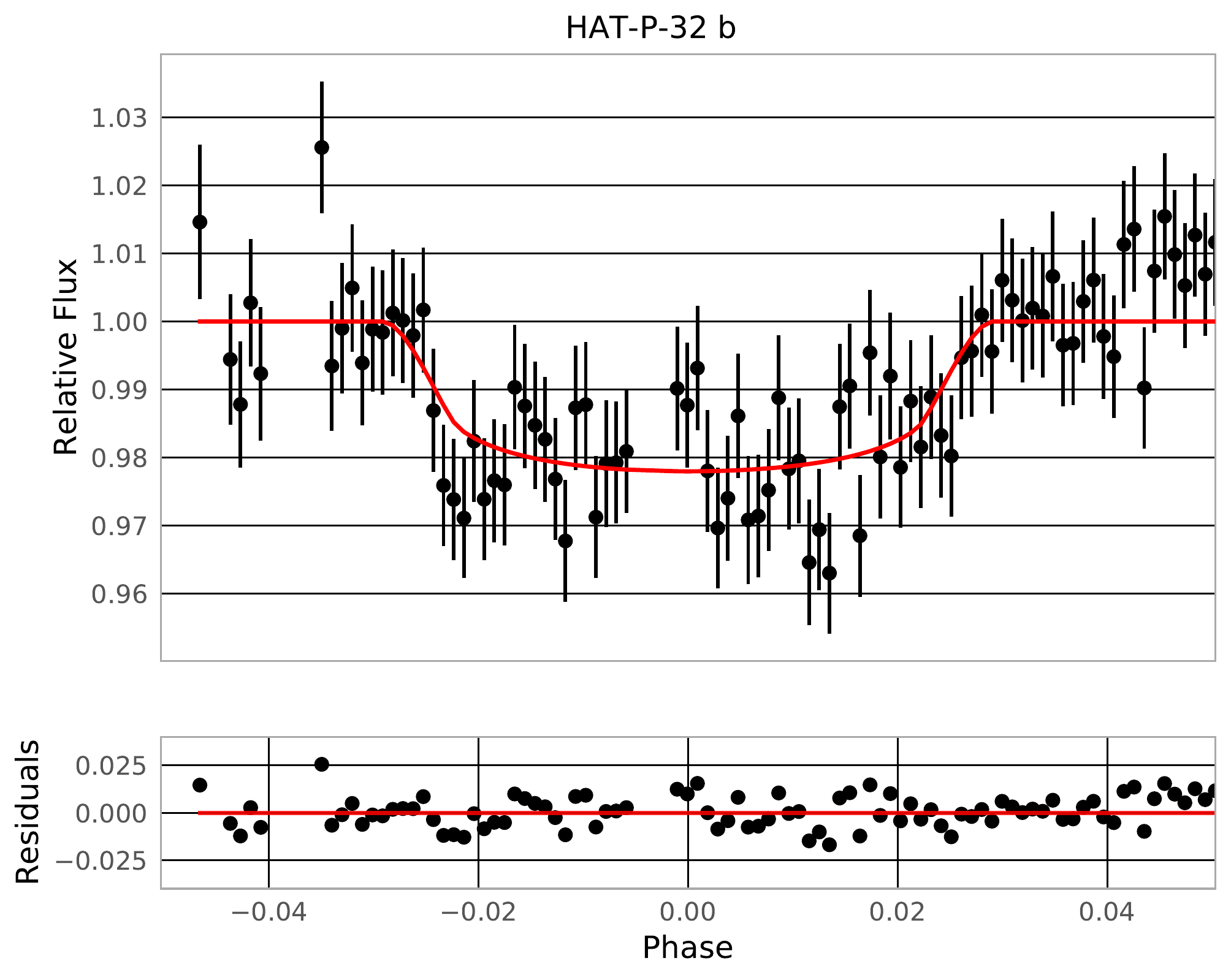}
  \end{minipage}
  \begin{minipage}[b]{0.5\linewidth}
    \centering
    \includegraphics[width=1\columnwidth]{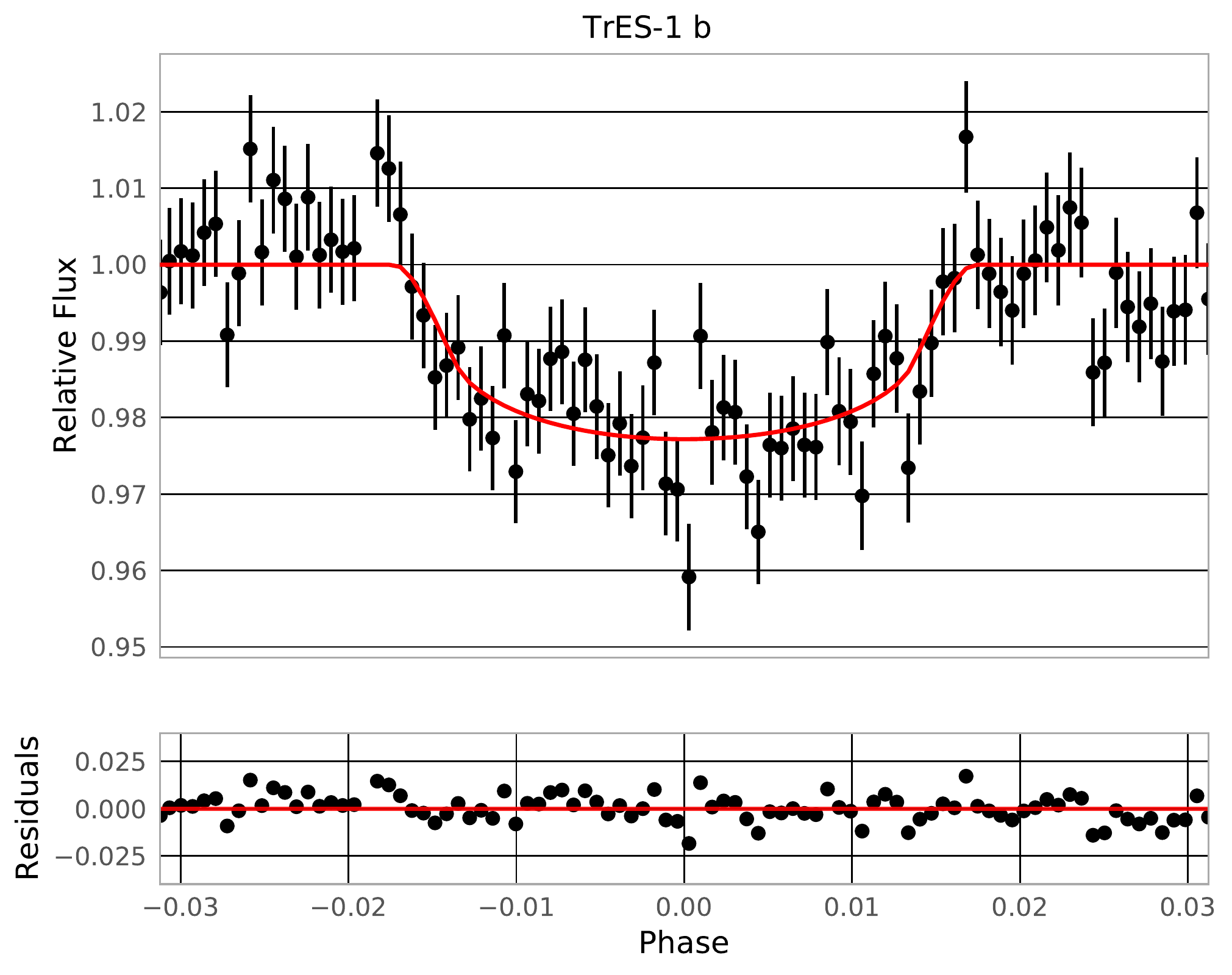}
  \end{minipage} 
 \caption{{{A sample of transits observed with a single 6-inch (15.24-cm) MicroObservatory telescope \citep{sadler01}. Despite most of these targets being relatively dim, the MicroObservatory can still achieve high observational precision (Table~\ref{tab:MO_14}).}}}
 		\label{fig:MOscopes1}
\end{figure*}

\begin{figure*}[htbp] 
  \begin{minipage}[b]{0.5\linewidth}
    \centering
    \includegraphics[width=.8\columnwidth]{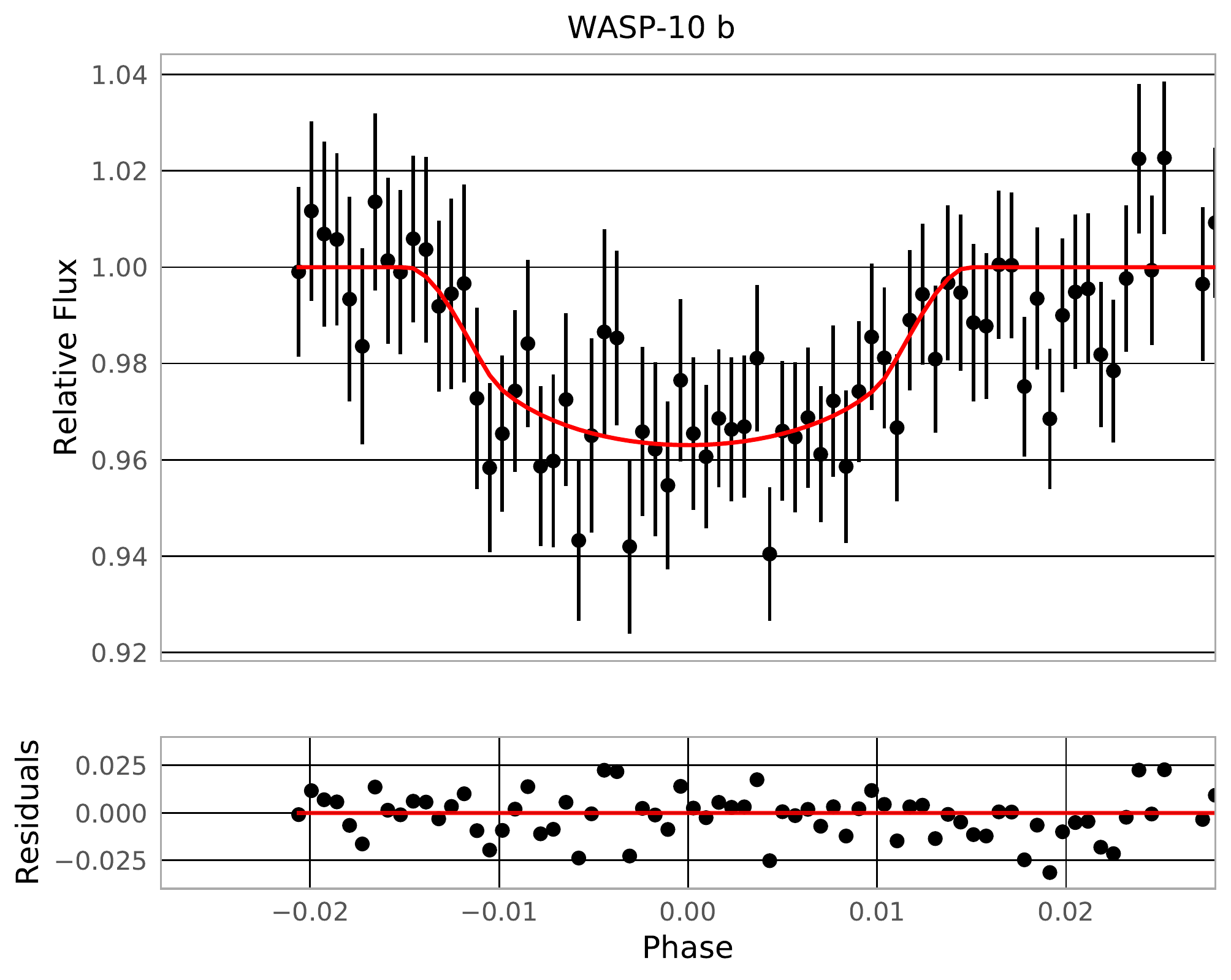}
  \end{minipage}
  \begin{minipage}[b]{0.5\linewidth}
    \centering
    \includegraphics[width=.8\columnwidth]{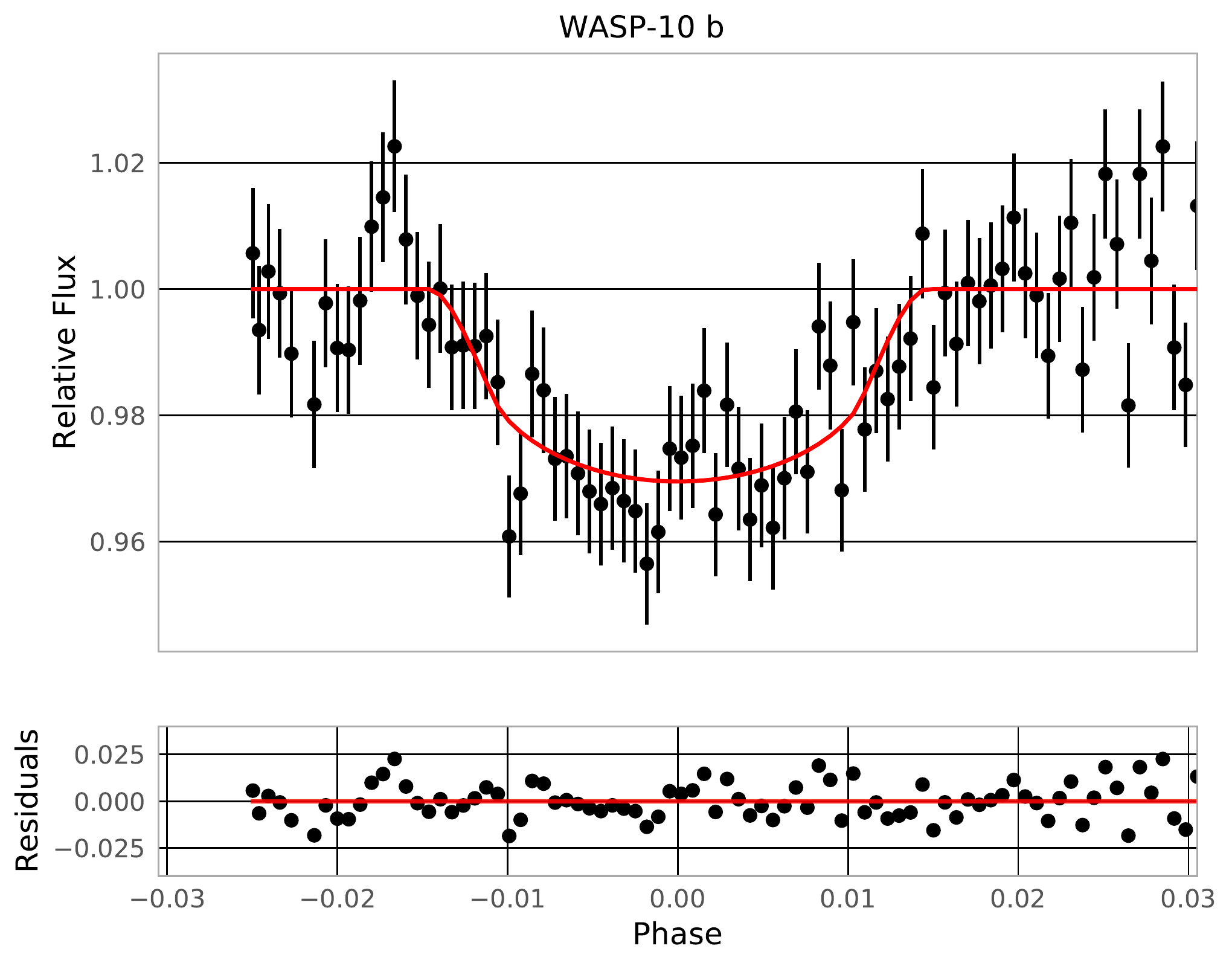}
  \end{minipage} 
  \begin{minipage}[b]{0.5\linewidth}
    \centering
    \includegraphics[width=.8\columnwidth]{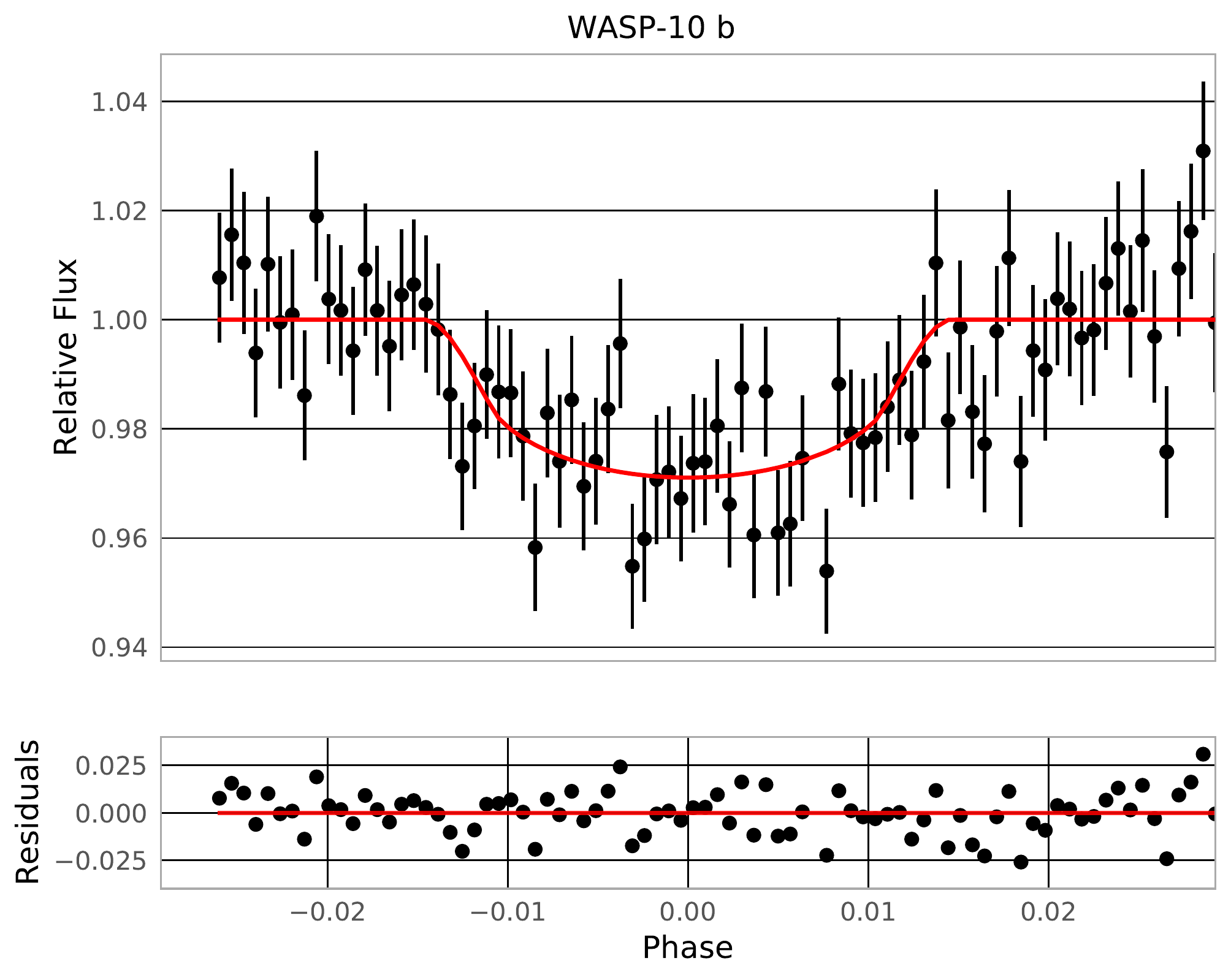}
  \end{minipage}
  \begin{minipage}[b]{0.5\linewidth}
    \centering
    \includegraphics[width=.8\columnwidth]{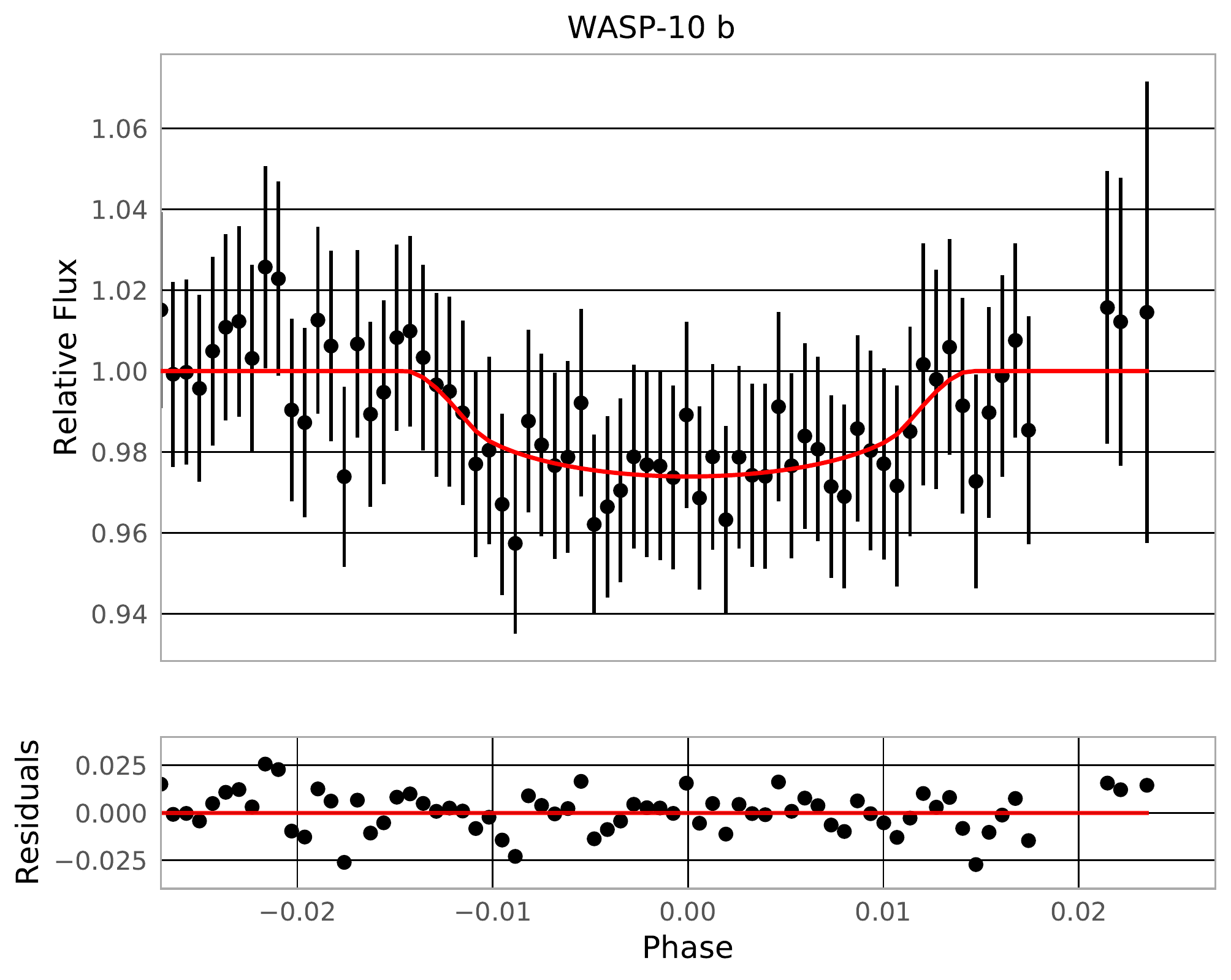}
  \end{minipage} 
    \begin{minipage}[b]{0.5\linewidth}
    \centering
    \includegraphics[width=.8\columnwidth]{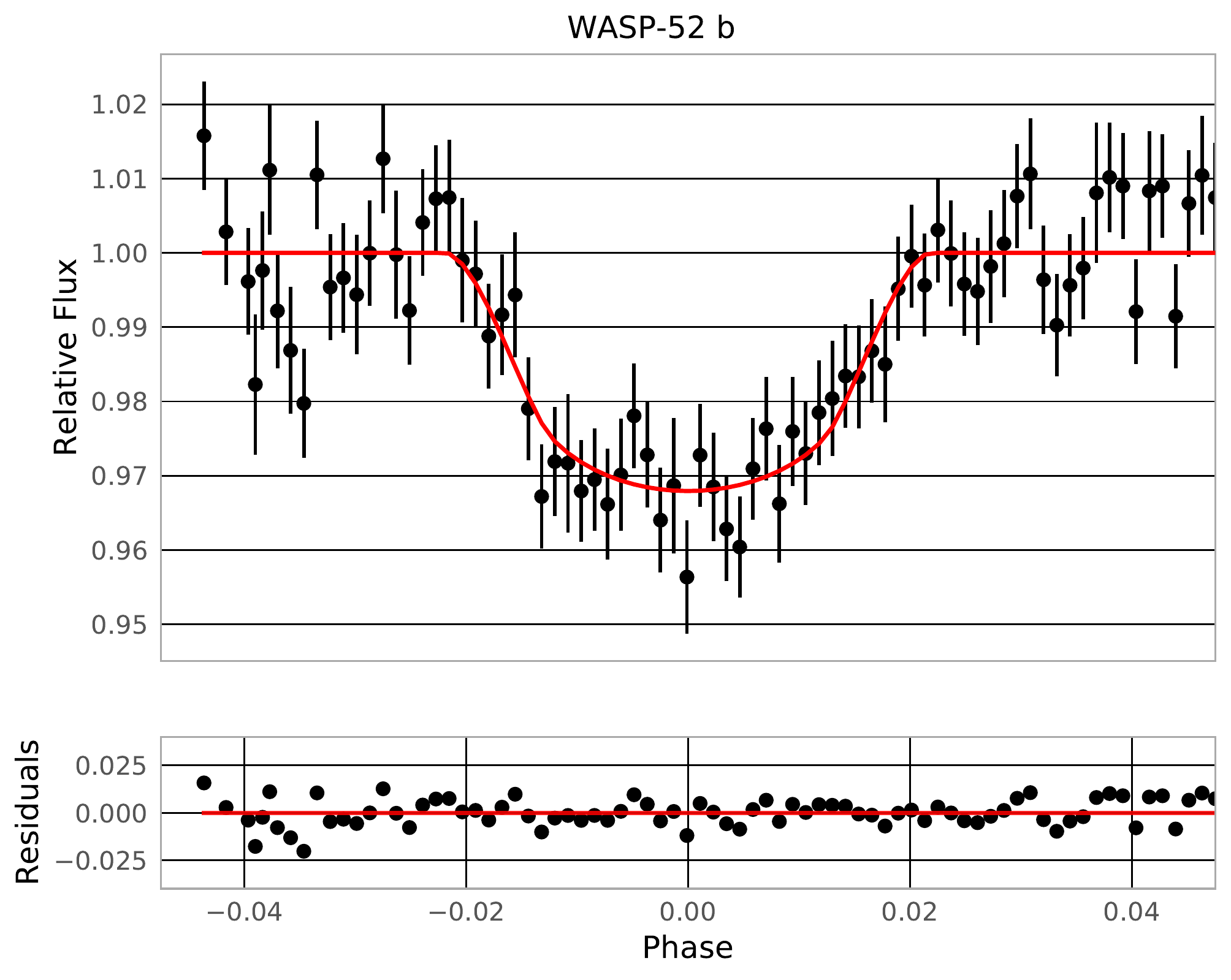}
  \end{minipage}
  \begin{minipage}[b]{0.5\linewidth}
    \centering
    \includegraphics[width=.8\columnwidth]{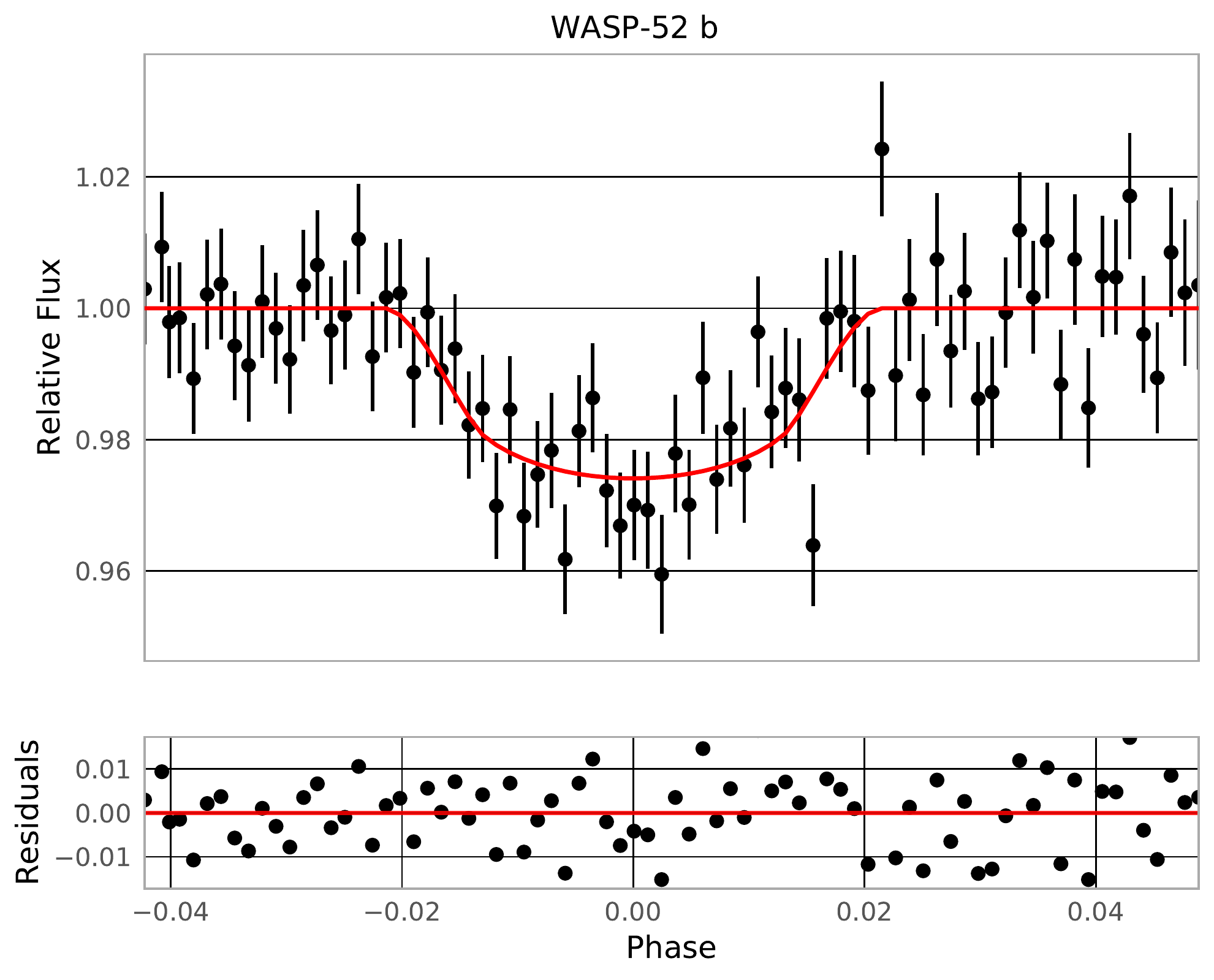}
  \end{minipage} 
    \begin{minipage}[b]{0.5\linewidth}
    \centering
    \includegraphics[width=.8\columnwidth]{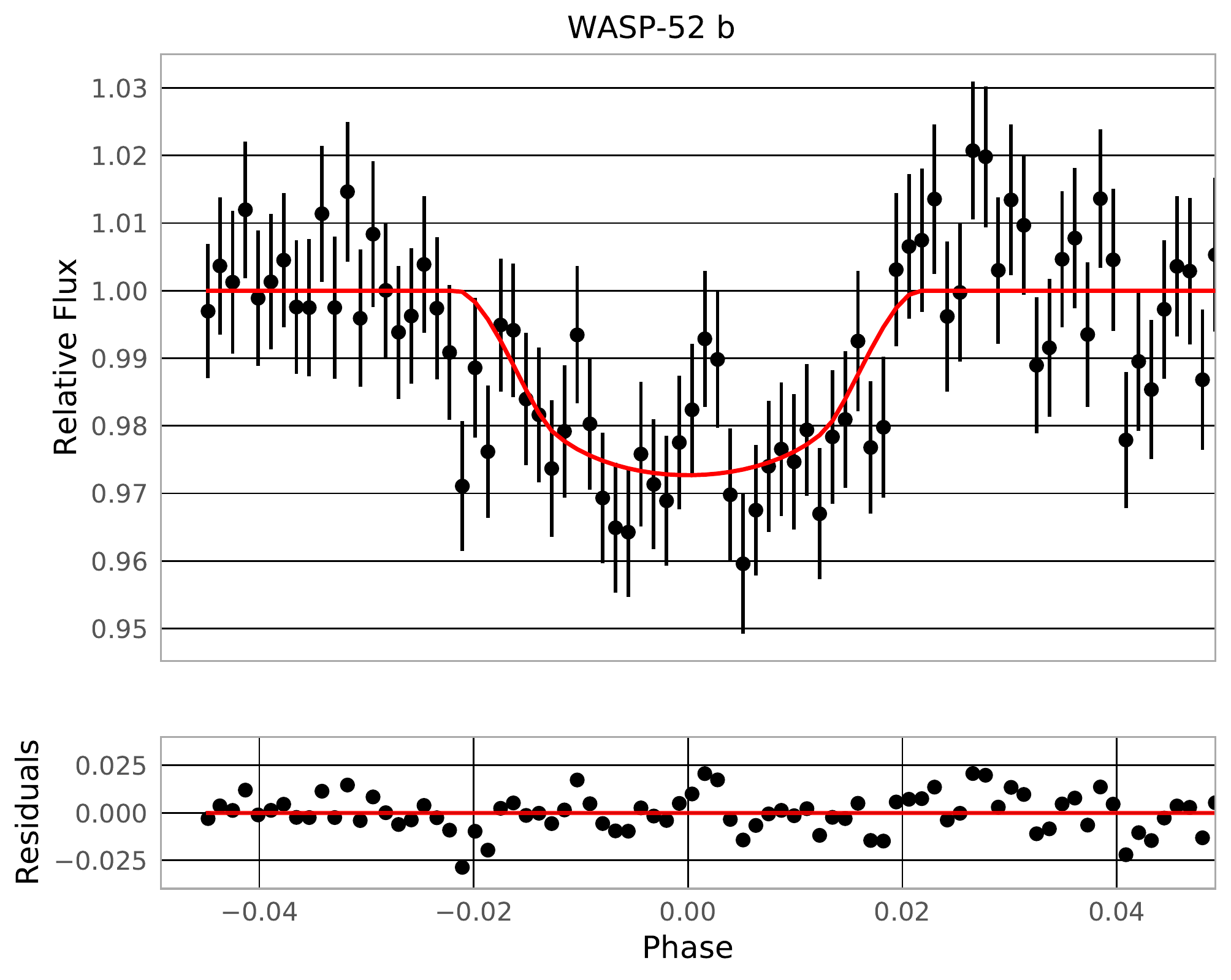}
  \end{minipage} 
    \begin{minipage}[b]{0.5\linewidth}
    \centering
    \includegraphics[width=.8\columnwidth]{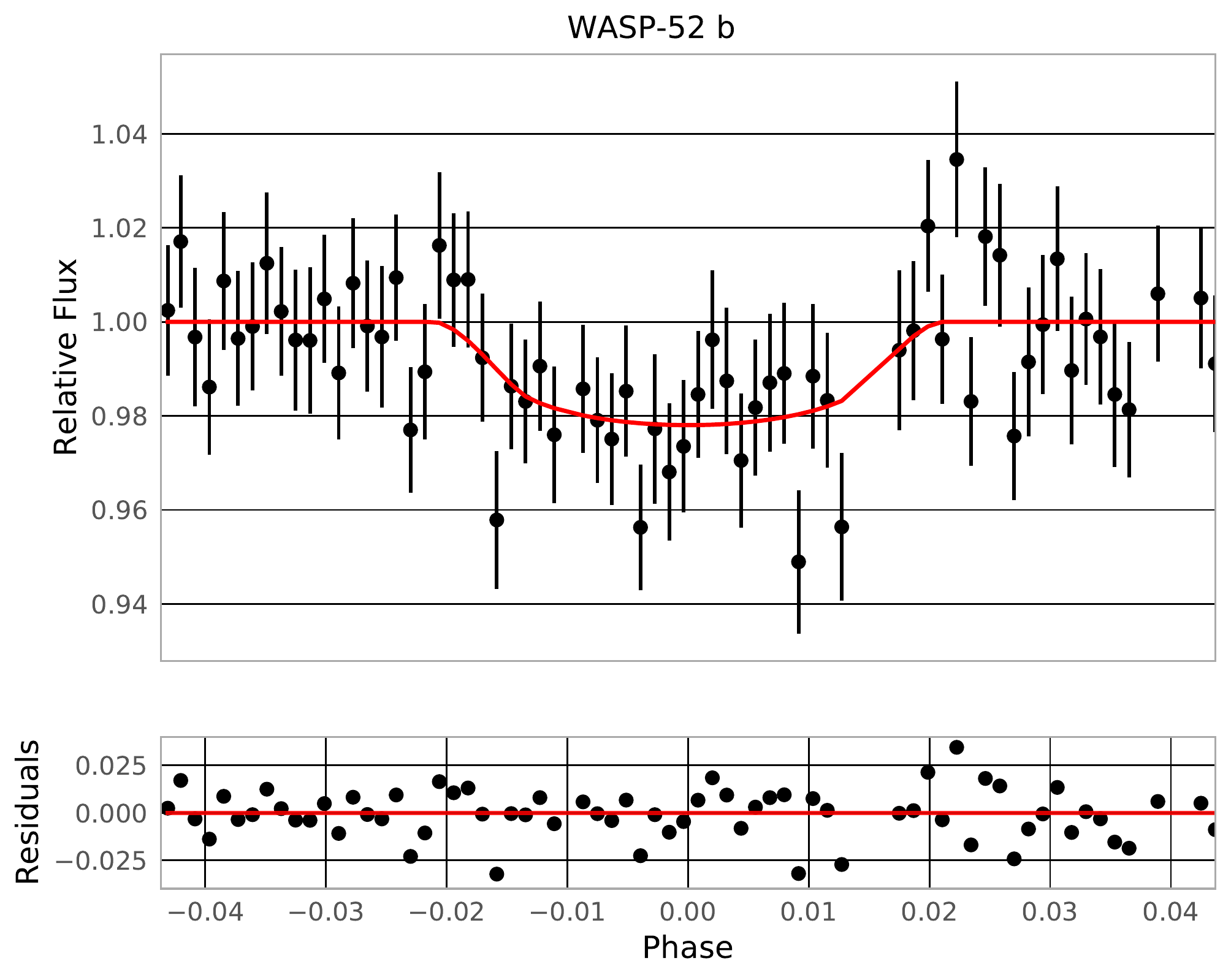}
  \end{minipage} 
 \caption{{{Same as Figure~\ref{fig:MOscopes1}.}}}
 		\label{fig:MOscopes2}
\end{figure*}

\begin{figure*}[htbp] 
  \begin{minipage}[b]{0.5\linewidth}
    \centering
        \includegraphics[width=.7\columnwidth]{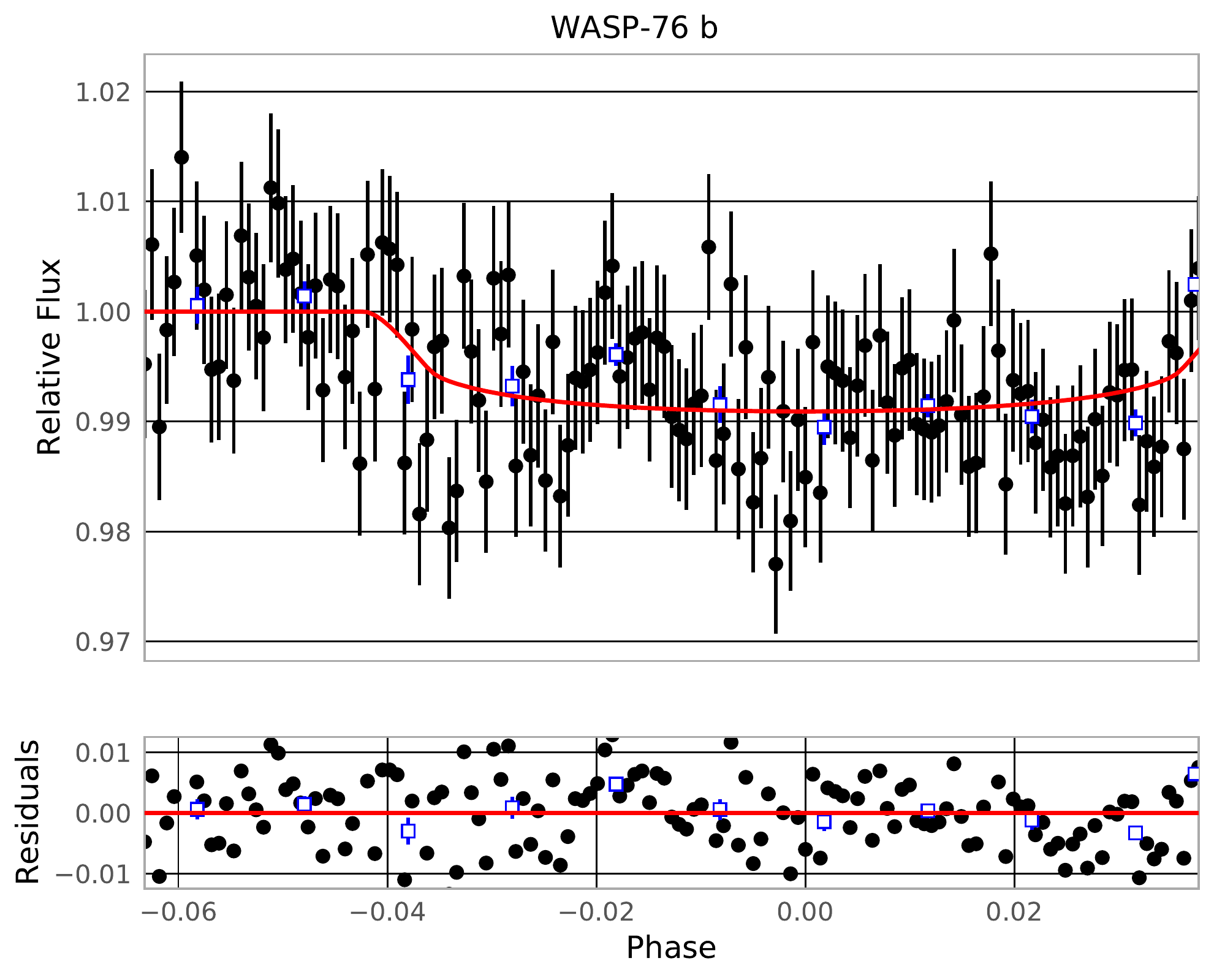}
  \end{minipage}
    \begin{minipage}[b]{0.5\linewidth}
    \centering
        \includegraphics[width=.7\columnwidth]{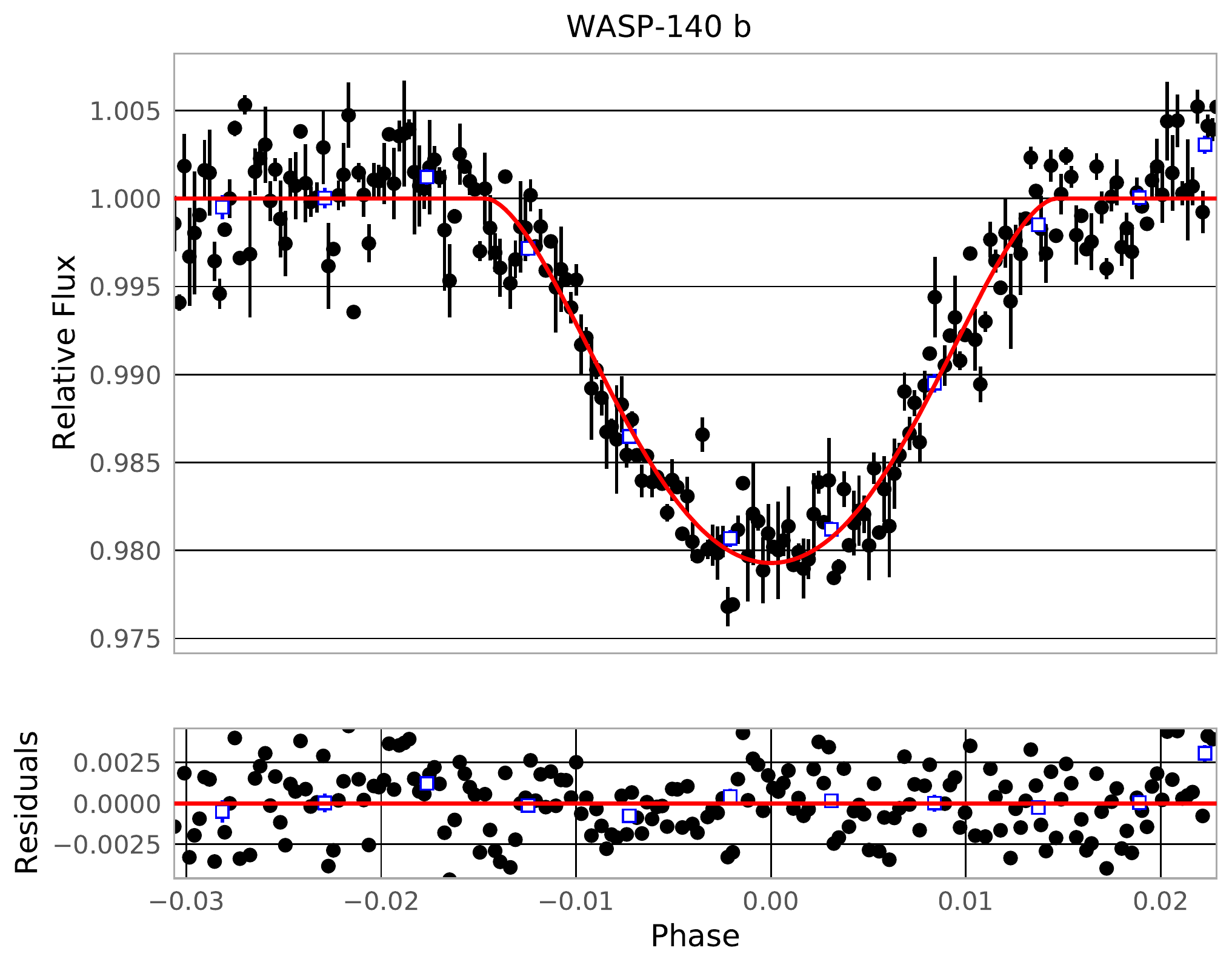}
  \end{minipage}
    \begin{minipage}[b]{0.5\linewidth}
    \centering
    \includegraphics[width=.7\columnwidth]{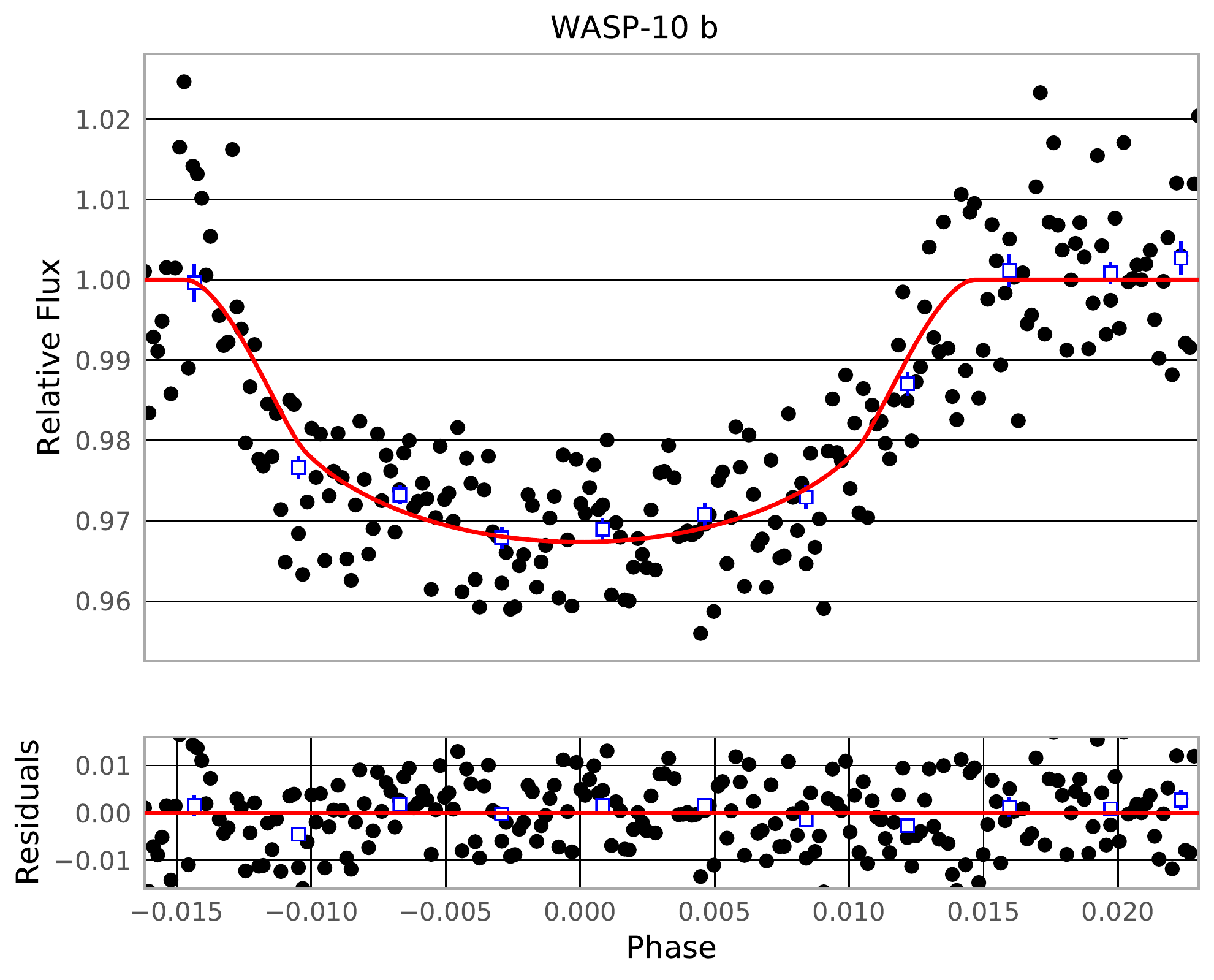}
  \end{minipage}
  \begin{minipage}[b]{0.5\linewidth}
    \centering
        \includegraphics[width=.7\columnwidth]{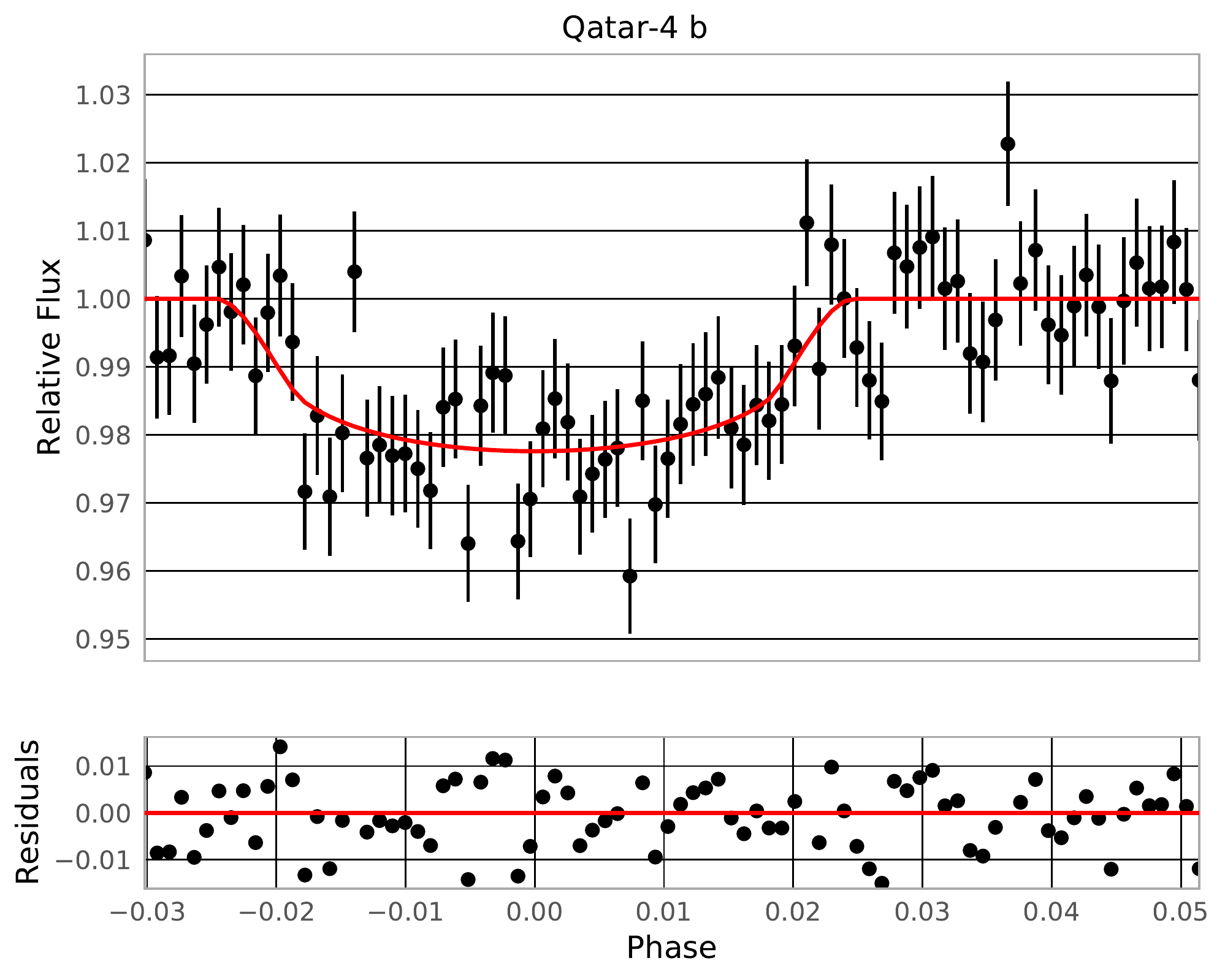}
  \end{minipage} 
    \begin{minipage}[b]{0.5\linewidth}
    \centering
        \includegraphics[width=.7\columnwidth]{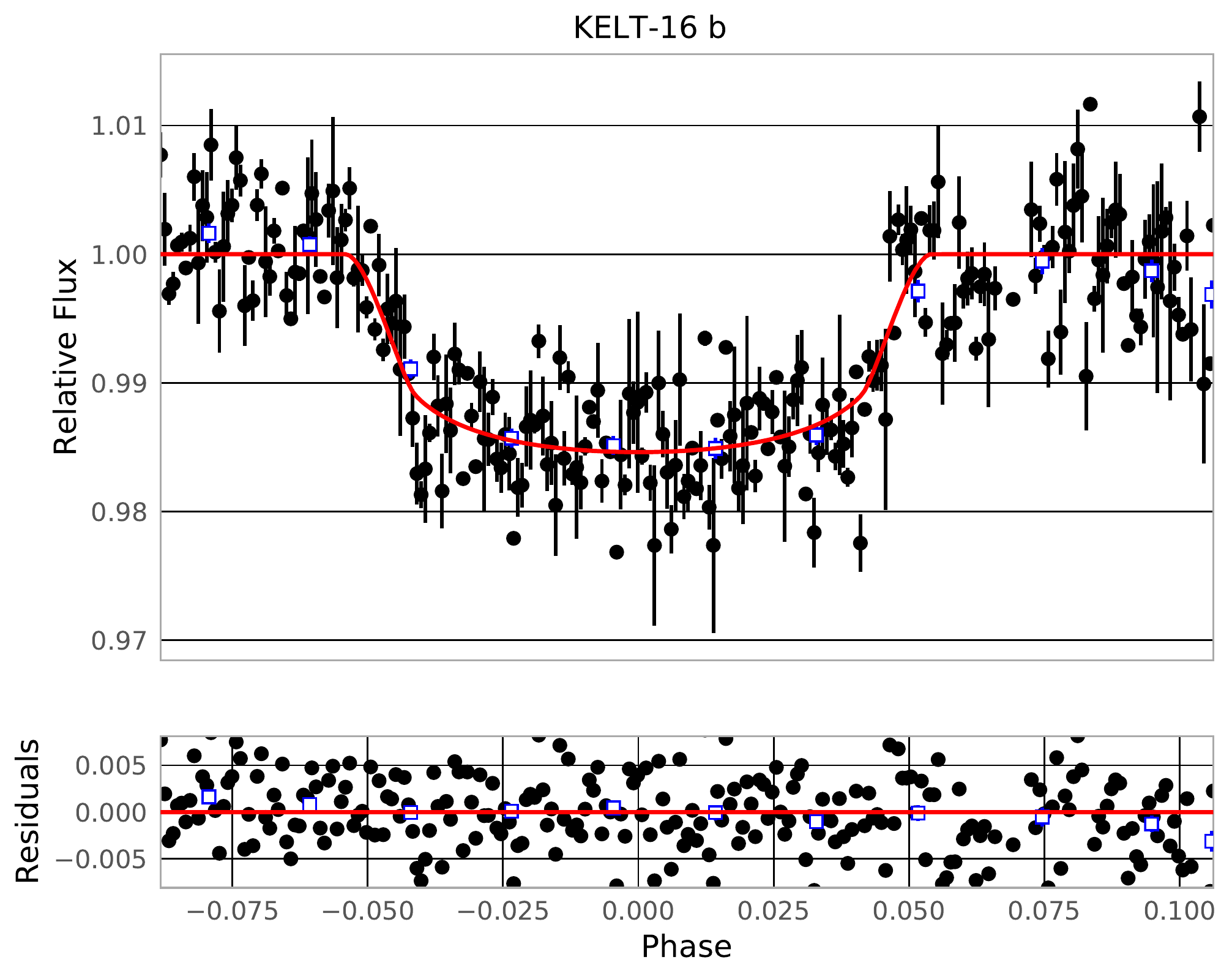}
  \end{minipage} 
  \begin{minipage}[b]{0.5\linewidth}
    \centering
    \includegraphics[width=.7\columnwidth]{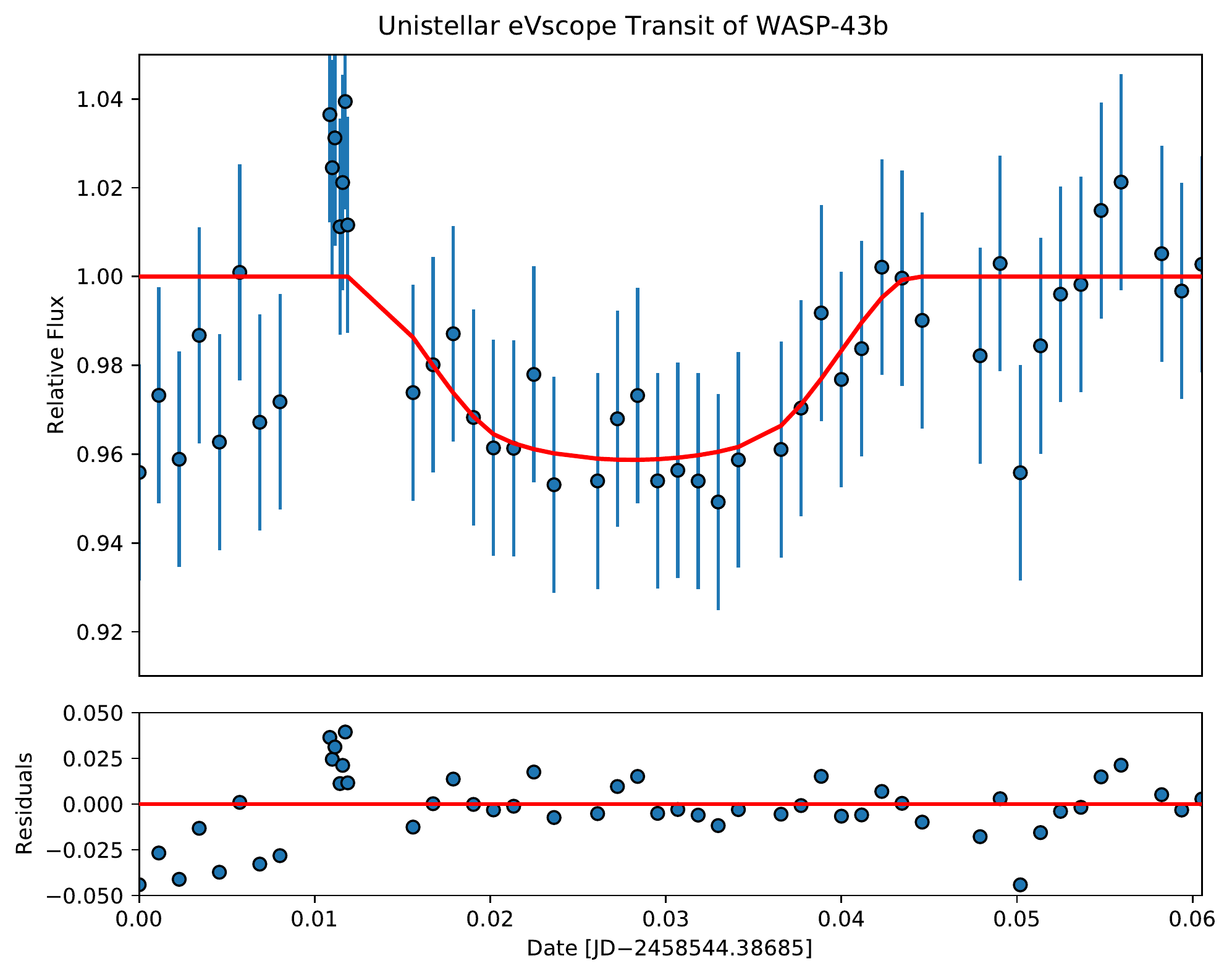}
  \end{minipage}
  \begin{minipage}[b]{0.5\linewidth}
    \centering
        \includegraphics[width=.7\columnwidth]{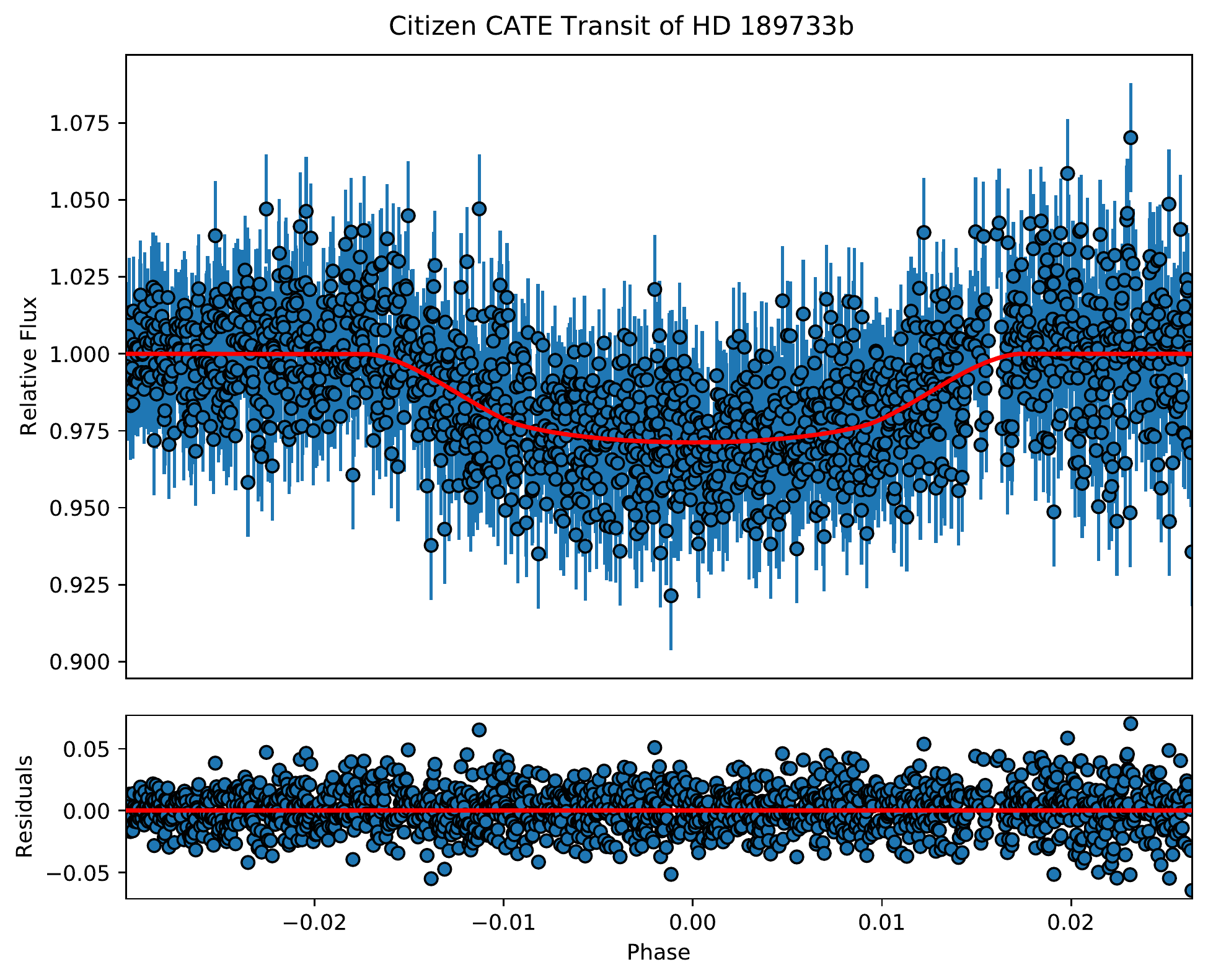}
  \end{minipage} 
    \begin{minipage}[b]{0.5\linewidth}
    \centering
    \caption{{{Sample transiting exoplanets observed with small telescopes. $First\; Row$: Las Cumbres Observatory 0.4-m observations of WASP-76b (9.5~V-mag) and WASP-140b (11.1~V-mag).
$Second\; Row$: WASP-10b (12.7~V-mag; $Left$) observed with a 16-inch (40.64-cm) Sierra Remote Observatory robotic telescope and Qatar-4b (13.60~V-mag; $Right$) observed with a 14-inch (35.56-cm) telescope. $Third\; Row$: KELT-16b (11.898~V-mag; $Left$) observed with an 11-inch (27.94-cm) telescope and WASP-43b (V-mag = 12.4) observed with a 4.5-inch (11.43-cm) Unistellar eVscope \citep{marchis20}. $Bottom$: HD~189733b (V-mag = 7.648) observed with a 3.15-inch (8-cm) Citizen CATE \citep{penn17, penn19}. All the datasets, except WASP-43b, were operated and/or reduced by amateur astronomers and the first five lightcurves were fully reduced with EXOTIC (Section~\ref{sec:EXOTIC}). {{Blue squares are the bins of 10 datapoints to help aid the eye.}}}}}
  \end{minipage} 
 		\label{fig:smallscopes}
\end{figure*}

{{To aid the citizen science community, we have developed the EXOplanet Transit Interpretation Code (EXOTIC\footnote{https://github.com/blaserethan/EXOTIC}), a complete transit data reduction tool written in Python3 and aimed at the amateur astronomer. EXOTIC can start with either raw fits files or a pre-reduced timeseries. If raw fits files are used, EXOTIC can calibrate them with flats, darks, and biases, and then perform multi-object optimal aperture photometry. The ideal aperture size, sky annulus (for background count estimation subtraction), and comparison star selection is determined by minimizing the residual scatter in the data after performing a least-squares fit using the \texttt{scipy.optimize.least$\textunderscore$squares} package \citep{scipy} to the data with a \citet{mandelagol02} model transit lightcurve. These optimal raw photometric data  (i.e., the combination of the comparison star, aperture size, and sky annulus that produces the least scatter) are then jointly fit by a  model lightcurve and an exponential airmass correction function \citep[e.g.,][]{swain10} with a Markov Chain Monte Carlo \citep[MCMC; e.g.,][]{ford05}. The priors for the MCMC are automatically scraped from the NASA Exoplanet Archive and the limb-darkening parameters are generated from EXOFAST \citep{eastman13}; all priors are then confirmed by the user. EXOTIC then calculates the 1-$\sigma$ uncertainties on the mid-transit time and transit depth {{(R$_{\mathrm{p}}$/R$_{\mathrm{s}}$)$^{2}$}} from the MCMC posteriors. EXOTIC can alternatively start with pre-reduced data (time-varying flux) and fit it with a model lightcurve. It also can reduce data in real-time to provide a useful visualization, e.g., for star parties. The EXOTIC reduction code has been fully-tested on five different observing platforms and was used to generate all of the data in Figures~\ref{fig:MOscopes1} and \ref{fig:MOscopes2} as well as the first five lightcurves in Figure~\ref{fig:smallscopes}.}}

{{Using EXOTIC, we find that even a single 6-inch (15.24-cm) telescope can produce high precision data (Table~\ref{tab:MO_14} and Figs.~\ref{fig:MOscopes1} and \ref{fig:MOscopes2}). For the relatively dim 12.29~V-mag HAT-P-32b, it achieves precisions up to 1.02~min for the mid-transit time, a 0.05\% transit depth precision {{$\Delta$(R$_{\mathrm{p}}$/R$_{\mathrm{s}}$)$^{2}$}} (a 37.58$\sigma$ detection), and a 0.60\% residual RMS scatter, despite the target drifting $\sim$300~pixels on the focal plane over the course of the observation.}}

\begin{table*}
  \caption{{{Measured Uncertainties for 14 Transiting Exoplanet Observations with a Single 6-inch (15.24-cm) MicroObservatory Telescope}}}
  \label{tab:MO_14}
  \begin{center}
    \resizebox{1\textwidth}{!}{%
\begin{tabular}{l|c|c|c|c|c|c}
\hline
\hline
\textbf{Planet} & \textbf{Host Star} & \textbf{Residual RMS} & \textbf{Mid-transit Time $T_{0}$ } & \textbf{Mid-transit Time} & \textbf{Transit Depth} & \textbf{Transit Depth} \\
\textbf{Name} & \textbf{V-mag} & \textbf{Scatter [\%/min]} & \textbf{[BJD$\_{TDB}$ $-$ 2400000.5]} & \textbf{Uncertainty $\Delta T_{0}$ [min]}& \textbf{(R$_{\mathrm{p}}$/R$_{\mathrm{s}}$)$^{2}$ [\%]} &\textbf{Uncertainty $\Delta$(R$_{\mathrm{p}}$/R$_{\mathrm{s}}$)$^{2}$ [\%]} \\
\hline
CoRoT-2b & 12.57 & 0.75 & 58746.1678 & 4.27 & 2.59 & 0.35 \\
HAT-P-20b & 11.339 & 0.73 & 58131.1368 & 4.38 & 1.66 & 0.31 \\
HAT-P-20b & 11.339 & 0.52 & 58818.3441 & 5.52 & 1.67 & 0.3 \\
HAT-P-32b & 11.289 & 0.83 & 58739.3143 & 5.5 & 1.81 & 0.29 \\
HAT-P-32b & 11.289 & 0.6 & 58107.213 & 1.02 & 1.95 & 0.05 \\
TrES-1b & 11.76 & 0.7 & 58722.2486 & 3.54 & 1.77 & 0.23 \\
WASP-10b & 12.7 & 1.06 & 58789.2378 & 8.33 & 2.02 & 0.54 \\
WASP-10b & 12.7 & 1.18 & 58721.2001 & 5.15 & 2.86 & 0.48 \\
WASP-10b & 12.7 & 0.95 & 58755.2135 & 4.62 & 2.36 & 0.33 \\
WASP-10b & 12.7 & 1.12 & 58758.3113 & 5.59 & 2.24 & 0.36 \\
WASP-52b & 12.0 & 1.27 & 58771.3086 & 4.82 & 1.95 & 0.4 \\
WASP-52b & 12.0 & 0.97 & 58764.3076 & 4.7 & 2.42 & 0.36 \\
WASP-52b & 12.0 & 0.86 & 58757.3088 & 4.07 & 2.3 & 0.34 \\
WASP-52b & 12.0 & 0.7 & 58038.1527 & 3.41 & 2.85 & 0.37 \\
\end{tabular}}
\end{center}
\end{table*}

\section{Demonstrating and Forecasting the Capability of Small Telescopes}\label{sec:network}
\begin{figure}
\centering
\includegraphics[width=.8\columnwidth]{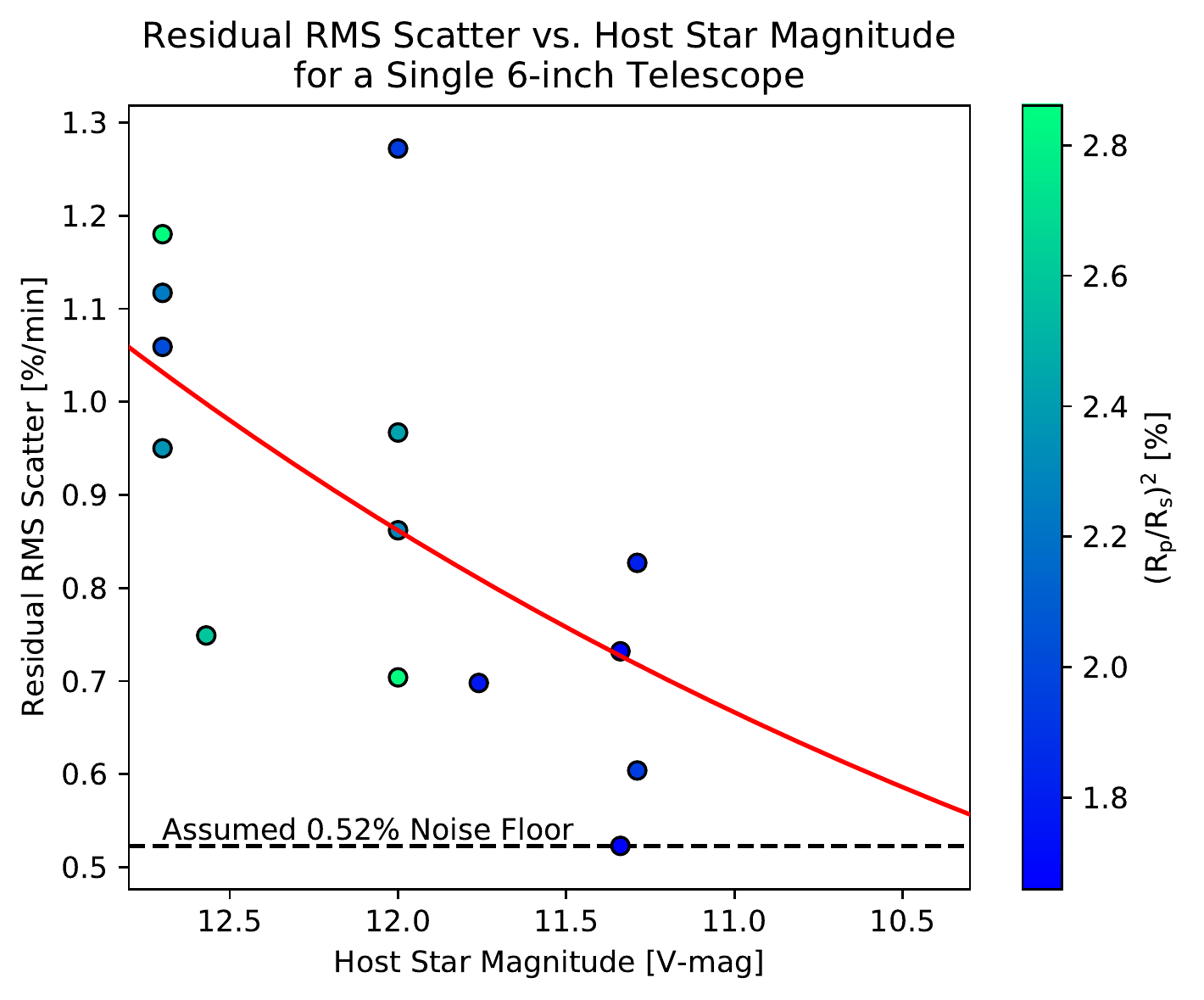}
\includegraphics[width=.8\columnwidth]{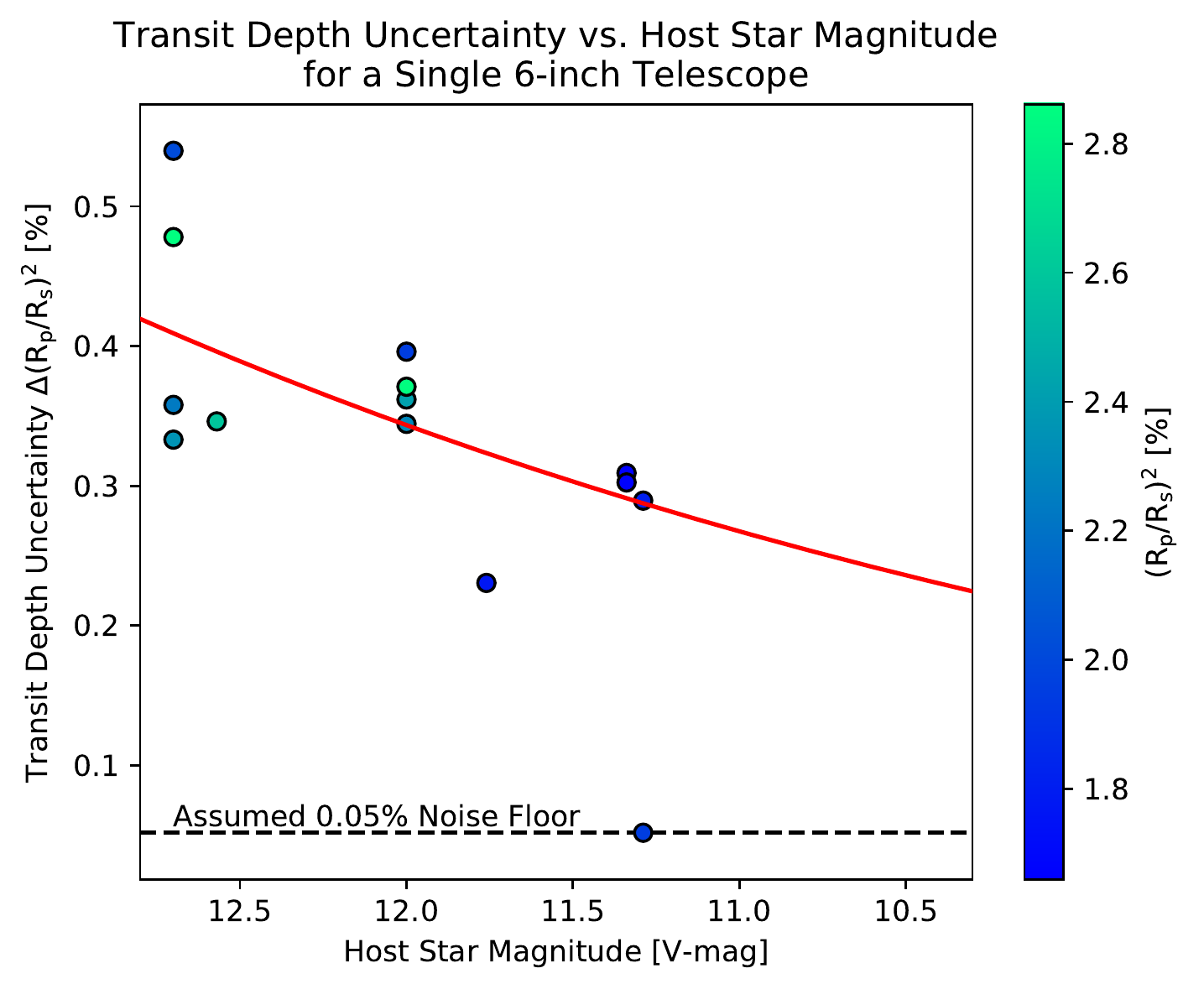}
\includegraphics[width=.8\columnwidth]{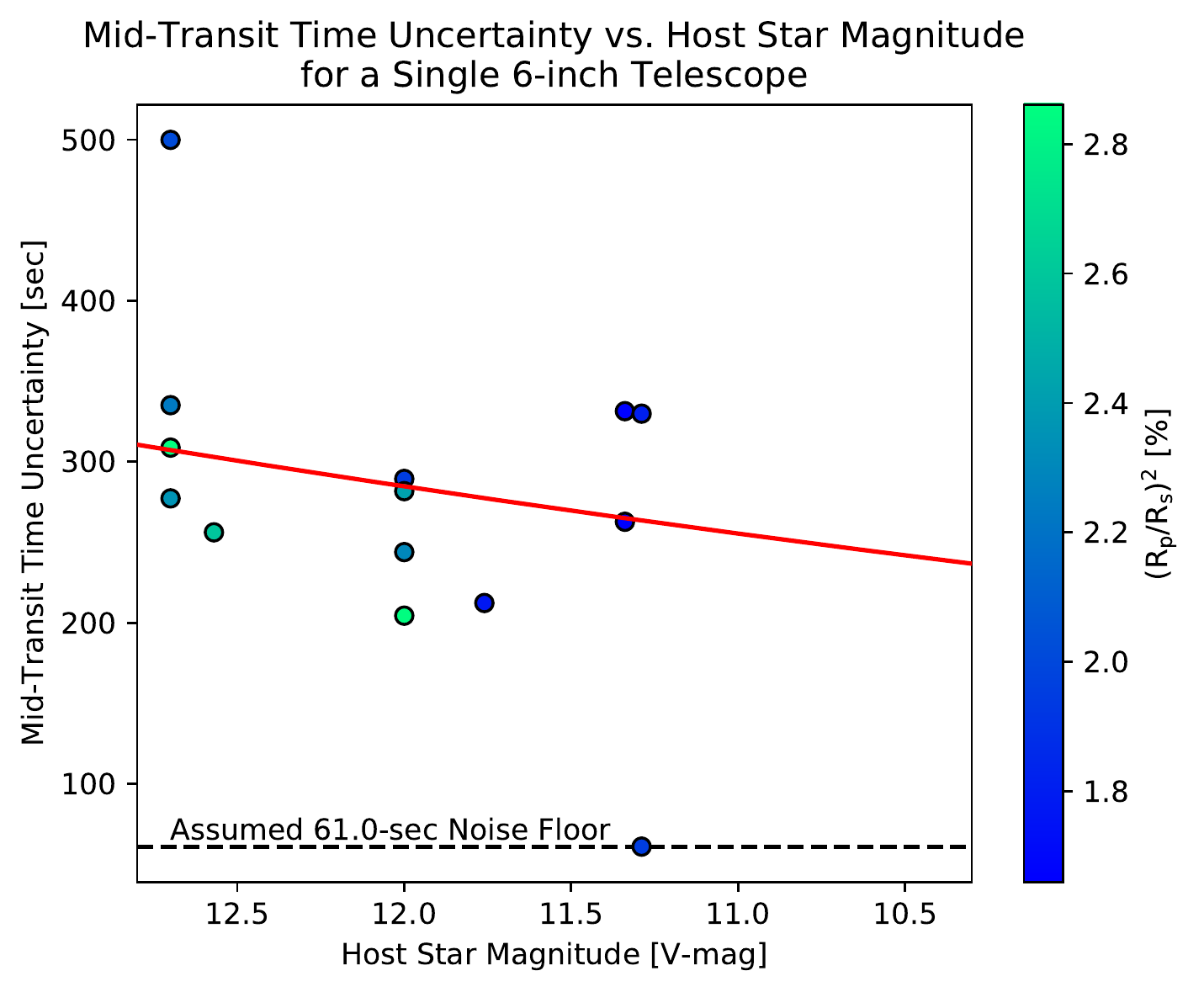}
\caption{{{Here we present the measured residual RMS scatter ($Top$), transit depth uncertainty {{$\Delta$(R$_{\mathrm{p}}$/R$_{\mathrm{s}}$)$^{2}$}} ($Middle$), and mid-transit time uncertainty ($Bottom$) as a function of an exoplanet's host star V-magnitude and transit depth {{(R$_{\mathrm{p}}$/R$_{\mathrm{s}}$)$^{2}$}} for 14 transit observations with a single 6-inch (15.24-cm) telescope. We fit each dataset with a function (red line) that assumes that each precision scales with the square root of the number of photons and conservatively adopt a noise floor that is equal to the minimum measured values. Thus we empirically determine how each parameter varies with an exoplanet's host star V-magnitude.}}}
\label{fig:MO_survey}
\end{figure}

{{By analyzing the 14 transit observations by a MicroObservatory telescope (Figs.~\ref{fig:MOscopes1} and \ref{fig:MOscopes2}), we forecast the capability of a single 6-inch telescope (adopted here as a typical aperture size operated by a citizen scientist) by empirically deriving the measured residual RMS scatter, mid-transit time uncertainty, and transit depth uncertainty {{$\Delta$(R$_{\mathrm{p}}$/R$_{\mathrm{s}}$)$^{2}$}} as functions of its host star's V-magnitude (Fig.~\ref{fig:MO_survey}). These relationships assume that the host star brightness is the largest determiner of the observed precision and that, to first order, these uncertainties scale proportionally to the square root of the number of photons, as indicated in Figure~\ref{fig:MO_survey}. We also conservatively adopt systematic noise floors that are equal to the {{highest}} observed precisions (0.52\% on the RMS scatter, 0.05\% on the transit depth {{$\Delta$(R$_{\mathrm{p}}$/R$_{\mathrm{s}}$)$^{2}$}}, and 1.02-min on the mid-transit time). {{We also conservatively remove these minimum values from our fits as they achieve comparatively much higher precision than the other measurements, in particular the transit depth and mid-transit time uncertainties, and might not necessarily reflect typical observations by a single 6-inch telescope. We also find that all three of these uncertainties are not dependent upon the measured planet-to-star radius ratio and}} that binning in time does not appreciably increase the precision of the MicroObservatory. For example, a single telescope achieves an average precision of {{0.83\% per minute}} for HAT-P-32b (Fig.~\ref{tab:MO_14}); extrapolating this precision over the entire 187.23-minute transit \citep{wang19b}, one would expect it to measure HAT-P-32b's transit depth with a precision of {{{{$\Delta$(R$_{\mathrm{p}}$/R$_{\mathrm{s}}$)$^{2}$}} = 0.06\%}}, however it only achieves a precision of {{0.29\%}}. Therefore we conservatively assume that the transit duration is largely negligible in determining the measured transit precision.}}

\begin{figure}
\centering
\includegraphics[width=1\columnwidth]{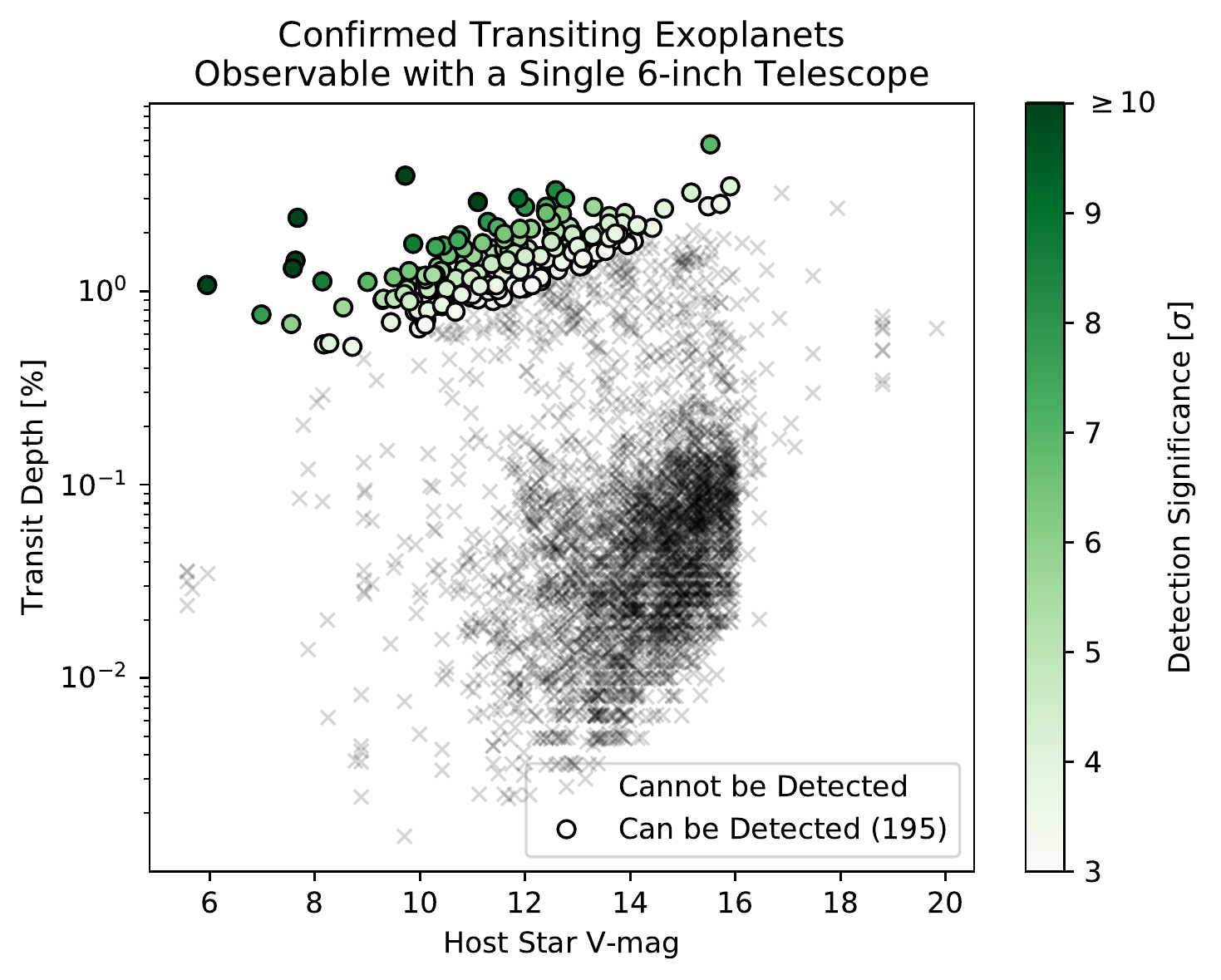}
\includegraphics[width=1\columnwidth]{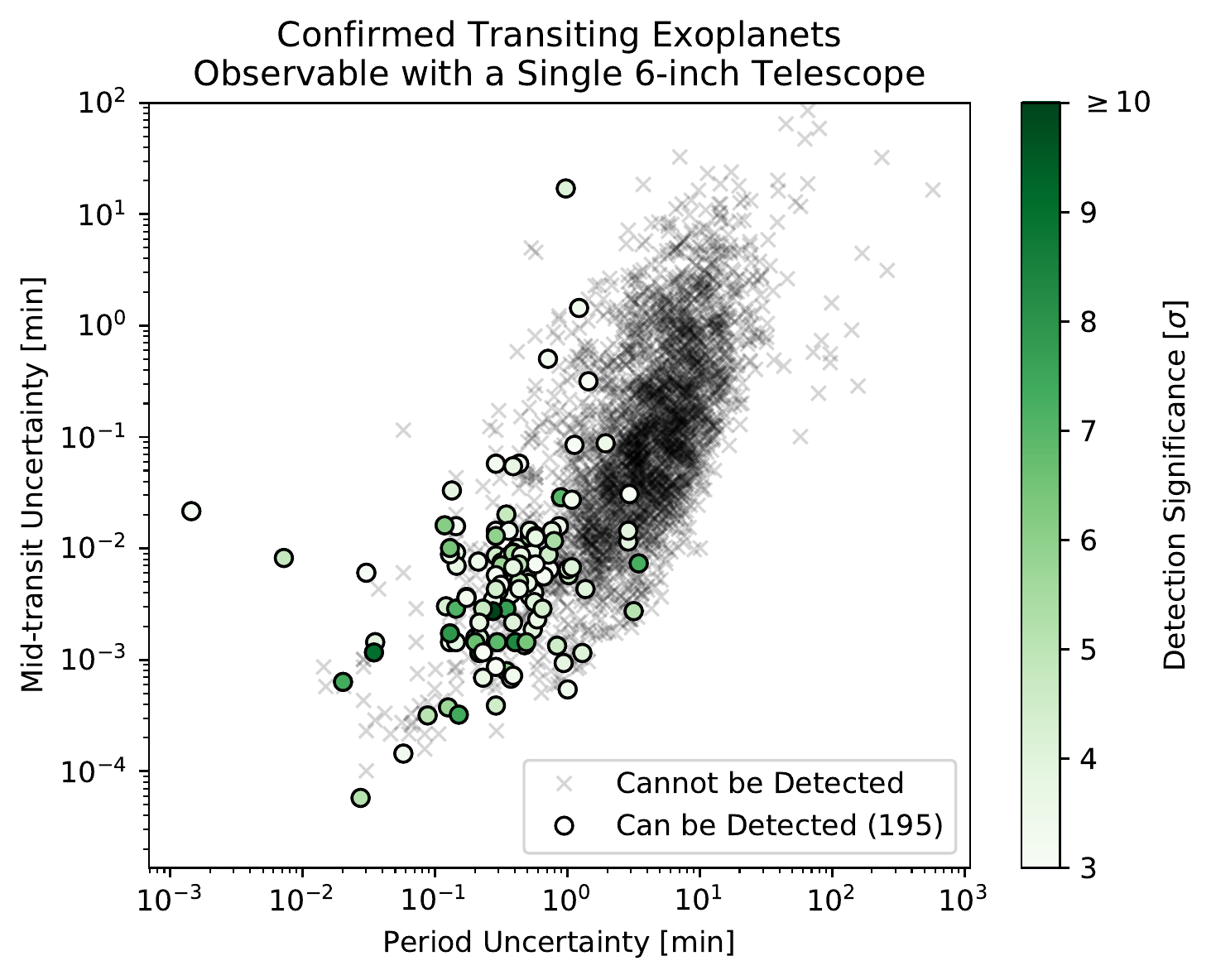}
\caption{{{Using the empirically-derived functions in Figure~\ref{fig:MO_survey}, we determine that a single 6-inch telescope can observe the transit of 195 currently-known exoplanets to $\geq$3$\sigma$ (circles), some of with have large timing uncertainties. Here we plot these targets' transit depth {{(R$_{\mathrm{p}}$/R$_{\mathrm{s}}$)$^{2}$}} versus their host star V-magnitude ($Top$) and their mid-transit uncertainty versus their period uncertainty ($Bottom$) in context with the other confirmed exoplanets that cannot be detected by these small telescopes (grey Xs).}}}
\label{fig:confirmed_obs_6inch}
\end{figure}

{{Using these empirically-derived functions, we estimate that a single 6-inch telescope can measure to 3$\sigma$ the transits of 195 of the currently-known transiting exoplanets listed on the NASA Exoplanet Archive (Fig.~\ref{fig:confirmed_obs_6inch}). Thus, even now, there are many targets that are accessible with a small telescope. {{We anticipate that this number will only increase as more planets are discovered by TESS, which is predicted to discover hundreds, if not thousands, of exoplanets that feature large transit depths orbiting bright host stars \citep{sullivan15, barclay18}.}} We also find that of the top 20 currently-known targets (Tables~\ref{tab:deltaTestimates_JWST}--\ref{tab:deltaTestimates_Astro2020}) ranked by our Figure of Merit $FOM_{maint}$ (Eqn.~\ref{eqn:maint_FOM}), a single 6-inch telescope can measure the transit depth {{(R$_{\mathrm{p}}$/R$_{\mathrm{s}}$)$^{2}$}} of 5 of the potential JWST targets to 3$\sigma$ (Table~\ref{tab:deltaTestimates_JWST}), thereby providing more accurate transit timings for future follow-up. Assuming root-n scaling, we also estimate the number of 6-inch telescopes, {{when observing simultaneously,}} required to detect these transits. We find that a modest-sized network of {{16}} 6-inch telescopes operated by citizen scientists can increase the number of 3$\sigma$ detections from 5 to 8. Larger telescopes, or a larger network, would be required to monitor the remaining dimmer and/or shallower targets.}}

\begin{figure}
\centering
\includegraphics[width=.9\columnwidth]{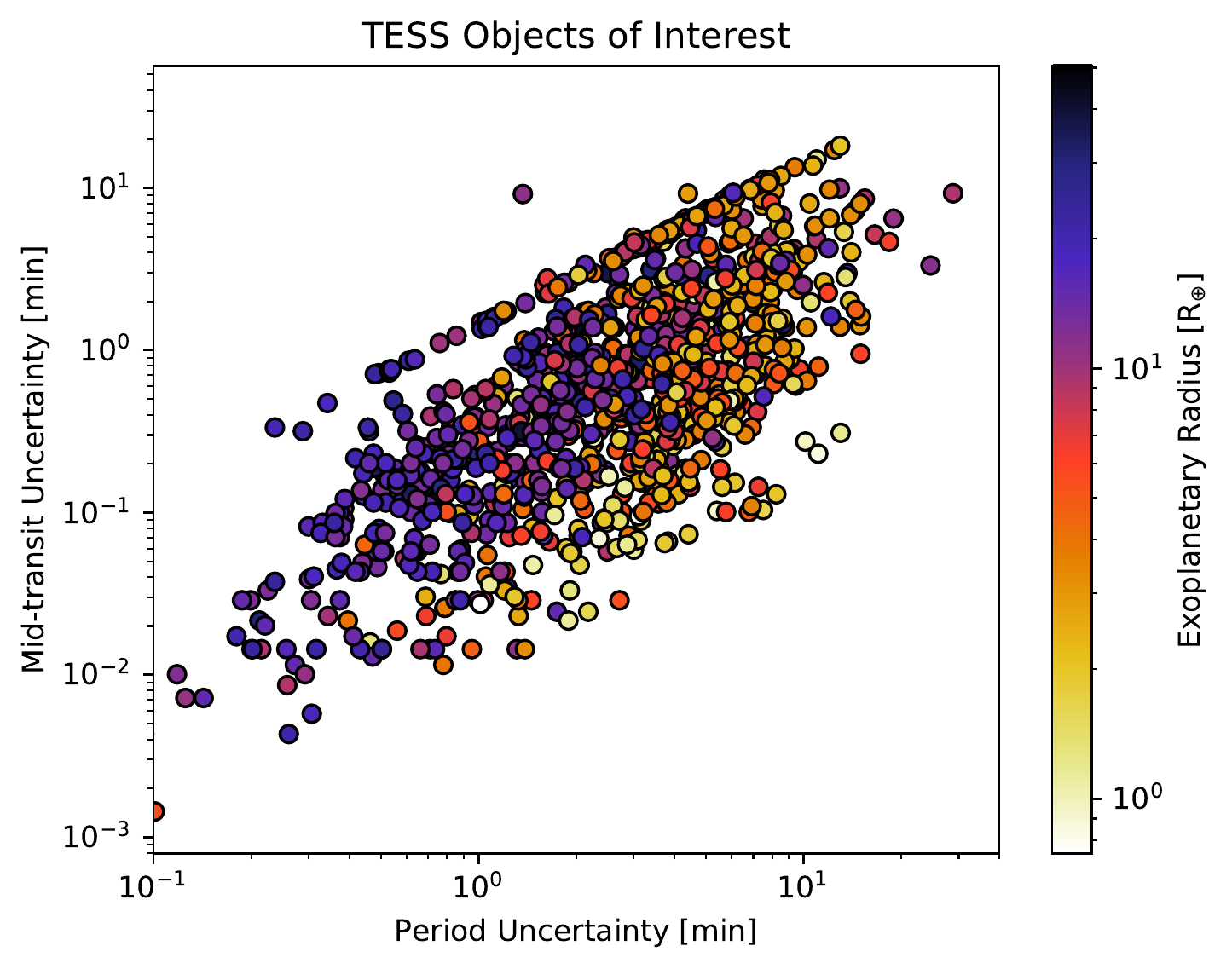}
\includegraphics[width=.9\columnwidth]{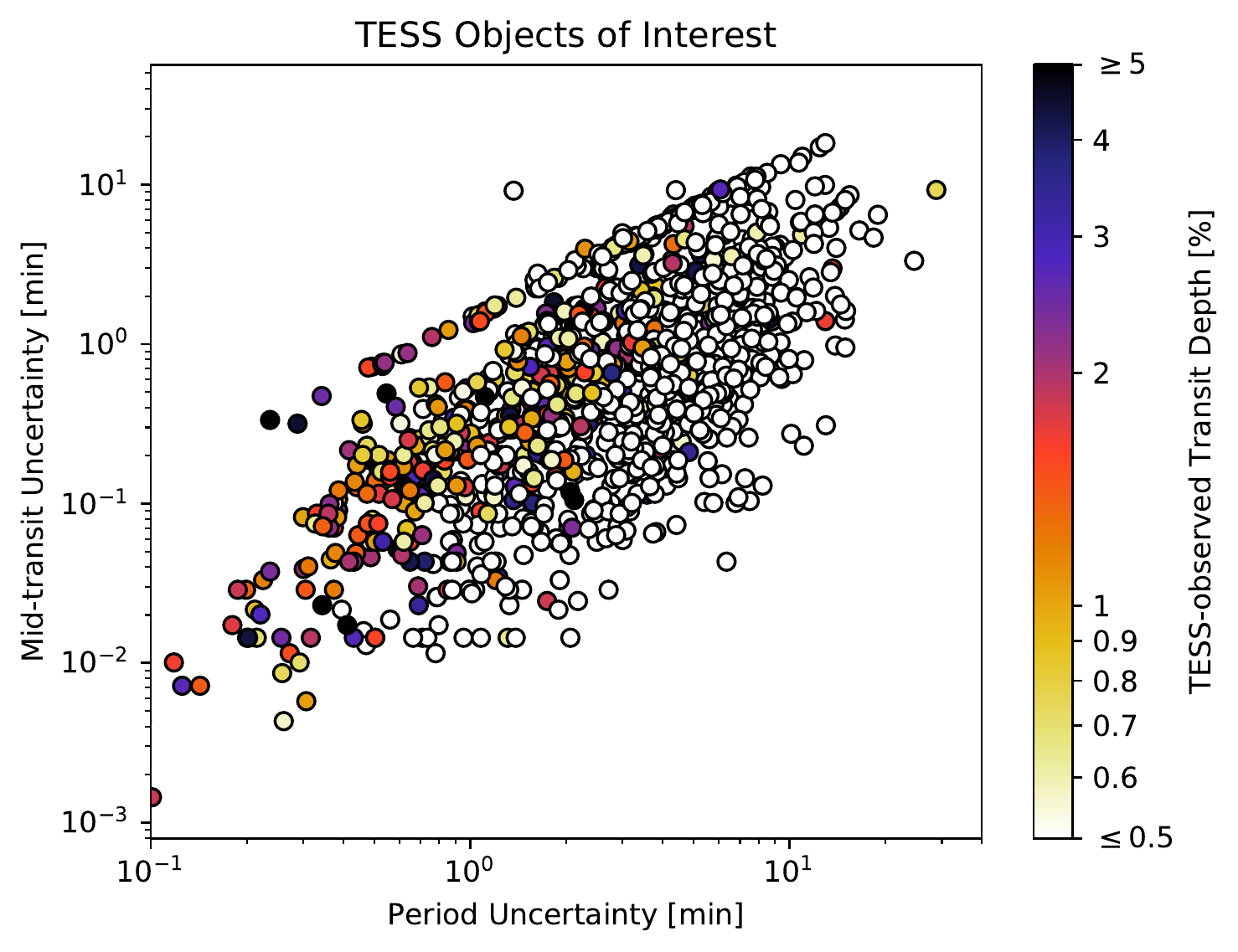}
\includegraphics[width=.9\columnwidth]{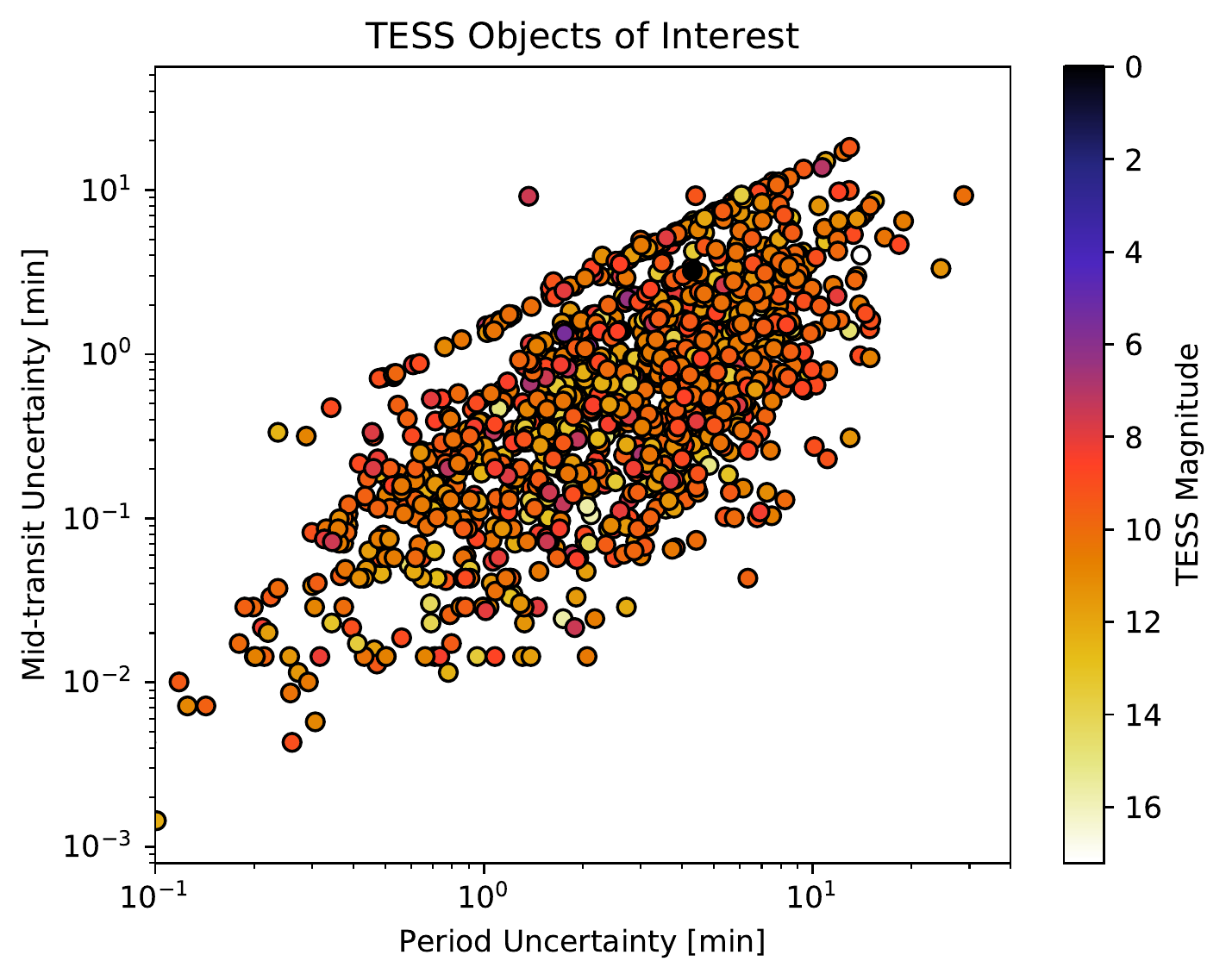}
\caption{Measured mid-transit uncertainty as a function of orbital period uncertainty and planet radius ($top$), transit depth ($middle$), and TESS mag ($bottom$) of the current TESS Objects of Interest. These targets feature large transits and bright host stars. Some planets, particularly those with Earth-sized radii, have large ($\sim$10~minute) uncertainties and could be accessible by smaller telescopes (Figs.~\ref{fig:MOscopes1}--\ref{fig:confirmed_obs_6inch} and Table~\ref{tab:MO_14}). The structure along the top of each plot in the form of a line of points where $\Delta T_{mid} \approx \Delta P$ \citep[compare to Fig.~1 in][]{kane18}. }
\label{fig:TOIs}
\end{figure}

At the conclusion of TESS's primary mission, it is predicted that there will be thousands of transiting exoplanets that are both bright and have large spectral modulation and are therefore ideal for detailed spectroscopic characterization with large platforms \citep{zellem17}. This prediction is supported by the population of the current TOIs (Fig.~\ref{fig:TOIs}), which feature planets with large transit depths and bright host stars and, in some cases, large uncertainties in their orbital period and mid-transit time. {{Some of these targets, due to their bright host stars and large transit depths, are accessible to smaller ($\leq$1-m) telescopes.}} By comparing Figure~\ref{fig:TOIs} with Figures~\ref{fig:MOscopes1}--\ref{fig:confirmed_obs_6inch} and Table~\ref{tab:MO_14}, we can see that many of these targets can be observed with relatively-small telescopes at high statistical significance, alleviating the need for larger observatories to keep their transit times fresh, allowing them to follow-up dimmer targets or smaller transiting exoplanets instead.

\subsection{Advocating for a Network of Small Telescopes Operated by Citizen Scientists}
A network of smaller telescopes {{($\gtrsim${{16}} $\geq$6-inch telescopes)}} could rapidly respond to new discoveries and high priority bright targets with large transit depths and monitor them, allowing larger telescopes to spend their time on other targets (e.g., Earth-sized planets transiting dim M-dwarf stars). Citizen scientists, in particular, provide a unique opportunity to the professional astronomical community: a large number of observers who are eager to aid NASA's mission goals and contribute to the observational needs of the professional community \citep[e.g.,][]{croll11, wigginscrowston11, catlingroves12, croll12, franzonisauermann14, marshall15, kuchner16, burdanov18, collins18, watson18}.

\subsection{Examples of Successful Citizen Science Efforts}
Specific programs that have leveraged observations by citizen scientists include the OSIRIS-REx mission’s Target Asteroids! (TA!) and the Astronomical League’s companion Target NEOs! (TNEO!) measure much needed astrometry and photometry over a wide range of phase angles \citep{hergenrotherhill13, hergenrother13, hergenrother14, lauretta17}.

A similar telescope network has demonstrated success in supporting observations of asteriod occultation events as well as photometric light curves of asteroids leading to the discovery of moons around asteroids \citep{timerson13, pravec16}, the shape and size of large main-belt asteroids \citep{hanus17}, as well as the first ring around an asteroid \citep{bragaribas14}.

In addition, the Kilodegree Extremely Little Telescope follow-up network (KELT-FUN) employs citizen scientists, students, and professionals to confirm KELT transit detections \citep{collins18} via increased precision and cadence, higher spatial resolution to search for stellar blends, and spectroscopy to determine the stellar type and properties of transit hosts. KELT-FUN has helped discover $>$20 transiting planets.

Lastly, with the increasing involvement of amateur astronomers in professional-amateur exoplanet collaborations, the American Association of Variable Star Observers (AAVSO) created an Exoplanet Section\footnote{https://www.aavso.org/exoplanet-section} in November 2015. Since then, AAVSO members have participated in a major Hubble study whose purpose was to characterize the atmosphere of some 15 confirmed exoplanets, they have been members of the KELT-FUN team, and more recently, they have played a key role in the TESS mission. With the latter for example, amateurs are providing follow-up observations of exoplanet candidates in order to help distinguish false positives from true exoplanet transits. An AAVSO Exoplanet Database has also been created as a repository for observations of confirmed exoplanets whose purpose is to help refine their ephemerides, as well as to support TTV studies. Similar efforts are being undertaken by other amateur astronomy groups, such as the British Astronomical Association's Exoplanet Division \footnote{https://britastro.org/section$\textunderscore$front/15474}. 

\subsection{Science Returns from a Transiting Exoplanet Citizen Science Network}
\subsubsection{Providing Accurate Transit Ephemerides}\label{sec:providing_ephemerides}
A small-telescope network {{($\gtrsim${{16}} $\geq$6-inch telescopes)}} operated by citizen scientists dedicated to exoplanet ephemerides maintenance could collect observational data that would be provided to the science community for planning efficient future observations. (This facet in particular is inspired by the Czech Astronomical Society's Transiting Exoplanets and Candidates\footnote{http://var2.astro.cz/EN/tresca/index.php} program and Exoplanet Transit Database\footnote{http://var2.astro.cz/ETD/} \citep{etd10}.) If a particular exoplanet's host star is bright enough, its transit depth is deep enough, and its transit duration is long enough, then a transiting exoplanet is accessible, sometimes at high statistical significance, to smaller telescopes ($\leq$1-m; Fig.~\ref{fig:smallscopes}).

\begin{figure}[htbp]
		\includegraphics[width=1\columnwidth]{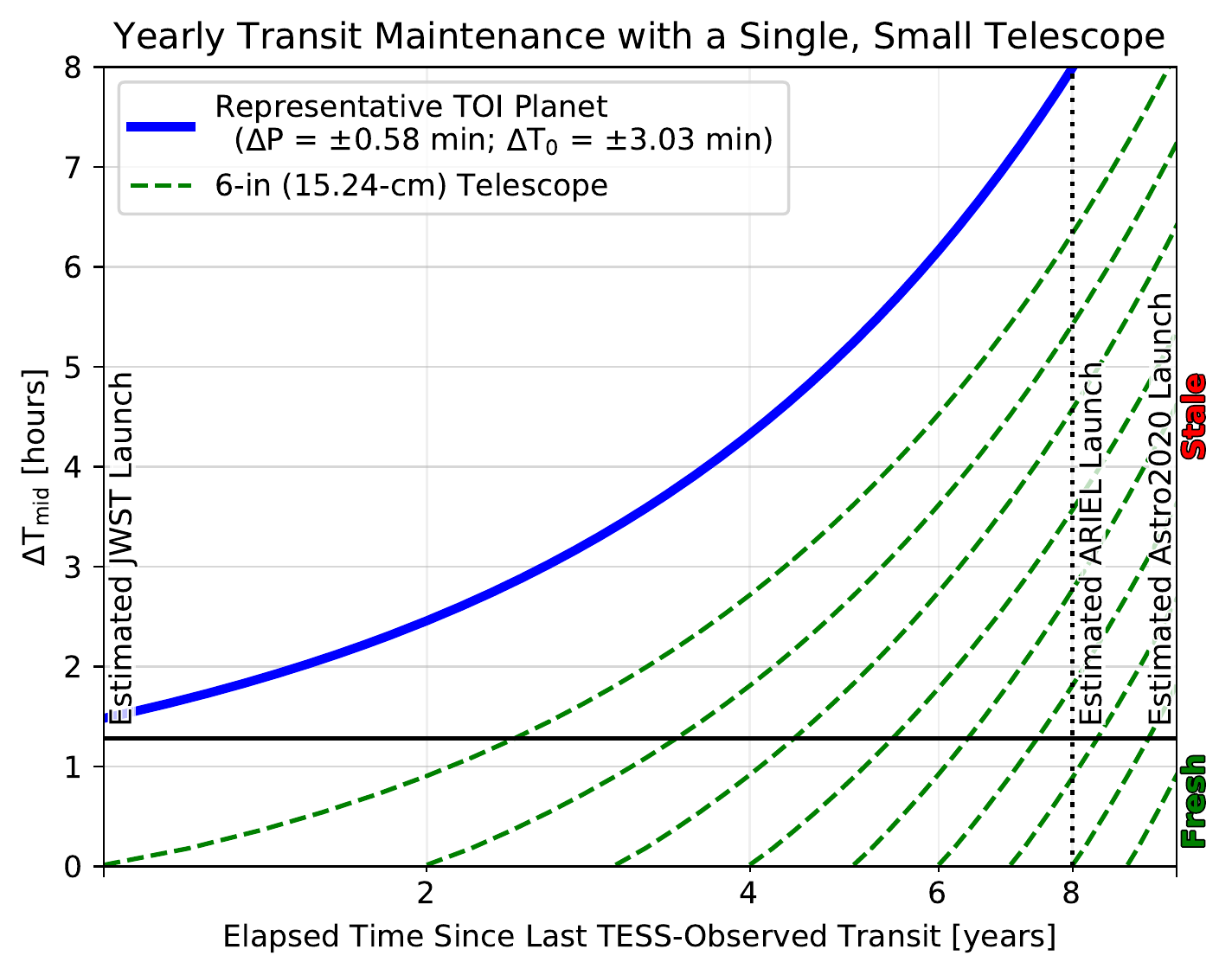}
		\caption{The mid-transit uncertainty as a function of the elapsed time since the last TESS observation of the Representative TOI Planet (thick blue solid line) and  {{cumulative observations with a single 6-inch telescope, repeated yearly (dashed green lines). Yearly, routine transit maintenance by a single 6-inch telescope can keep the Representative TOI Planet fresh for about a year at a time.}}}
	\label{fig:repplanet}
\end{figure}

{{We estimate the impact of a single 6-inch telescope performing ephemeris maintenance on the Representative TOI Planet via routine, yearly observations. We first forward-propagate the Representative TOI mid-transit uncertainty using Equation~\ref{eqn:dt_next} and estimate the uncertainty on the mid-transit time with a single 6-inch telescope (Fig.~\ref{fig:MO_survey}). Using these two measured mid-transit times and uncertainties (one from TESS, the other from a single 6-inch telescope) and the orbital period and uncertainty measured by TESS, we solve Equation~\ref{eqn:t_next} for the planet's period $P$ and mid-transit time $T_{0}$ and associated uncertainties using a Markov Chain Monte Carlo \citep[MCMC; e.g.,][]{ford05}. We then simulate an observing campaign of yearly revisits by the single 6-inch telescope, taking into account its cumulative {{(previous)}} observations. We present the results of these simulations in Figure~\ref{fig:repplanet} and find that even a single 6-inch telescope can powerfully keep the Representative TOI Planet fresh for $\sim$1~year, ensuring efficient follow-up by JWST, ARIEL, and an Astro2020 mission.}}

We next perform a {{detailed analysis of the 1000-planet population used by the CASE team for their independent study of ARIEL's capabilities \citep{zellem19b}. This target list is approximately the same size as the ARIEL mission Tier-1 sample \citep{tinetti16, tinetti18, edwards19}, contains the currently-known transiting exoplanets and those predicted to be discovered with TESS \citep{sullivan15}, and is chosen by a figure of merit that prioritizes planets with large scale heights orbiting bright stars \citep[Eqn.~\ref{eqn:transit_FOM};][]{cowan15, zellem17, goyal18, kempton18, morgan18}.}} 

{{To estimate the time saved by a citizen science ground-based monitoring effort, we adopt the mid-transit times, orbital periods, and their associated uncertainties from the NASA Exoplanet Archive \citep{akeson13} for the known planets. For the predicted TESS targets, we estimate each planet's mid-transit and period uncertainties from the current confirmed TESS-discovered exoplanets. We find that the mid-transit and period uncertainties of these planets do not scale with their host stars' V-magnitudes (Fig.~\ref{fig:TESS_Vmag_uncertainties}), {{potentially}} suggesting that these objects are in a systematic-dominated regime. {{Assuming the photon-limited regime, one would reasonably expect the measured precisions to scale with the number photons and thus the number of repeat observations. However, since the number of observations (by TESS and by follow-up observations from the TFOP) that went into determining these values are not immediately available to us, we have conservatively adopted the median uncertainties of 64.66~sec for $\Delta T_{0}$ and 26.35~sec for $\Delta P$, regardless of the host star's V-magnitude.}}

\begin{figure}[htbp]
		\includegraphics[width=1\columnwidth]{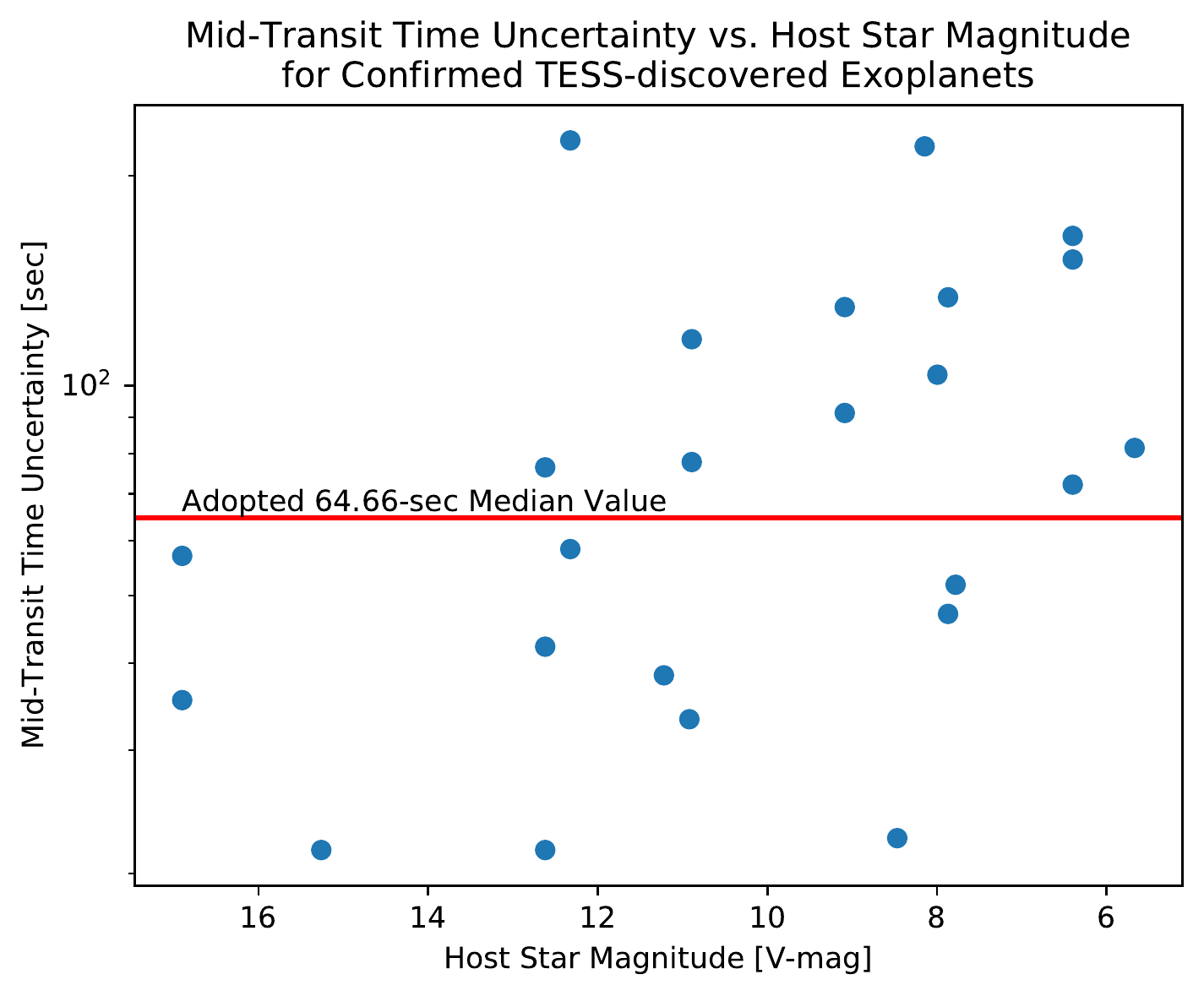}
		\includegraphics[width=1\columnwidth]{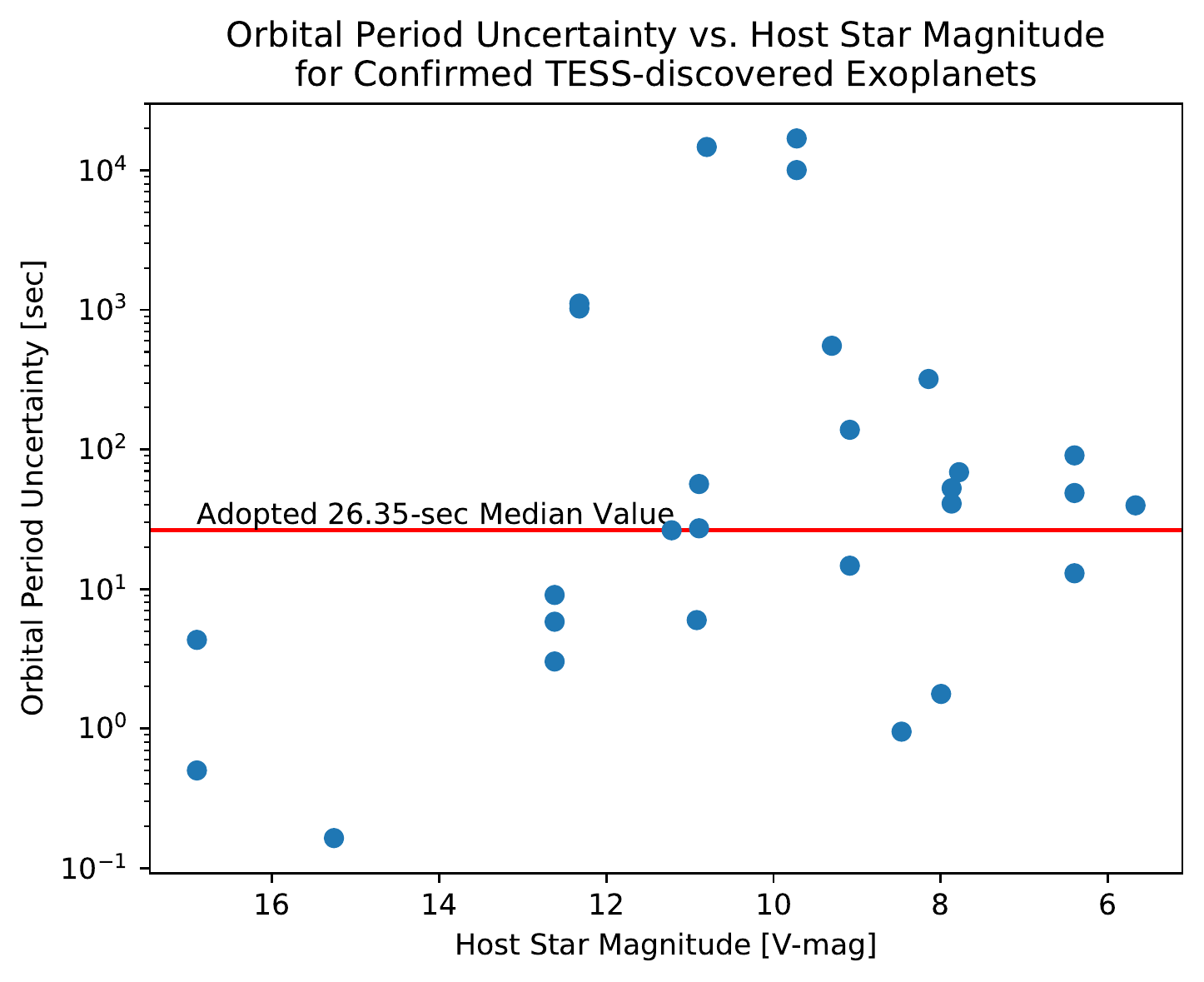}
		\caption{{{Similar to Figure~\ref{fig:MO_survey}, except for the confirmed TESS-discovered exoplanets. These planets seem to be in a systematic-dominated regime, as evidenced by their non-correlation between the mid-transit and orbital period uncertainties and the host star V-magnitude. Therefore, we conservatively adopt the median uncertainties of {{64.66~sec}} for $\Delta T_{0}$ and {{26.35~sec}} for $\Delta P$, regardless of the host star's V-mag. }}}
	\label{fig:TESS_Vmag_uncertainties}
\end{figure}

{{For each predicted TESS target, we adopt an ephemeris of the end of TESS's nominal mission lifetime ($\sim$2020). Then, using Equation~\ref{eqn:dt_next}, we propagate the uncertainty of each planet's mid-transit time at six different epochs: the beginning and end of the nominal missions for JWST, ARIEL, and an Astro2020 Decadal mission, such as HabEx \citep{gaudi18}, LUVOIR \citep{luvoir18}, or the Origins Space Telescope \citep{battersby18}. For each planet, we conservatively adopt a covariance term $\Delta P \Delta T_{0} = 0$ to provide a lower-limit estimate. We then simulate a mid-transit measurement of a single 6-inch telescope at each of these six epochs {{by estimating their mid-transit uncertainties from their host star V-magnitude, as}} described in detail in Section~\ref{sec:network} {{and Figure~\ref{fig:MO_survey}}} and use a MCMC to solve for a new mid-transit ephemeris (similar to the procedure described in detail in Section~\ref{sec:providing_ephemerides}), taking into account the 6-inch telescope observation as well as either the ephemeris reported on the NASA Exoplanet Archive for the real planets or the median TESS uncertainties (Fig.~\ref{fig:TESS_Vmag_uncertainties}) for the predicted TESS targets. The time saved by a small telescope is then the difference between the transit time uncertainty $\Delta T_{mid}$ propagated from values listed on the NASA Exoplanet Archive or simulated from TESS and the new reduced uncertainty due to an observation by a single 6-inch telescope. We estimate that even a single 6-inch telescope performing transit maintenance {{could observe the transit of 188 exoplanets to 3$\sigma$ and thus}} save $\sim$10,000 days for both JWST and ARIEL (Fig.~\ref{fig:timesaved_1000}, Top Left) {{while a network of 16 6-inch telescopes could observe 507 exoplanets and therefore save $\sim$5000~days or JWST during a 200 planet survey and $\sim$20,000~days for ARIEL's 1000-planet survey (Fig.~\ref{fig:timesaved_1000}, Bottom Left). The ``jumps'' in the left plots in Figure~\ref{fig:timesaved_1000} represent planets that have large ephermeris unceratinties listed on the NASA Exoplanet Archive. Therefore, a single observation by one 6-inch telescope can greatly decrease these uncertainties, resulting in a large savings of time. If we remove the {{two (top row) and three (bottom row)}} exoplanets with the largest ephemeris uncertainties from the target list, we still find that a single 6-inch telescope can save future missions $\sim$5~days {{while a network of 16 6-inch telescopes can save future missions $\sim$80~days}} (Fig.~\ref{fig:timesaved_1000} right column).}}

\begin{figure*}
\centering
  \begin{minipage}[b]{0.45\linewidth}
    \centering
    \includegraphics[width=1\columnwidth]{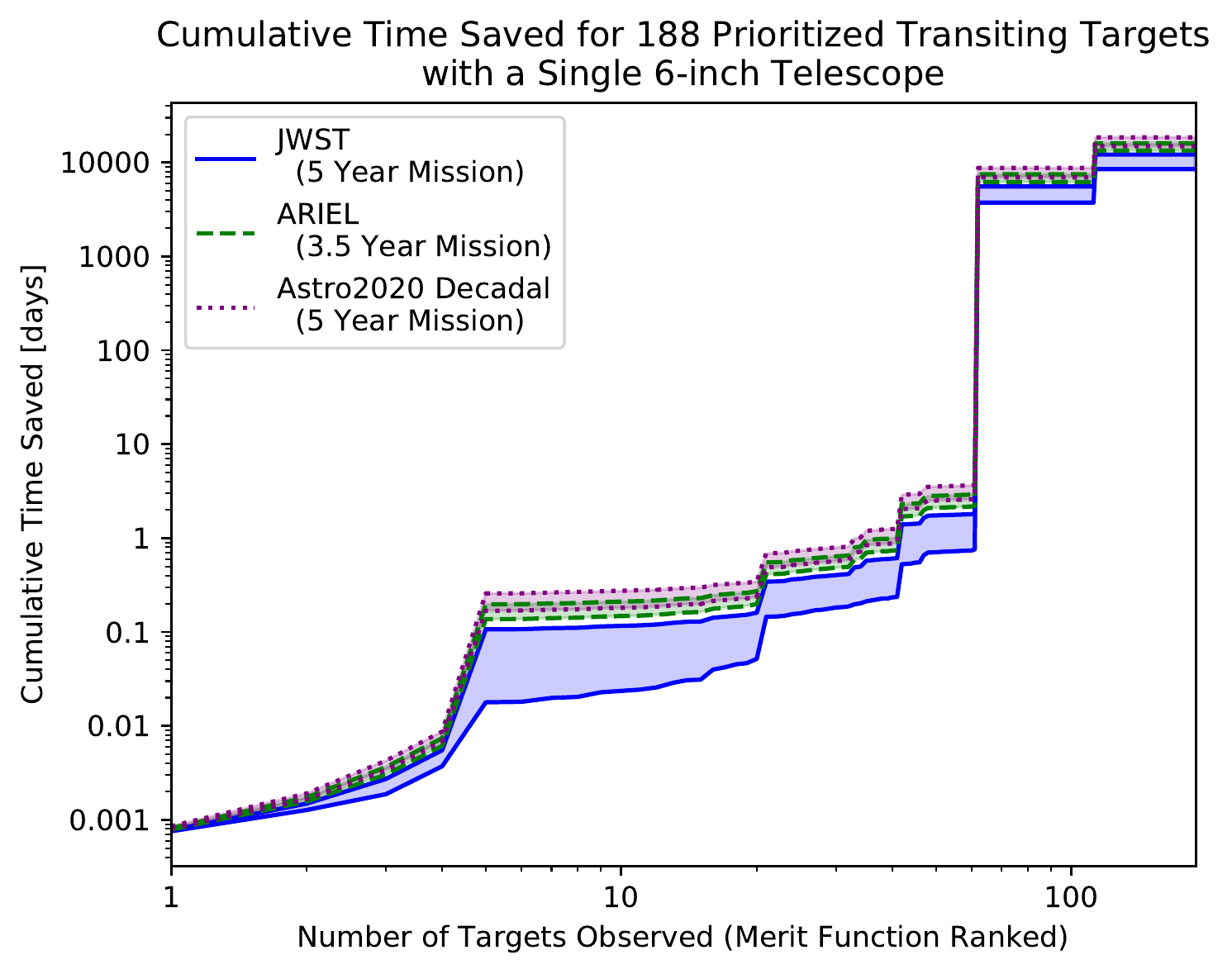}
  \end{minipage}
  \begin{minipage}[b]{0.45\linewidth}
    \centering
    \includegraphics[width=1\columnwidth]{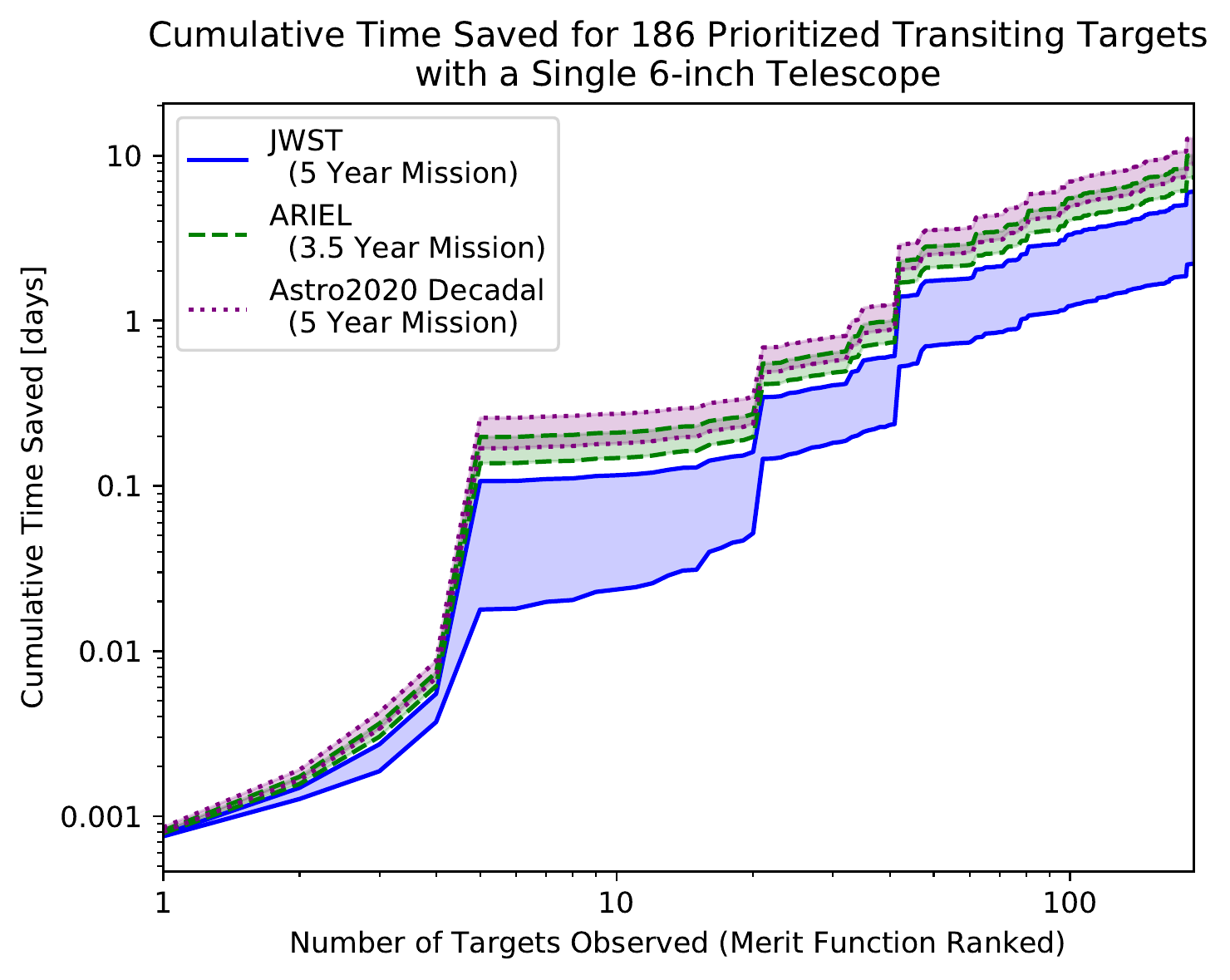}
  \end{minipage} 
  \begin{minipage}[b]{0.45\linewidth}
    \centering
    \includegraphics[width=1\columnwidth]{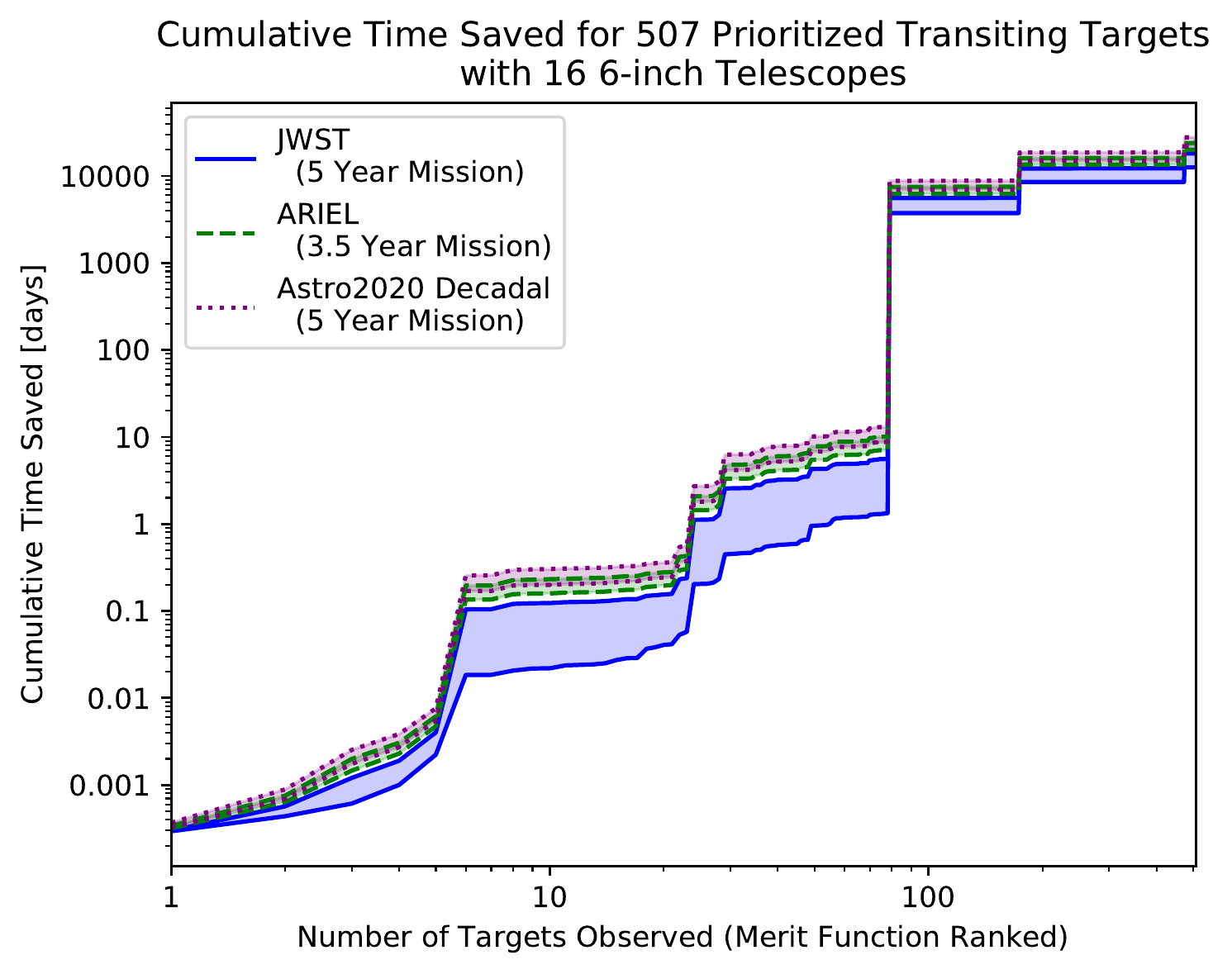}
  \end{minipage}
  \begin{minipage}[b]{0.45\linewidth}
    \centering
    \includegraphics[width=1\columnwidth]{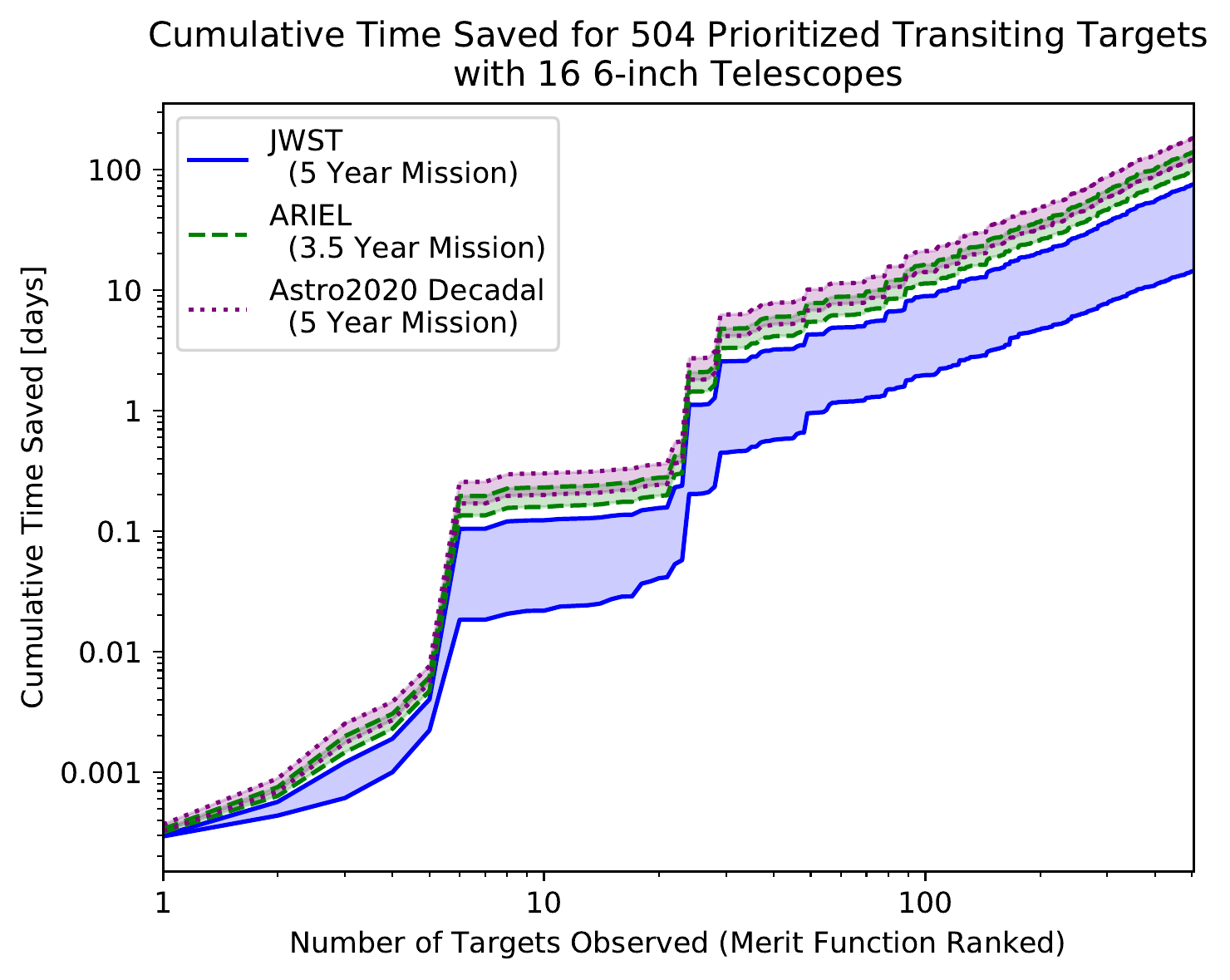}
  \end{minipage} 
\caption{The cumulative time saved with single-visit transit maintenance on a single 6-inch telescope {{(top row) and a network of 16 6-inch telescopes (bottom row)}} for JWST (solid blue lines), ARIEL (dashed green lines), and an Astro2020 Decadal Mission (dotted purple lines). {{The targets are drawn from}} the {{1000}} targets best-suited for transit spectroscopy \citep[targets are relatively ranked by their scale height and host star magnitude {{and are a mix of the currently-known exoplanets and those predicted to be discovered with TESS}};][]{zellem17} {{and are then filtered for 3$\sigma$ detections of the transit.}} The shaded regions show the time saved over the entire lifetime of each mission. Transit maintenance with even a single 6-inch telescope can save {{on the order of $\sim$10,000 days for both JWST and ARIEL {{while a network of 16 6-inch telescopes could save on the order $\sim$5000~days and $\sim$20,000~days, respectively.}} {{The comparatively large ``jumps'' in the data ({{left column}}) are where the published transit times have large uncertainties, therefore a single 6-inch telescopic observation can significantly refine the times for these targets. When the {{two (top row) and three (bottom row)}} targets with the largest uncertainties are removed from this target list ({{right column}}), we more conservatively find that the time saved {{by a single 6-inch telescope}} is on the order of $\sim$5~days, {{while the time saved by a network of 16 6-inch telescopes is on the order of $\sim$80~days.}}}}  }}}
\label{fig:timesaved_1000}
\end{figure*}

\subsubsection{Measuring Planetary Masses and Discovering Additional Companions}
If citizen scientists continue to monitor targets over a large baseline, they could enable the measurement of transit timing variations (TTVs), the discovery of new exoplanets, and aid in the calculation of the masses of known planets \citep[e.g.,][]{agol05, holmanmurray05, dalba16}. Past detections of transit timing variations have primarily come from the Kepler mission due to its long-term photometric monitoring of the same field of stars for 4 years \citep{borucki10}. One large constraint on the search for timing variations is TESS' significantly shorter baseline compared to Kepler. Only select areas of the sky will have data for more than 27.4~days and, based on prior studies, $\sim$20 transits are generally needed to ensure a unique orbit solution when interpretting TTVs \citep{nesvorny08}. This requirement limits the TTV search to planets in compact orbits if they have only one sector of TESS data. {{For multi-planet systems, the average period of an inner planet is $\sim$10 days \citep{mulders18} suggesting follow-up is necessary even to achieve three transit measurements to confirm the planet let alone detect a TTV.}}
 
One of the primary objectives for the TESS Followup Program (TFOP) is to help achieve TESS's Level 1 requirement to measure masses for 50 planets smaller than 4~$R_{\oplus}$\footnote{https://tess.mit.edu/followup/}. We estimate that this radius constraint corresponds to a mass of 15$^{+7}_{-5}$~$M_{\oplus}$ based on a mass-radius relation derived from a Gaussian Process \citep[using the \texttt{gpy} Python package;][]{gpy2014} of planets with measured masses (M$_{s}<$2~M$_{\odot}$), radii, and orbital periods (1$\le$P$\le$100~days) from the NASA Exoplanet Archive \citep{akeson13}. The Gaussian Process uses 524 exoplanets to determine a mass-radius relationship and uncertainty \citep{pearson19}. We estimate that a 15~M$_{\oplus}$ planet with a short orbital period corresponds to a radial velocity semi-amplitude signal on the order of $\sim$few m/s, which currently pushes the limits of modern instruments \citep[e.g.,][]{fischer16, delisle18}.
 
However, a network of smaller telescopes can provide additional transit measurements to contribute to the TFOP and community efforts {{\citep[e.g., KOINetwork;][]{vonessen18, freudenthal19}}} to determine planetary masses via TTV measurements (Figs.~\ref{fig:ttv_grid} and \ref{fig:rv_grid_masked}), thus helping to achieve one of TESS's primary objectives. Therefore, even if all of the transiting exoplanets observed by TESS have ephemeris precisions high enough where they are not at risk of becoming stale, a network of small telescopes are still necessary to establish long baselines to search for TTVs or additional transiting planets.

{{The fraction of false positives planet candidates can be reduced by requiring at least three self-consistent transit detections instead of a single one \citep{petigura13, burke14, rowe15, coughlin16}. Even if the mutual inclination of multi-planet systems yield non-transiting companions their orbit and mass can still be constrained through TTV measurements of at least one transiting planet \citep[e.g.,][]{pearson19}.}}

\begin{figure}
\centering
\includegraphics[width=1\columnwidth]{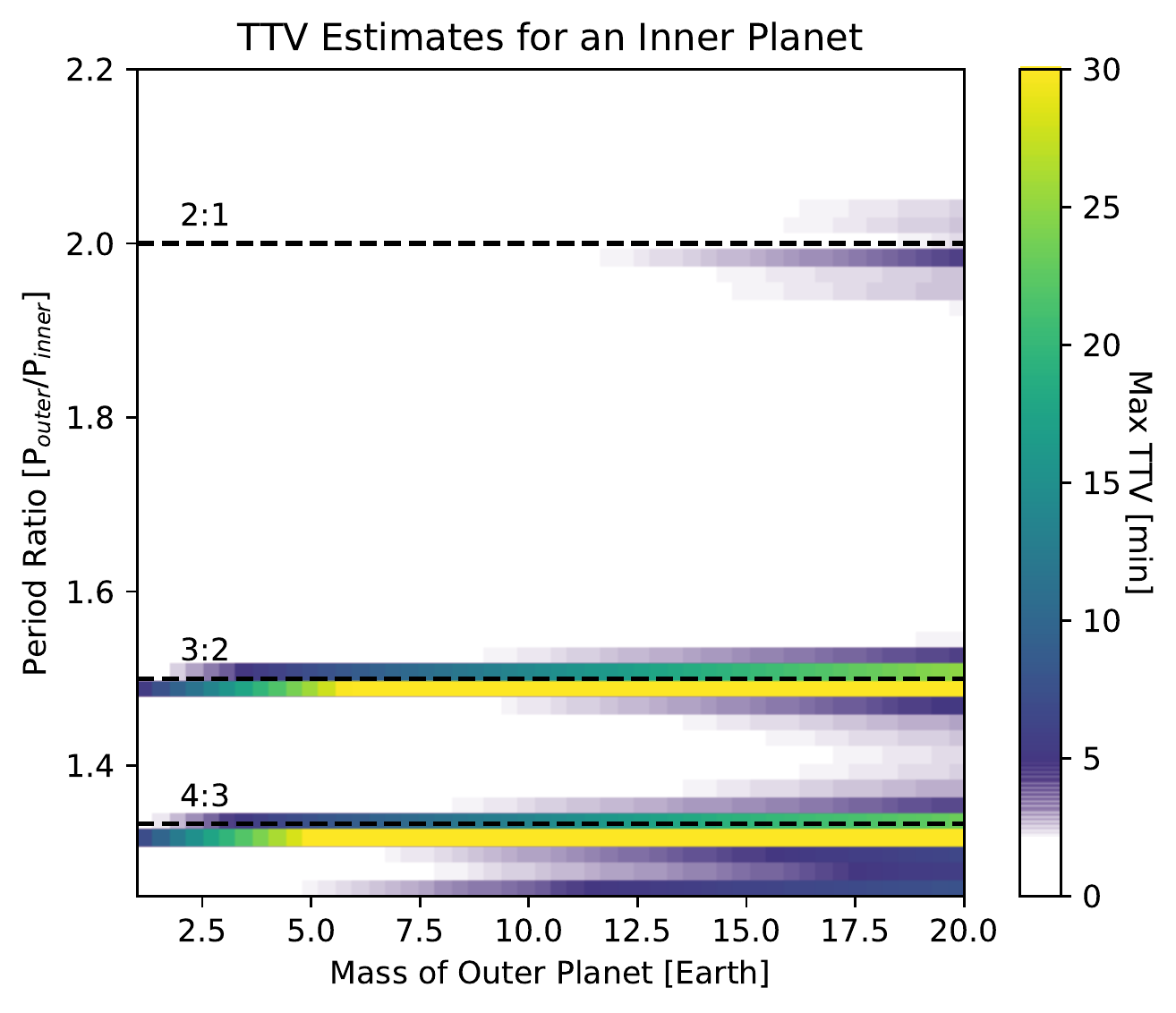}
\caption{Transit timing variations of an inner most planet estimated with various companion planets. The N-body simulations\textsuperscript{$\dagger$} use a stellar mass of 1.26~M$_{\odot}$ and the innermost planet is the Representative TOI Planet with an orbital period of 3.85288~days with a mass of 0.23~M$_{Jupiter}$. The fixed parameters are median values computed from the TESS TOI catalog. The second, outer companion planet in the system has varying masses and orbital periods. Each grid point is an N-body simulation run for 180~days, with the colors being mapped to the maximum perturbation after performing a linear fit to the transit times of the inner planet.}
\small\textsuperscript{$\dagger$} https://github.com/pearsonkyle/Nbody-AI
\label{fig:ttv_grid}
\end{figure}

\begin{figure}
\centering
\includegraphics[width=1\columnwidth]{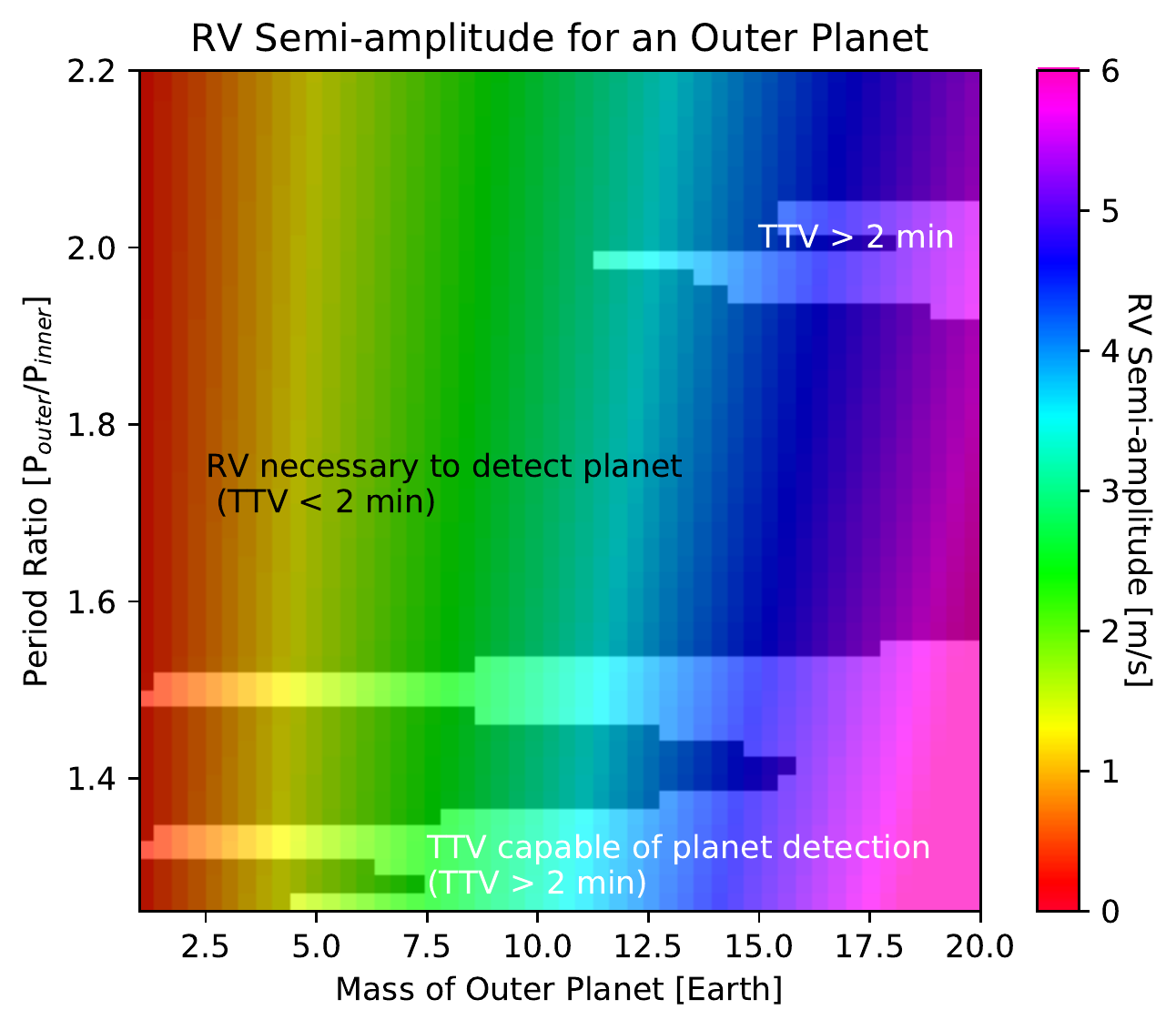}
\caption{Estimated radial velocity semi-amplitude signals for an outer planet with 1--20~M$_{\oplus}$ and an orbital period 1.25--2.2$\times$ the length of its inner planet, here assumed to be the Representative TOI Planet (3.85288~days). Currently, Earth-sized planets are difficult to detect with radial velocity instruments since the signals are only a $\sim$few m/s. However, it is possible to detect Earth-sized planets using TTV measurements but only at certain orbit configurations, particularly those near close orbital resonances. The darker region is based on Figure~\ref{fig:ttv_grid} and masks a portion of the parameter space where TTVs are $>$2~minutes and thus could be accessible with smaller ground-based telescopes. Unmasked regions are where the TTV signal is $<$2~minutes, thereby requiring radial velocity measurements.}
\label{fig:rv_grid_masked}
\end{figure}
 
The transit timing variation of existing multi-planet systems is estimated and reported in Table \ref{tab:ttvestimates}. This subset of planets is ranked by the maximum TTV amplitude calculated from 20 simulated transit epochs. Only planets in the NASA Exoplanet Archive \citep{akeson13} with periods less than 60~days and host stars $<$2~M$_{\odot}$ are used. These targets could benefit from TESS high cadence observations or follow-up observations in order to constrain the parameters of the perturbing planet. For example, a single 6-inch telescope could achieve a mid-transit precision $\Delta T_{0}$ of 1.02~minutes for even a 11.289~V-mag star (Fig.~\ref{tab:MO_14}), allowing the opportunity for even relatively-small telescopes to measure TTVs (particularly those listed in Table~\ref{tab:ttvestimates}) to high significance. {{Multi-planet systems are common with somewhere between $\sim$50-70$\%$ of exoplanets being in a multi-planet system. The other fraction could be single planet systems or multi-planet systems with higher mutual inclinations. Not all multi-planet architectures yield observable TTVs though (Fig.~\ref{fig:ttv_grid}). Based on Kepler data, the spacing between adjacent planets follows a wide distribution with a peak at an orbital period ratio of 1.8 \citep{mulders19}. Observational biases can reduce this ratio since it's easier to measure a TTV for a more compact orbit than for those far apart.}}

\begin{table*}
  \caption{TTV Estimates for Exoplanet Systems That Are Ideal for TTV Measurements with Small Ground-based Observatories}
  \label{tab:ttvestimates}
  \begin{center}
    \leavevmode
    \resizebox{\textwidth}{!}{%
    \begin{tabular}{l|l|l|l|l| l|l|l|l} \hline \hline
    Planet & Planet Mass & Period & Eccentricity & Stellar Mass  & V-Mag
& TTV & Discovery & Parameter \\
     & [M$_{\oplus}$] & [day] & & [M$_{\odot}$] &  & [min] & Paper  & Reference \\
    \hline
Kepler-30 b & 11.44 & 29.334 & 0.046 & 0.99 & 15.403 & 746.9 &
\cite{Fabrycky2012} & \cite{Sanchi-Ojeda2012} \\
KOI-142 b & 8.58 & 10.954 & 0.026 & 0.96 & 13.11 & 189.5 &
\cite{Nes2013} & \cite{Nes2013} \\
K2-266 e & 8.3 & 19.482 & 0.009 & 0.69 & 11.808 & 156.0 &
\cite{Rodriguez2018} & \cite{Rodriguez2018} \\
Kepler-9 c & 54.35 & 38.91 & 0.005 & 1.07 & 13.803 & 127.8 &
\cite{Holman2010} & \cite{Holman2010} \\
K2-266 d & 9.4 & 14.697 & 0.007 & 0.69 & 11.808 & 122.8 &
\cite{Rodriguez2018} & \cite{Rodriguez2018} \\
Kepler-223 c & 5.1 & 9.846 & 0.003 & 1.12 & 15.344 & 77.0 &
\cite{Rowe2014} & \cite{Mills2016} \\
Kepler-9 b & 80.09 & 19.24 & 0.011 & 1.07 & 13.803 & 69.9 &
\cite{Holman2010} & \cite{Holman2010} \\
Kepler-223 b & 7.4 & 7.384 & 0.008 & 1.12 & 15.344 & 51.5 &
\cite{Rowe2014} & \cite{Mills2016} \\
TRAPPIST-1 f & 0.68 & 9.207 & 0.004 & 0.08 & 18.8 & 32.3 &
\cite{Gillon2016} & \cite{Gillon2017} \\
Kepler-36 b & 4.45 & 13.84 & 0.002 & 1.07 & 11.866 & 30.8 &
\cite{Carter2012} & \cite{Carter2012} \\
Kepler-11 f & 1.91 & 46.689 & 0.001 & 0.96 & 14.2 & 26.1 &
\cite{Lissauer2011} & \cite{Lissauer2013} \\
K2-285 d & 6.5 & 10.456 & 0.003 & 0.83 & 12.03 & 21.0 & \cite{Palle2018}
& \cite{Palle2018} \\
Kepler-30 c & 638.81 & 60.323 & 0.0 & 0.99 & 15.403 & 20.2 &
\cite{Fabrycky2012} & \cite{Sanchi-Ojeda2012} \\
K2-32 d & 10.3 & 31.715 & 0.002 & 0.86 & 12.31 & 19.7 & \cite{Dai2016} &
\cite{Petigura2017} \\
TRAPPIST-1 g & 1.34 & 12.353 & 0.004 & 0.08 & 18.8 & 19.7 &
\cite{Gillon2016} & \cite{Gillon2017} \\
Kepler-36 c & 7.95 & 16.239 & 0.001 & 1.07 & 11.866 & 19.6 &
\cite{Carter2012} & \cite{Carter2012} \\
Kepler-117 b & 29.87 & 18.796 & 0.003 & 1.13 & 14.274 & 16.7 &
\cite{Rowe2014} & \cite{Bruno2015} \\
Kepler-223 e & 4.8 & 19.726 & 0.005 & 1.12 & 15.344 & 15.7 &
\cite{Rowe2014} & \cite{Mills2016} \\
Kepler-79 e & 4.13 & 81.066 & 0.0 & 1.17 & 13.914 & 12.9 &
\cite{Rowe2014} & \cite{Jontof2014} \\
K2-24 c & 15.4 & 42.339 & 0.001 & 1.07 & 11.07 & 11.6 &
\cite{Petigura2016} & \cite{Petigura2018} \\
K2-24 b & 19.0 & 20.89 & 0.003 & 1.07 & 11.07 & 11.5 &
\cite{Petigura2016} & \cite{Petigura2018} \\
K2-32 c & 12.1 & 20.66 & 0.001 & 0.86 & 12.31 & 11.4 & \cite{Dai2016} &
\cite{Petigura2017} \\
KOI-94 c & 15.57 & 10.424 & 0.003 & 1.28 & 12.205 & 10.9 &
\cite{Weiss2013} & \cite{Weiss2013} \\
Kepler-11 e & 7.95 & 32.0 & 0.001 & 0.96 & 14.2 & 10.8 &
\cite{Lissauer2011} & \cite{Lissauer2013} \\
Kepler-11 d & 7.31 & 22.684 & 0.001 & 0.96 & 14.2 & 9.7 &
\cite{Lissauer2011} & \cite{Lissauer2013} \\
Kepler-79 c & 6.04 & 27.403 & 0.001 & 1.17 & 13.914 & 7.7 &
\cite{Xie2013} & \cite{Jontof2014} \\
Kepler-79 d & 6.04 & 52.09 & 0.0 & 1.17 & 13.914 & 7.6 & \cite{Rowe2014}
& \cite{Jontof2014} \\
K2-285 e & 10.7 & 14.763 & 0.001 & 0.83 & 12.03 & 6.5 & \cite{Palle2018}
& \cite{Palle2018} \\
TRAPPIST-1 d & 0.41 & 4.05 & 0.005 & 0.08 & 18.8 & 6.5 &
\cite{Gillon2016} & \cite{Gillon2017} \\
Kepler-18 c & 17.16 & 7.642 & 0.002 & 0.97 & 13.549 & 6.1 &
\cite{Cochran2011} & \cite{Cochran2011} \\
K2-285 c & 15.68 & 7.138 & 0.001 & 0.83 & 12.03 & 6.1 & \cite{Palle2018}
& \cite{Palle2018} \\
Kepler-11 b & 1.91 & 10.304 & 0.001 & 0.96 & 14.2 & 5.7 &
\cite{Lissauer2011} & \cite{Lissauer2013} \\
Kepler-11 c & 2.86 & 13.024 & 0.001 & 0.96 & 14.2 & 5.1 &
\cite{Lissauer2011} & \cite{Lissauer2013} \\
Kepler-51 b & 2.22 & 45.154 & 0.0 & 1.04 & 14.669 & 4.7 &
\cite{Steffen2013} & \cite{Masuda2014} \\
Kepler-79 b & 10.9 & 13.484 & 0.001 & 1.17 & 13.914 & 4.5 &
\cite{Xie2013} & \cite{Jontof2014} \\
Kepler-18 d & 16.53 & 14.859 & 0.0 & 0.97 & 13.549 & 4.0 &
\cite{Cochran2011} & \cite{Cochran2011} \\
Kepler-223 d & 8.0 & 14.789 & 0.001 & 1.12 & 15.344 & 3.8 &
\cite{Rowe2014} & \cite{Mills2016} \\
KOI-94 d & 106.15 & 22.343 & 0.0 & 1.28 & 12.205 & 2.7 &
\cite{Weiss2013} & \cite{Weiss2013} \\
K2-285 b & 9.68 & 3.472 & 0.002 & 0.83 & 12.03 & 2.6 & \cite{Palle2018}
& \cite{Palle2018} \\
K2-266 c & 0.29 & 7.814 & 0.001 & 0.69 & 11.808 & 2.3 &
\cite{Rodriguez2018} & \cite{Rodriguez2018} \\
TRAPPIST-1 e & 0.62 & 6.1 & 0.004 & 0.08 & 18.8 & 2.1 &
\cite{Gillon2016} & \cite{Gillon2017} \\
Kepler-101 c & 3.18 & 6.03 & 0.001 & 1.17 & 13.8 & 2.0 & \cite{Rowe2014}
& \cite{Bonomo2014} \\
    \end{tabular}}
  \end{center}
\end{table*}

\subsubsection{Helping Confirm Long-period Planets}
This network of observers could also support the TFOP effort to continue to monitor TESS fields, particularly those which will be observed by TESS for only 27.4~days, helping to confirm planets with long orbital periods (P$>$27.4~days) which would have only one TESS-observed transit. {{A citizen scientist could be directed to a particular field which contains both a known transiting planet in need of maintenance and a potential long-period planet, since small telescopes are capable of having large fields-of-view (e.g., a 6-inch MicroObservatory image is 0.94$\degr$ by 0.72$\degr$, 3.4$\times$ the angular area of the full Moon). Thus an observer could both perform ephemeris maintenance on the known transit planet while simultaneously monitoring the other host star for potential transits.}}

\subsubsection{Spatially-resolving Stellar Blends}
TESS' pixel scale is 21~arcsec, raising the possibility that an object that appears to be a single star in a TESS image is rather a blend of multiple stars (Fig.~\ref{fig:stellarblends}). Therefore due to the dilution by additional stars within the pixel, an observed transit would appear shallower than it truly is, thereby altering the physical interpretation of the planet \citep[e.g.,][]{crossfield12c, bergfors13, stevenson14_gemini, ciardi15, collins18, colon18}. While spatial resolution has typically been conducted on larger platforms, smaller telescopes could probe fields at $\gtrsim$5$\times$ higher spatial resolution than TESS, assuming seeing-limited observations on the order of 1--4~arcsec. 

{{But even achieving spatial resolutions that are plate-scale limited (rather than seeing-limited) are still useful for identifying stellar blends. For example in the top panel of Figure~\ref{fig:stellarblends}, we present a subset of a single image by a 6-inch MicroObservatory telescope of the stars TYC 3280-846-1 and TYC 3280-697-1, which have V-magnitudes of 9.83 and 10.68, respectively \citep{wenger00}. Despite the image not being seeing-limited (the plate scale is 5.21 arcsec/pixel) and imperfect tracking over the course of the exposure (causing the oblong shape of the stars), the MicroObservatory is able to distinctly resolve these two stars. However, when we bin this image to the spatial resolution of TESS (21 arcsec/pixel), the two stars, which are separated by 21.5~arcsec, blend together and appear as a single source. Assuming that TYC 3280-846-1 is similar in size to the Sun and hosts a Jupiter-sized transiting exoplanet, the MicroObservatory would observe a 1\% transit depth. However since TESS would blend the two stars, TESS would observe a transit depth of 0.3\% \citep[calculated using Eqn.~4 in][]{ciardi15}, and would potentially incorrectly classify this hot Jupiter as a sub-Neptune.}}

\begin{figure}
\centering
\includegraphics[width=1\columnwidth]{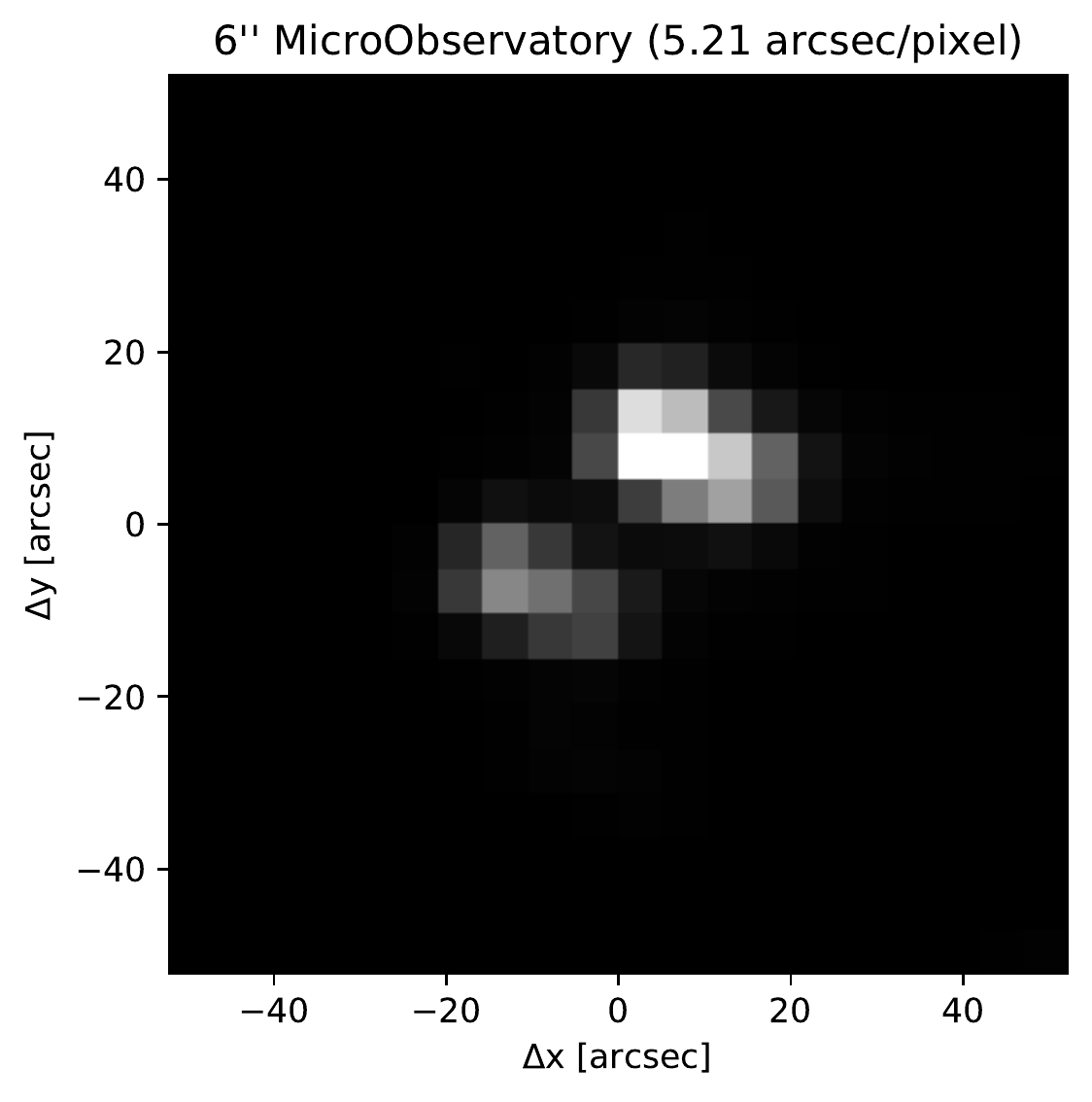}
\includegraphics[width=1\columnwidth]{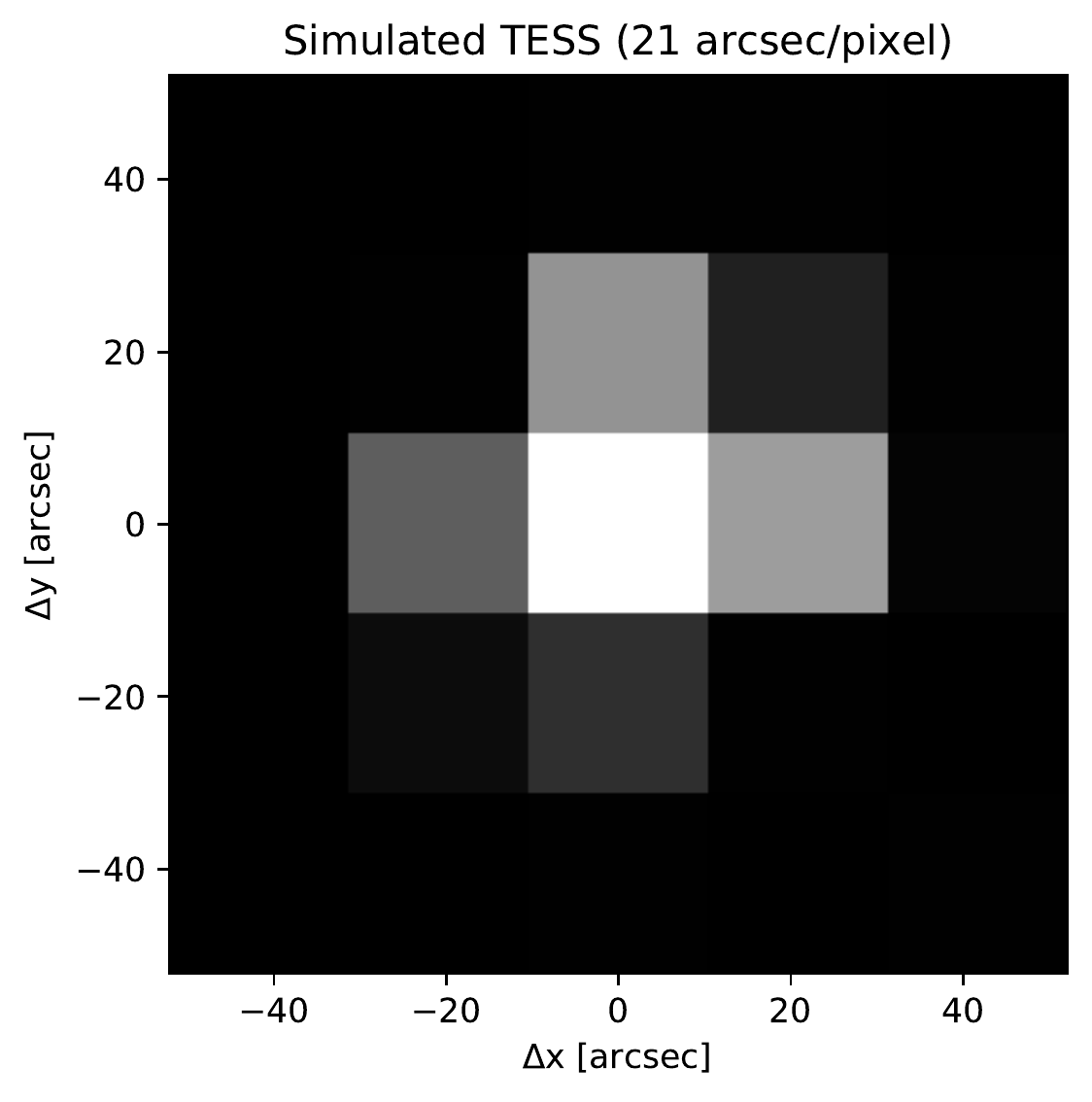}
\caption{{{This subset of a 6-inch MicroObservatory image ($Top$) clearly distinguishes between two stars, which are separated by 21.5~arcsec. However, due to TESS' relatively large plate scale (21~arcsec/pixel), these two stars blend together and appear as one source ($Bottom$). If a planet were to be discovered orbiting either of these stars, TESS would observe a diluted transit, thereby misidentifying the planet's radius and therefore its type.}}}
\label{fig:stellarblends}
\end{figure}

\subsubsection{Characterizing Epoch-to-epoch Stellar Variability}
{{If sufficiently large, stellar activity (spots and/or plages) can change the observed transit signal, thereby altering the retrieved physical properties of the planet itself \citep{alonso08, pont08, silvavalio08, czesla09, wolter09, agol10, berta11, carter11, desert11, sing11, narita13, fraine14, mccullough14, oshagh14, barstow15, damasso15, zellem15, zellem17, rackham17, rackham18, zzhang18}. Epoch-to-epoch host star variability has been monitored and quantified by long-term ground-based photometric observations. For example, data from the Tennessee State University Celestron-14 (0.36-m) AIT at Fairborn Observatory in southern Arizona \citep[e.g.,][]{henry99, eaton03} has been used to quantify the time-varying activity of the hot Jupiter host star HD~189733b. These data have then been used to confirm that the host star's activity does not impact the planet's phase curve \citep{knutson12} nor its multi-epoch eclipse measurements \citep{kilpatrick19} as observed with Spitzer/IRAC. While it is estimated that epoch-to-epoch stellar variations will not influence the observed planetary signal in most cases, if the observations are at sufficiently-high precision (as is typically the case for bright targets), the transit depth is large, and the star is active enough, stellar variability could impact the observed transit signal \citep[see Fig.~3 in][]{zellem17}.}} {{Epoch-to-epoch stellar activity $\gtrapprox$0.2\% could begin to effect high precision observations (e.g., with JWST) of transit depths on the order of $\sim$1\% \citep[see Fig.~3 in][]{zellem17}.}}

{{Fortunately, small telescopes have the capability of high precision observations even for relatively dim stars. From Figure~\ref{fig:MO_survey} $Top$ we estimate that for a 11.3~V-mag star, a 6-inch telescope could achieve a per-minute precision of 0.67\%$\pm$0.12\%. {{Note that this value and its associated uncertainty is calculated from {{four total observations by the same 6-inch telescope}}: {{two independent observations}} of HAT-P-20b and {{two independent observations}} of HAT-P-32b {{taken over four different nights}} (Table~\ref{tab:MO_14}) and thus takes into account night-to-night variations.}} A 6-inch telescope could therefore measure 2.1\% variability to 3$\sigma$ on a 11.3~V-mag star in only a minute of observing. {{Using the observations of WASP-52b taken over four separate nights by the same 6-inch telescope, we estimate that a single 6-inch telescope can achieve a per-minute precision of 0.95\%$\pm$0.21\% for a 12 V-mag star; similarly using the observations of WASP-10b taken over four separate nights by the same 6-inch telescope, we estimate a per-minute precision of 1.08\%$\pm$0.08\% for a 12.7 V-mag star.}} Thus, an amateur astronomer could monitor exoplanet host stars for time-varying flux suggesting stellar variability before or after their transit observations and could improve their measured precision by binning over time.

\section{Results and Discussion}
A network of small telescopes {{(here defined as $\gtrsim${{16}} $\geq$6-inch telescopes; see Section~\ref{sec:network} and Table~\ref{tab:MO_14})}} can crucially aid future transiting exoplanet observations (with, e.g., HST JWST, ARIEL, Astro2020 Decadal, and larger ground-based platforms; Figs.~\ref{fig:repplanet}--\ref{fig:timesaved_1000}) by providing up-to-date and accurate transit ephemerides of planets discovered with ground- and space-based surveys. Transit maintenance with even a single 6-inch telescope can save up to $\sim$10,000 days for both JWST and ARIEL {{whereas a network of 16 6-inch telescopes could save on the order $\sim$5000~days for a 200-planet JWST survey and $\sim$20,000~days for a 1000-planet ARIEL survey}} (Fig.~\ref{fig:timesaved_1000}). Thus while transit ephemeris uncertainties do not prevent the accomplishment of any of these missions, precise mid-transit time predictions would enable each mission to act more efficiently and potentially achieve even more science.

This network, in addition to helping confirm long-period ($P>27$~days) planets and spatially-resolving stellar blends, could use transit timing variations in multi-planet systems to measure planetary masses, thereby contributing to the achievement of one of TESS' Level 1 Requirements, and discover new planets. Our N-body simulations estimate that TTVs could be as large as 30~minutes or more (0.02~days), which can be achieved with high significance with even a single 6-inch ground-based telescope \citep[Figs.~\ref{fig:MOscopes1} and \ref{fig:MOscopes2} and Table~\ref{tab:MO_14}; see also][]{fowler19}.

{{Lastly, this network of small telescopes could monitor exoplanet host stars for variability due to, e.g., spots and plages. Epoch-to-epoch variability on the order of even 0.2\% can effect JWST observations of exoplanets with a $\sim$1\% transit depth \citep{zellem17}. We estimate that a single 6-inch telescope could measure the variability of a 11.3 V-mag star with a per-minute precision of 0.67\%$\pm$0.12\%.}}

\section{Ramifications for Other Platforms and Surveys}
While transit maintenance could be conducted with TESS via an extended mission \citep{bouma17, huang18, dalba19} or professional observatories \citep[e.g.,][]{benneke17}, a large ground-based collaboration of smaller telescopes ($\le$1-m) could follow-up hundreds of bright targets with large transit depths. Such a network would free-up precious time on TESS, other missions, and larger ground-based telescopes that could be instead dedicated to observing new fields (e.g., the ecliptic), comparatively under-sampled fields (e.g., the 27.4-day fields), or areas of the sky with high-priority targets (e.g., a potentially-habitable planet orbiting a Solar-type or dim M-dwarf star).

Given that this small telescope network could keep the ephemerides fresh for transiting exoplanets {{ with large transit depths and bright host stars}}, it would also be complementary to current transit {{discovery}} surveys and efforts, such as CoRoT \citep{auvergne09}, HATNet \citep{bakos07hatnet, bakos13}, KELT \citep{pepper03, pepper07, pepper12}, Kepler \citep{borucki10}, K2 \citep{howell14}, MASCARA \citep{talens17}, NGTS \citep{wheatley18}, PANOPTES \citep{guyon14}, QES \citep{alsubai11}, TERMS \citep{kane18}, TrES \citep{alonso04}, WASP \citep{pollacco06}, and XO \citep{mccullough05}. In addition to ephemerides maintenance, this small telescope network could free up the time for these other surveys to continue their own discovery {{programs}} or {{confirm}} high-priority targets (e.g., potentially habitable planets around dim M-dwarf stars). The small telescope citizen science network could also join in the efforts of these other surveys to increase the total collection area and world-wide coverage, enabling the pursuit of small and/or dim transits or long-duration transits.

The small telescope network advocated for here would also be complementary to the current on-going TESS Follow-Up Observation Program\footnote{https://heasarc.gsfc.nasa.gov/docs/tess/followup.html} (TFOP), which is eliminating TESS candidate false positives and confirming true TESS planetary systems. The TFOP utilizes ground-based time series photometry to eliminate blended eclipsing binaries, spectroscopy to obtain stellar parameters, high resolution imaging to identify potential binary systems, and precision radial velocities to determine the masses of the planets. The TFOP effort will help the community identify some of the best TESS planets for atmospheric characterization from the ground or with space facilities like JWST and ARIEL.

{{In addition, the small telescope citizen science network would monitor $all$ known transiting exoplanets {{accessible to them (e.g., large planets with bright host stars;}} Tables~\ref{tab:deltaTestimates_JWST}--\ref{tab:deltaTestimates_Astro2020}) to ensure efficient follow-up observations with large platforms, in addition to the other science cases described above.}} Without continued follow-up, the ephemerides of {{all the transiting}} planets will eventually grow stale \citep[see][for a detailed study on TESS targets going stale]{dragomir19}. The work advocated for here {{complements}} the work of the TFOP–-so that the mid-transit times will be known sufficiently for efficient future observations. The TFOP is necessary to identify the true planetary systems detected by TESS while a small telescope network would maintain the quality of the ephemerides to ensure the suitability and efficiency of future characterization observations with JWST, ARIEL, and beyond.

Thus, the ground-based network of smaller ($\le$1~m) telescopes described here would be complementary to current and future ground- and space-based exoplanetary efforts. {{We find that a network of 16 6-inch telescopes could measure to 3$\sigma$ the transits of 507 of the 1000 exoplanets expected to be spectroscopically characterized by future missions (e.g., ARIEL) and thus save up to $\sim$20,000~days of observational overhead.}} This network would {{also}} allow TESS to revisit comparatively less-sampled fields (27.4-day fields) or even new fields (e.g., near the ecliptic). While not impacting the achievement of the prime missions of JWST, ARIEL, or an Astro2020 Decadal, this network would also help ensure the efficient use of these great observatories and potentially enable more science return. This network would also build upon the legacies of Kepler, K2, CoRoT, and other ground-based transit surveys by keeping their ephemerides fresh and extending their measurement baselines to probe for TTVs.

\section{Conclusions}
TESS will revolutionize the field of exoplanet science by providing 10,000+ targets \citep{barclay18} for future transiting exoplanet atmospheric characterization missions (e.g., JWST, ARIEL, and an Astro2020 Decadal Mission, such as HabEx, LUVOIR, or Origins). However, the uncertainty in predicting the time of the next predicted transit for any one of these targets can be significant---larger than the transit duration itself---thereby limiting the efficient use of these large observatories.

We find here that follow-up observations with small ($\le$1-m) telescopes can provide strong constraints on estimating future transits and increase the observational efficiency of these great observatories. We therefore advocate the establishment of a network of small telescopes operated by citizen scientists to provide accurate transit times to the professional astronomer community to help plan upcoming observations. One such proposed network is an open collaboration leveraging small telescopes operated by citizen scientists to provide accurate ephemerides, as described in detail in \citet{zellem19, zellem20} , and another is the ExoClock program \footnote{https://www.exoclock.space}.

\section*{Acknowledgments}
Part of the research was carried out at the Jet Propulsion Laboratory, California Institute of Technology, under contract with the National Aeronautics and Space Administration. Copyright 2020. All rights reserved.

We thank the JPL Exoplanet Science Initiative for partial support of this work.

This material is based upon work supported by NASA under cooperative agreement award number NNX16AC65A.  Any opinions, findings, and conclusions or recommendations expressed in this material are those of the authors and do not necessarily reflect the views of the National Aeronautics and Space Administration.

This research has made use of the NASA Exoplanet Archive, which is operated by the California Institute of Technology, under contract with the National Aeronautics and Space Administration under the Exoplanet Exploration Program.

{{This research has made use of the SIMBAD database, operated at CDS, Strasbourg, France.}}

RTZ would like to thank Padi Boyd, Knicole Colon, Sam Halverson, Stella Kafka, Tiffany Kataria, David Latham, and Kevin Stevenson for their helpful discussions.

D. Dragomir acknowledges support provided  by NASA through Hubble Fellowship grant HST-HF2-51372.001-A awarded by the Space Telescope Science Institute, which is operated by the Association of Universities for Research in Astronomy, Inc., for NASA, under contract NAS5-26555.

\bibliography{references}

\end{document}